\documentclass[fleqn,usenatbib]{mnras}

\usepackage{anyfontsize} 
\usepackage{textgreek}
\usepackage{xspace}
\usepackage{amsmath}	
\usepackage{xargs}
\usepackage{multirow}
\usepackage{txfonts}
\usepackage{hyperref}
\usepackage{ulem}
\usepackage{xargs}

\usepackage[LGR,T1]{fontenc}

\DeclareRobustCommand{\VAN}[3]{#2}
\let\VANthebibliography\thebibliography
\def\thebibliography{\DeclareRobustCommand{\VAN}[3]{##3}\VANthebibliography}

\usepackage{graphicx}	
\usepackage{amsmath}	
\usepackage{subcaption}
\usepackage{makecell}

\newcommand{\msun}{\ensuremath{\rm M_\odot}\xspace} 
\newcommand{\kms}{\ensuremath{\rm km\,s^{-1}}\xspace} 
\newcommand{\jwst}{\textit{JWST}\xspace}
\newcommandx{\forbiddenEL}[6][1=O,2=III,3=,4=,5=,6=]{\text{[{#1}\,{\sc{#2}}]{#3}{#4}{#5}{#6}}\xspace}

\newcommand{\OIII}{\forbiddenEL[O][iii]}
\newcommandx{\OIIIL}[1][1=5007]{\forbiddenEL[O][iii][\textlambda][#1]}

\newcommand{\Halpha}{\text{H\,\textalpha}\xspace}

\title[Low-mass black holes at high $z$]{JADES reveals a large population of low mass black holes at high redshift}

\author[S. Geris et al.]{\parbox{\textwidth}{
Sophia Geris,$^{1,2}$\thanks{E-mail: ssjg2@cam.ac.uk}
Roberto Maiolino,$^{1,2,3}$
Yuki Isobe,$^{1,2,4}$
Jan Scholtz,$^{1,2}$
Francesco D'Eugenio$^{1,2}$,
Xihan Ji$^{1,2}$,
Ignas Juodžbalis$^{1,2}$,
Charlotte Simmonds$^{1,2}$,
Pratika Dayal$^{5,25}$,
Alessandro Trinca$^{6,7,8}$,
Raffaella Schneider$^{7, 8, 9, 10}$,
Santiago Arribas$^{11}$,
Rachana Bhatawdekar$^{12}$,
Andrew J. Bunker$^{13}$,
Stefano Carniani$^{14}$,
Stéphane Charlot$^{15}$,
Jacopo Chevallard$^{13}$,
Emma Curtis-Lake$^{16}$,
Benjamin D. Johnson$^{17}$,
Eleonora Parlanti$^{14, 18}$,
Pierluigi Rinaldi$^{19}$,
Brant Robertson$^{20}$,
Sandro Tacchella$^{1, 2}$,
Hannah Übler$^{21}$,
Giacomo Venturi$^{14}$,
Christina C. Williams$^{22}$,
Joris Witstok$^{23, 24}$}\vspace{0.4cm}
\\
\parbox{\textwidth}{
$^{1}$Kavli Institute for Cosmology, University of Cambridge, Madingley Road, Cambridge, CB3 0HA, United Kingdom\\
$^{2}$Cavendish Laboratory - Astrophysics Group, University of Cambridge, 19 JJ Thomson Avenue, Cambridge, CB3 0HE, United Kingdom\\
$^{3}$Department of Physics and Astronomy, University College London, Gower Street, London WC1E 6BT, UK\\
$^{4}$Waseda Research Institute for Science and Engineering, Faculty of Science and Engineering, Waseda University, 3-4-1, Okubo, Shinjuku, Tokyo 169-8555, Japan\\
$^{5}$Kapteyn Astronomical Institute, University of Groningen, PO Box 800, 9700 AV Groningen, The Netherlands\\
$^{6}$Como Lake Center for Astrophysics, DiSAT, Università degli Studi dell’Insubria, via Valleggio 11, 22100, Como, Italy\\
$^{7}$INAF/Osservatorio Astronomico di Roma, Via Frascati 33, 00040 Monte Porzio Catone, Italy\\
$^{8}$
INFN, Sezione Roma I, Dipartimento di Fisica, “Sapienza” Università di Roma, Piazzale Aldo Moro 2, I-00185, Roma, Italy\\
$^{9}$Dipartimento di Fisica, “Sapienza” Università di Roma, Piazzale Aldo Moro 5, I-00185 Roma, Italy\\
$^{10}$Sapienza School for Advanced Studies, Viale Regina Elena 291, 00161 Roma, Italy\\
$^{11}$Centro de Astrobiolog\'ia (CAB), CSIC–INTA, Cra. de Ajalvir Km.~4, 28850- Torrej\'on de Ardoz, Madrid, Spain\\
$^{12}$European Space Agency (ESA), European Space Astronomy Centre (ESAC), Camino Bajo del Castillo s/n, 28692 Villanueva de la Cañada, Madrid, Spain\\
$^{13}$Department of Physics, University of Oxford, Denys Wilkinson Building, Keble Road, Oxford OX1 3RH, UK\\
$^{14}$Scuola Normale Superiore, Piazza dei Cavalieri 7, I-56126 Pisa, Italy\\
$^{15}$Sorbonne Universit\'e, CNRS, UMR 7095, Institut d'Astrophysique de Paris, 98 bis bd Arago, 75014 Paris, France\\
$^{16}$Centre for Astrophysics Research, Department of Physics, Astronomy and Mathematics, University of Hertfordshire, Hatfield AL10 9AB, UK\\
$^{17}$Center for Astrophysics $|$ Harvard \& Smithsonian, 60 Garden St., Cambridge MA 02138 USA\\
$^{18}$ Max-Planck-Institut für extraterrestrische Physik (MPE), Gießenbachstraße 1, 85748 Garching, Germany \\
$^{19}$ Steward Observatory, University of Arizona, 933 North Cherry Avenue, Tucson, AZ 85721, USA \\
$^{20}$Department of Astronomy and Astrophysics University of California, Santa Cruz, 1156 High Street, Santa Cruz CA 96054, USA\\
$^{21}$Max-Planck-Institut f\"ur extraterrestrische Physik (MPE), Gie{\ss}enbachstra{\ss}e 1, 85748 Garching, Germany\\
$^{22}$NSF National Optical-Infrared Astronomy Research Laboratory, 950 North Cherry Avenue, Tucson, AZ 85719, USA\\
$^{23}$Cosmic Dawn Center (DAWN), Copenhagen, Denmark\\
$^{24}$Niels Bohr Institute, University of Copenhagen, Jagtvej 128, DK-2200, Copenhagen, Denmark\\
$^{25}$ Canadian Institute for Theoretical Astrophysics, 60 St George St, University of Toronto, Toronto, ON M5S 3H8, Canada \\
}
}

\date{Accepted XXX. Received YYY; in original form ZZZ}

\pubyear{\the\year{}}

\begin{document}
\label{firstpage}
\pagerange{\pageref{firstpage}--\pageref{lastpage}}
\maketitle

\begin{abstract}
JWST has revealed a large population of active galactic nuclei (AGN) in the distant universe, which are challenging our understanding of early massive black hole seeding and growth. We expand the exploration of this population to lower luminosities by stacking
$\sim 600$ NIRSpec grating spectra 
from the JWST Advanced Deep Extragalactic Survey (JADES) at $3<z<7$, in bins of redshift, [OIII]5007 luminosity and equivalent width, UV luminosity and stellar mass. 
In various stacks, we detect a broad component of H$\alpha$ without a counterpart in [OIII], implying that it is not due to outflows but is tracing the Broad Line Region (BLR) of a large population of low-luminosity AGN not detected in individual spectra. We also consider the possible contribution from Supernovae (SNe) and Very Massive Stars and conclude that while this is very unlikely, we cannot exclude some potential contribution by SNe to some of the stacks. The detection, in some stacks, of high [OIII]4363/H$\gamma$, typical of AGN, further confirms that such stacks reveal a large population of AGN. We infer that the stacks probe black holes with masses of a few times $10^6~M_\odot$ accreting at rates $L/L_{Edd}\sim 0.02-0.1$, i.e. a low mass and dormant parameter space poorly explored by previous studies on individual targets. We identify populations of black holes that fall within the scatter of the local $M_{BH}-M_{*}$ scaling relation, indicating that there is a population of high-z BHs that are not overmassive relative to their host galaxies and which have been mostly missed in previous JWST observations. Yet, on average, the stacks are still overmassive relative the local relation, with some of them 1--2 dex above it. We infer that the BH mass function (BHMF) at $3<z<5$ rises steeply at low masses. The BHMF is consistent with models in which BHs evolve through short bursts of super-Eddington accretion.
\end{abstract}




\section{Introduction}

Understanding the formation and evolution of galaxies and their black holes is a complex and multilayered problem in astrophysics but is central to our understanding of the Universe. Supermassive black holes (SMBHs) play a critical role in galaxy evolution, as during their accretion phases they provide feedback onto the host galaxy, affecting star formation and the properties and distribution of the interstellar and circumgalactic medium \citep[e.g.][]{bourne_recent_2023, Fabian2012}, possibly leading to the correlations observed in the local Universe between the black holes themselves and their host galaxies \citep{2015ApJ...813...82R, kormendy_coevolution_2013, greene_intermediate-mass_2020}. 
Understanding how such black holes were originally formed and how they managed to grow so massive is one of the main problems of modern astrophysics  \citep[e.g.][]{rees_emission_1978,volonteri_origins_2021}.

Observations of massive black holes at high redshift have started to constrain the possible seeding and subsequent growth mechanisms. The Sloan Digital Sky Survey (SDSS) was the first survey to detect quasars at z $\geq$ 6 \citep{fan_survey_2001, fan_2003} with black hole masses $M_{BH} \sim 10^{9}~\msun$, or even higher, which are extraordinarily large for objects existing less than 1 billion years after the Big Bang. 
Similar SMBHs at $z\sim6-7$ were found by multiple other surveys carried out subsequent to SDSS \citep{inayoshi_assembly_2020, 2007MNRAS.379.1599L, 2007AJ....134.2435W, 2012AJ....143..142M,Banados2016,Mortlock2011,Fan2023ARA&A}.
These early Universe discoveries started to challenge theoretical models, not only because of the large black hole (BH) masses at such early epochs, but also because these systems were found to significantly deviate from local scaling relations \citep[e.g.,][]{pensabene_alma_2020}. 
Various seeding and growth scenarios have been proposed to explain the early emergence of massive black holes, such as 
 direct collapse black holes (DCBHs), Population III remnants accreting via super-Eddington bursts, and rapid merging of stars and stellar remnants associated with dense stellar clusters in the cores of galaxies \citep[e.g.][]{madau_massive_2001, greene_intermediate-mass_2020,volonteri_relative_2009,agarwal_unravelling_2013, visbal_identifying_2018,Partmann2025,Rantala2025}.
 
 However, until the launch of \jwst, observations were limited to high redshift luminous massive quasars, or AGN in the local Universe or relatively low redshift. Thus, potential probes of black hole seeding and growth mechanisms could not be fully utilised. 

\jwst has sparked a revolution in AGN discoveries, characterising well over 100 AGN in the early Universe, with luminosities much lower than the previously observed quasars.

Specifically, spectroscopic observations with \jwst NIRSpec MOS (Multi-Object Spectroscopy), IFU (Integral Field Unit) and NIRCam Grism have revealed that there is a large population of accreting black holes at $z>4$, coeval with the population of bright quasars already discovered but with smaller luminosities \citep{kocevski_hidden_2023, ubler_ga-nifs_2023, Kocevski2025, harikane_jwstnirspec_2023, matthee_little_2024, maiolino_small_2024, maiolino_jades_2024, juodzbalis_dormant_2024, 2024A&A...684A..24P, scholtz_jades_2025, juodzbalis_2025, 2025arXiv250522567M, 2024arXiv240815615M, juodzbalis_2025, 2025arXiv250504609T, 2025arXiv250502896L, 2025arXiv250408039L,greene_uncover_2023,kokorev_uncover_2023,furtak+2024}. These observations include Type-1 AGN, characterised by broad Balmer lines without a corresponding broad component of the forbidden lines (hence the broad lines are associated with a Broad Line Region, rather than an outflow), and type 2 AGN which are identified using narrow emission line diagnostics.

Many studies are using these observations to advance toward understanding black hole seeds \citep[e.g.][]{ isobe_jades_2025, maiolino_pristine_2025,Schneider2023,Natarajan2024,Regan2024}. However,  
one problematic aspect of past observations is the shortage of detections and characterization of black holes with masses around $10^6~M_\odot$ or lower, i.e. approaching the masses that might be similar to the expected seeds in scenarios like the Direct Collapse scenarios. Finding and characterizing such objects is essential to understanding the origin of SMBHs, as it may be the closest we can get to directly observing black holes in their infancy.

Additionally, almost all of the \jwst observed AGNs have BHs that are overmassive relative to their host galaxies, lying well above the local $M_{BH}-M_{*}$ scaling relations \citep[e.g][]{ubler_ga-nifs_2023,harikane_jwstnirspec_2023,maiolino_jades_2024,juodzbalis_dormant_2024,Marshall2024,
Li2025JWSThosts}.

It has been suggested that such an offset relative to the local relation is primarily driven by a very large scatter in the relation combined with selection effects, in the sense that more massive black holes are easier to detect as they can reach higher luminosities \citep{2025ApJ...981...19L,juodzbalis_2025}. Even if the offset is driven by selection effects, the existence of black holes nearly as massive as their host galaxies in the early universe is anyway a remarkable finding, as it implies that the Universe manages to create such massive monsters in relatively small galaxies in the early Universe, and highly informative on the seeding and early growth processes. Yet, it is also important to understand the overall underlying distribution, and therefore the effect of selection effects. Searching for black holes with low masses can help in this process.

The main issue in detecting intermediate mass black holes is that current surveys may lack the sensitivity to detect faint broad Balmer lines associated with black hole masses lower than the bulk of the population currently found (e.g., $\sim 10^7-10^8~ M_\odot$). 
Stacking of large numbers of spectra may help to increase the signal-to-noise of the combined spectra and possibly enable the detection of such faint broad lines.
Stacking techniques have already been used by many studies to explore the average properties of high redshift galaxies \citep{hayes_average_2025, hu_characterizing_2024, roberts-borsani_between_2024, glazer_stacking_2025, johnson_h2_2025, 2024MNRAS.535.1796B} and AGN \citep{isobe_jades_2025, juodzbalis_2025, scholtz_jades_2025, 2024arXiv240611997K} in much more detail than available from individual spectra. In this paper, we present detections of the missing population of low-mass black holes in the JADES survey \citep{eisenstein_overview_2023}, identified using stacks of the NIRSpec MOS spectra. These AGN signatures are confirmed by the presence of faint broad components around H$\alpha$ without a counterpart in the forbidden lines.

The paper is organized as follows. In Section \ref{sec:data} we describe the data and criteria used to select our sample. In Section \ref{sec:stack} and \ref{sec:continuum} we describe our stacking methods, and in Section \ref{sec:line} our methods for emission line fitting are explained. Section \ref{sec:evidence} presents the final stacked spectra and investigates evidence of small accreting black holes within our sample. Section \ref{sec:diagnostics} presents narrow line diagnostics of our stacks to help confirm whether we have detected AGN, and Section \ref{sec:agn_props} describes the properties of these AGN. In Section \ref{sec:bhmf} we investigate the implications of our findings on the black hole mass function. Sections \ref{sec:SNEs} and \ref{sec:stars} explore the possibility that faint broad lines in the stacks could be driven by core collapse supernovae or very massive stars, respectively. Section \ref{s.discussion} discusses our findings and we summarise our conclusions in Section \ref{sec:conclusions}. 
Throughout this work we assume a flat \textLambda CDM cosmology with $\Omega_{\rm m}\ \text{=}\ 0.315$, $H_{0}= 67.4$ km s$^{-1}$ Mpc$^{-1}$ \citep{2020A&A...641A...6P} and all reported magnitudes are in the AB system.

\section{Data processing and sample selection}
\label{sec:data}
The spectra used in the stacks we present come from the \jwst Advanced Deep Extragalactic Survey \citep[JADES;][]{eisenstein_overview_2023, bunker+2024, Rieke+23} from data release 3 \citep{deugenio_jades_2024}, using the NIRSpec \citep{jakobsen_2022} spectra from the micro-shutter assembly \citep[MSA;][]{ferruit_near-infrared_2022}. JADES targets the GOODS-S and GOODS-N fields and combines deep imaging and spectroscopy with NIRCam and NIRSpec. The survey consists of spectra of $\sim$5000 objects across the GOODS-N and GOODS-S fields spanning redshifts from 0.6 to 14. The JADES NIRSpec spectra were reduced by the NIRSpec GTO Team \citep{2018SPIE10704E..0QA}, using the data reduction pipeline developed by the ESA NIRSpec Science Operations Team \citep{ferruit_near-infrared_2022}, the details of which are described in \citet{deugenio_jades_2024, 2023NatAs...7..622C, 2024A&A...690A.288B}. In order to maximise our SNR (signal to noise ratio) in the stack to search for Type-1 AGN, we use the 3 pixel extraction of the 1D spectra, which are optimised for unresolved sources such as broad line regions of Type-1 AGN. The parent sample of spectra is mostly the same as that described in Section 2 of \citet{isobe_jades_2025}, apart from some more specific filtering of galaxies that is described below.

The first stage in composing the stacked spectrum was to exclude all known AGN from the JADES survey; both the Type-1 AGN from \cite{juodzbalis_2025}, and the type 2 from \cite{scholtz_jades_2025}. The properties of the brightest and most common lines in the JADES spectra were measured (including flux, equivalent width (EW), continuum flux, and refined redshift measured from \OIII $\lambda$5007 and H$\alpha$) and used to select the desired sources that meet the following criteria:

\begin{enumerate}
    \item have redshift flags 6, 7, or 8, meaning that redshifts have been reliably determined;
    \item are in the redshift range of $3<z<7$;
    \item  H$\alpha$ is detected with a SNR $> 10$ -- this is because this line is the focus of our analysis and also ensures a reliable refinement of the redshift through its profile;
    \item SNR $>$ 3 on the \OIII $\lambda$5007 EW and flux (these values will be used to normalise the spectra);
    \item have available measurements of stellar mass from \cite{2024MNRAS.527.6139S};
\end{enumerate}

We set the upper redshift threshold to ensure that the H$\alpha$ emission is covered by the wavelength range of the NIRSpec instrument (i.e., $5.3 \mu m$). The wavelength range for the gratings can be potentially extended \citep{DEugenio_QSO1}, but this introduces contamination from the second order, which may potentially introduce spectral features that might result in artificial broad components of the lines; therefore, we have not used the wavelength-extended spectra.  
The lower redshift limit (z$>$3) ensures that we probe a redshift range well populated by the JADES selection function and also not previously studied from ground in terms of H$\alpha$ emission.
We finally note that the requirement on the SNR ratio of \OIII $\lambda$5007 detections is also needed, so that this line can be used to rule out outflows if a broad component is found in H$\alpha$.

We focus on the medium resolution grating spectra ($R \sim 1000$). The prism spectra ($R \sim 100$) are potentially also useful for detecting very broad wings, but the strongly varying resolution with wavelength ($\sim 100 -\sim 325$) of the prism makes it difficult to identify real broad lines from resolution-broadened lines arising from different redshifts, and it also makes it difficult to compare the width of the broad lines with \OIII which is a shorter wavelength and therefore observed with much lower spectral resolution. The high resolution ($R \sim 2700$) grating spectra are only available in G395H and most of them are truncated \citep[][their figure 6]{jakobsen_2022}. We also select sources with measurements of stellar mass and \OIII EW available, so that the stacks can be split into bins of these values (which will be described further in Section \ref{sec:evidence}). The stellar masses were measured using the spectral energy distribution (SED) fitting code, \textsc{Prospector} \citep{2021ApJS..254...22J}, as described in \cite{2024MNRAS.527.6139S}. 

Our selection criteria result in 576 sources, of which 424 are in the redshift range $3<z<5$ and 152 in the redshift range $5<z<7$.

\section{Stacking methods}
\label{sec:stack}
Our stacking method builds on the techniques already presented in \citet{isobe_jades_2025}. The first step to prepare the individual spectra for the stacking process is to combine the data from the three medium resolution gratings (i.e., f100lp-G140M, f170lp-G235M, and f290lp-G395M), to obtain one spectrum covering the entire NIRSpec wavelength range for each source. For simplicity, in the spectral region where two gratings overlap, we always select data from the reddest grating. This choice ensures the highest SNR, because in the spectral region of overlap between two gratings, the reddest grating has 2$\times$ lower spectral resolution, and thus 40\% higher SNR. The spectra are then shifted to rest frame using the redshifts re-measured from the H$\alpha$ line. We then resample the spectra to a user-defined common wavelength grid using \textsc{SpectRes} \citep{spectres} which preserves the integrated flux and also accounts for resampled errors. The resolving power of the gratings changes significantly across NIRSpec's wavelength range. This means that the appropriate size for each wavelength bin, required to maintain the instrument's resolution while not oversampling, varies across the wavelength range. To define such a non-uniform wavelength grid, we use the  standard deviation, $\sigma$, of the point source line spread function (LSF) for medium resolution NIRSpec data \citep{deGraaff2024} to calculate the resolution at each wavelength, using the median redshift of the stack.  The bin size is then chosen as the half-width at half maximum of the line spread function at that wavelength. Although it is not clear whether the sources included in the sample are extended or point sources, the point source $\sigma$ values are conservatively used. This is because they are narrower than the nominal LSF values (which are for uniformly illuminated slit), so at worst, the wavelength bin will be narrower than required, which should have no detrimental effect to the resampling process. 

With the resampled spectra, the stack is created by taking the average of the flux at each wavelength bin. We produced stacks using the following methods: 

\begin{enumerate}
    \item taking the average flux with no normalisation or weighting 
    \item same as i) but normalising each spectrum by its $F_{\text{[O III]}}$ (\OIII $\lambda5007$ flux)
    \item same as i) but weighting by 1/rms$^{2}$
\end{enumerate}

Using no normalisation or weighting allows us to understand the average properties of the spectra. With this approach it may be that the spectrum and the detection of broad H$\alpha$ in the stacked spectrum is dominated by a few brightest sources; however, in the latter case, those few sources would be expected to have their broad H$\alpha$ detected individually and included in the Type-1 AGN sample of \cite{juodzbalis_2025}. To confirm that this is not the case, we investigate the possibility of the stack being dominated by a few strongest sources via the jackknife technique (see section \ref{sec:props}). To investigate the possibility that more active sources, with stronger nebular emission are dominating the average, we also make a stack that is normalised by $F_{\text{[O III]}}$.

We also create stacks weighted by 1/rms$^2$, which is an effective way to suppress data with poor S/N and which would increase the SNR of the final stack. We use the noise within each wavelength bin so that the weighting changes across the wavelength range. Because of this, we do not use the weighted stacks for determining emission line ratios e.g. in Section \ref{sec:diagnostics}. The errors on the flux in each wavelength bin for the stack that is not weighted or normalised and on the one that is normalised by \OIII flux are calculated by propagating the resampled errors from \textsc{SpectRes} through the standard error of the mean.
The errors for the weighted stack are calculated by including the weighting term in the standard error of the mean. Most of the analysis is performed on the unweighted and unnormalised stacks, but we also discuss some results from the weighted and normalised stacks in the appendices.

Sources are stacked in two redshift bins, $3<z<5$ and $5<z<7$, to explore possible redshift evolution.

We only choose two redshift bins to maintain a large number of sources in each stack for high SNR.

To explore the potential presence of H$\alpha$ in connection with different degrees of activity and properties of the galaxy population, we also stack (each time) the spectra in four bins of the following quantities: [OIII] luminosity, [OIII] Equivalent Width, UV absolute magnitude, and stellar mass; these stacks will be discussed more in detail in Section \ref{sec:props}.

\section{Continuum subtraction}
\label{sec:continuum}
The final stacks will not only be used to search for broad components to the H$\alpha$ line, but also to measure the relative fluxes of other emission lines for use on diagnostic diagrams (described in Section \ref{sec:diagnostics}). Therefore, in order to accurately measure the line fluxes, we also need to accurately model the continuum. 

Typically, the continuum remains nearly undetected in individual spectra in the medium grating data in the majority of sources. However, due to the increased SNR in the stacks, the continuum becomes more apparent and must be considered.
In addition to this astrophysical signal physically associated to each spectrum, stacking adds artificial steps where, as described above, we clip and splice the spectra from the different gratings. The steps are due to remaining flux-calibration issues between the gratings \citep[e.g.,][]{deugenio_jades_2024}. 
Additionally, JADES grating spectroscopy also includes extensive overlap of spectra coming from sources in other shutters, due to the crowded mask design \citep[e.g.,][]{bunker+2024}. Spectral overlaps may cause additional jumps and artificial continuum features at any wavelength location in all but the highest-priority targets which are protected to avoid overlaps.
Fortunately, the stacking procedure dilutes and smooths both splicing artifacts and spectral overlaps, making them easier to model.
To remove the continuum, we utilise the Penalized PiXel-Fitting \citep[pPXF;][]{cappellari_parametric_2004, cappellari_improving_2017, cappellari_full_2023} method to fit simultaneously the continuum and the emission lines. pPXF is implemented in Python and is used to extract the stellar and gas kinematics, as well as the stellar population of galaxies. However, for our purposes of accurately fitting the continuum in order to subtract it, we are not interested in the stellar kinematics or stellar-population properties (the latter would anyway be unreliable because of the spectral overlap with other sources).
To model the continuum, we utilise as input a set of stellar-population spectra
from the Flexible Stellar Population Synthesis models \citep{2009ApJ...699..486C, 2010ApJ...712..833C}, with MIST isochrones and MILES stellar atmospheres \citep{2016ApJ...823..102C, 2006MNRAS.371..703S}. We also use a 20\textsuperscript{th}-order additive Legendre polynomial, which accounts for the combined contribution of splicing artifacts and spectral overlaps. While this is a very high degree polynomial, we found that it was best fit to the stacked spectrum after trial and error. \cite{2025NatAs...9..280D} use a 6\textsuperscript{th}-order polynomial to model the continuum from NIRSpec prism, and since the medium resolution grating spectra used in our analysis have at least 3 times the number of spectral pixels, we argue that a polynomial of order 20\textsuperscript{th} is probably required.
We set the full width half maximum (FWHM) of the line spread function (LSF) of the templates to be a factor of 5 smaller the FWHM of the LSF that is found using the redshift value at the centre of the bin. The reason why we artificially increase the input spectral resolution is due to the
challenge of modelling stacked spectra with pPXF. Indeed, our stack includes galaxies at sufficiently different redshifts that the LSF at any given wavelength is a flux-weighted combination of the LSF of each stacked spectrum. Therefore, using the LSF for the average redshift is inadequate, as demonstrated by the fact
that pPXF could not accurately model the spectral shape of the (narrow) emission lines in the stack. Since we only use the continuum fit from pPXF and not the estimated line fluxes (we use our own method for estimating line fluxes described in Section \ref{sec:line}), the overestimation of the pPXF line width does not have an effect on our results. 

To mitigate the effects of the wavelength dependence of the LSF, emission lines with similar wavelengths are grouped together as follows:

\begin{enumerate}
    \item   $\left[\text{NeIII}\right]\lambda\lambda$3869,3968, H$\epsilon$3971, HeI $\lambda$3889, [OII]3726, [OII]3729, H$\delta$
    \item H$\gamma$, \OIII $\lambda$4363
    \item HeII $\lambda$4686, H$\beta$, \OIII$\lambda\lambda$4959,5007
    \item HeI $\lambda$5876
    \item $\left[\text{OI}\right]\lambda\lambda$6300,6364
    \item $\left[\text{NII}\right]\lambda\lambda$6548,6583, H$\alpha$, $\left[\text{SII}\right]$6716, $\left[\text{SII}\right]$6731.
\end{enumerate}

For doublets arising from the same upper level, their flux ratios are fixed by quantum mechanics. For the [NeIII] doublet, [NeIII]3869 has 3.32 times the flux of [NeIII]3968, for the \OIII doublet, \OIII $\lambda$4959 has 1/3 of the \OIII $\lambda$5007 flux, [OI]6364 has 1/3 of the [OI]6300 and [NII]6548 has 1/3 of the [NII]6583 flux (\citep{2023AdSpR..71.1219D}). These are approximate flux ratios are taken from PyNeb \citep{pyneb}. Figure \ref{fig:ppxf} in Appendix \ref{appendix:appendix_ppxf} shows an example of the continuum fit to the $3<z<5$ stack.

After using pPXF to subtract the continuum, we noticed that there were some regions around the emission lines that were overfit, likely resulting from the attempt of fitting the whole complex spectrum (with all issues discussed above) across the entire wavelength range. Therefore, before fitting a broad component to H$\alpha$ (described in Section \ref{sec:line}), we perform a finer linear fit on the continuum around the line ($\sim \pm 100$Å) and subtract this from the data.

\section{Emission line fitting}
\label{sec:line}
The goal for stacking the JADES spectra is to reveal very faint broad components around the H$\alpha$ line originating from the broad line region (BLR) of the AGN. Therefore, we fit two separate models to the H$\alpha$ emission line: a single Gaussian component describing the narrow line and a model with two Gaussian components describing the narrow and broad line regions.
In principle, more complex profiles might be more appropriate \citep[double Gaussian, power-law, exponential, Lorentizian][]{DEugenio_QSO1,Nagao2006,Cracco2016,maiolino_jades_2024}, but with the limited signal-to-noise on the broad H$\alpha$ of the final stack, these different profiles would be mostly degenerate. Therefore, we keep the fit of the broad component in the simplest possible form by only using one Gaussian.
In addition to the narrow H$\alpha$, we also fit a narrow component to the [NII]$\lambda\lambda$6548,6583 doublet with a single Gaussian component per emission line. We fixed the kinematics (line width and velocity offset) of the narrow H$\alpha$ and [NII]$\lambda\lambda$6548,6583 lines to be the same, assuming that the three narrow emission lines are all produced within gas having the same kinematics. As described above, the broad emission model has an additional broad Gaussian component for the H$\alpha$  emission line. In order to avoid degeneracies between the narrow and broad H$\alpha$ components, we constrained that the amplitude of the narrow component must be larger than the peak of the broad component and the broad component FWHM must be larger than the FWHM of the narrow component. As the systemic velocity of the narrow line and broad line region does not have to be necessarily the same, we allow the centroid of the broad component to move by $\pm 500$~\kms -- leaving the possibility of having a velocity offset will be important to test some specific scenarios.

To obtain initial conditions for the fit of the parameters, we use \texttt{curvefit} from SciPy library, to find the narrow line parameters. Bayesian inference is then used via the MCMC implementation in Python, \texttt{emcee} \citep{2013PASP..125..306F} to obtain the best fitting broad line parameters, as well as the final values for the narrow lines. Uniform priors are used with the \texttt{curvefit} best fit value approximately at the centre of the range for the narrow line components. The peak of the H$\alpha$ broad component is allowed to vary between 0 and the narrow line peak from \texttt{curvefit}, while the broad FWHM can fall between 100 and 5000~\kms.

We measure the line fluxes of other emission lines in the stack (e.g. those used in section \ref{sec:diagnostics}) using the same method as described above except without adding a second Gaussian component, keeping the velocity offset fixed at zero and constraining lines close in wavelength to have the same width.
\subsection{BLR detection criteria}
In order to distinguish between the two models outlined above (narrow only and narrow+broad), we use the Bayesian Information Criterion ($BIC$),

\begin{equation}
\label{bic}
BIC = \chi^{2} - k \ln n ,
\end{equation}

where $\chi^{2}$ is the $\chi$-squared of the fit, $k$ is the number of model parameters, and $n$ is the number of data points. The difference between the $BIC$ values of two models tests whether the better fit resulting from the additional parameters to the model (hence adding complexity to the model) is statistically significant. Both unnecessary complexity and poor fit to the data are penalised, resulting in a larger $BIC$. Therefore, a smaller $BIC$ represents a statistically better fitting model. $\Delta BIC>6$ is usually required as the bare minimum to confirm the best fitting model. However, given the various sources of uncertainties, we prefer to be conservative and impose a $\Delta BIC>10$. Therefore, in the case of broad line fitting we require

\begin{equation}
\Delta BIC = BIC_{\text{only narrow}} - BIC_{\text{broad+narrow}} > 10.
\label{bic10}
\end{equation}

\citep{2007MNRAS.377L..74L}. Using $\Delta BIC$ to detect broad lines is a common approach in other spectroscopic studies of AGN (e.g. \cite{juodzbalis_2025, maiolino_jades_2024}). Since the model only includes the emission lines and a linear continuum within the close vicinity of the lines (±100Å), $\Delta BIC$ is only a measure of the fit of the model to this small spectral region of the entire stacked spectrum.

We also require $\sim 3 \sigma$ detection of the broad component. The SNR is defined from the integral of the Gaussian that is fit to the line. 

As we are searching for faint, low-mass AGN that may have relatively narrow FWHM of the broad component, we do not put any strict constraints on the FWHM of the broad component, other than being larger than the narrow component. Despite this, as we will see in Section 
\ref{sec:props}, all stacks that comply with our detection criteria ($\Delta BIC$, S/N and without an [OIII] counterpart) have a broad component of H$\alpha$ close to or above 1000~\kms. 

\subsection{Ruling out outflows}

A broad component of the H$\alpha$ can originate either from the broad line region of an AGN or from galaxy-wide outflows, driven by either star-formation or AGN. Indeed, high velocity ionized gas in the form of outflows have been detected in the JADES spectra in H$\alpha$ \citep{carniani_outflows_2024}. However, these galaxy-wide outflows are seen in both the Balmer lines and the \OIII $\lambda$5007 emission line, which is an excellent tracer of ionised gas outflows \citep[e.g.,][]{2018MNRAS.481.1690C, carniani_outflows_2024}. Therefore, we search for a similar broad component seen in H$\alpha$ also in the \OIII$\lambda$$\lambda$4959,5007 doublet. We fitted the \OIII doublet with the same model as the H$\alpha$ emission line with FWHM and velocity offset fixed to those found from fitting the H$\alpha$ line. However, we allow the intensity of the Gaussian components to vary freely, fixing the \OIII $\lambda$5007/\OIII $\lambda$4969 ratio to 3. We also perform an additional fit where we let the kinematics of both the narrow and broad components of the \OIII emission line to vary freely relative to those of H$\alpha$. We confirm the presence of a BLR associated with an AGN if the $BIC$ of the model that has the broad component added to \OIII with the same velocity and width as that of H$\alpha$ is larger than the $BIC$ resulting from fitting an \OIII broad component with untied/free velocity and width. The scenario where there is a broad component in \OIII that is less broad than the H$\alpha$ narrow component would represent the presence of both an AGN and an outflow.

Finally, since a broad component in the H$\alpha$ profile
should also appear in the other Balmer lines, we add a broad component to H$\beta$ that has the same width and velocity offset as the H$\alpha$ component and compare the $BIC$ to a narrow only model fitted to H$\beta$. However, since the AGN we are searching for are faint (because they are not detected through their individual spectra), and H$\beta$ is intrinsically at least 3 times fainted than H$\alpha$, it is unlikely that any broad H$\beta$ component is detected (we confirm in Section \ref{sec:props} that no broad component is detected in H$\beta$), so our primary method for AGN identification is the presence of a significant $H\alpha$ broad that does not have a counterpart in \OIII $\lambda$$\lambda$4959,5007.

\begin{figure*}
    \centering
    \begin{subfigure}[b]{\textwidth}
        \includegraphics[width=\textwidth]{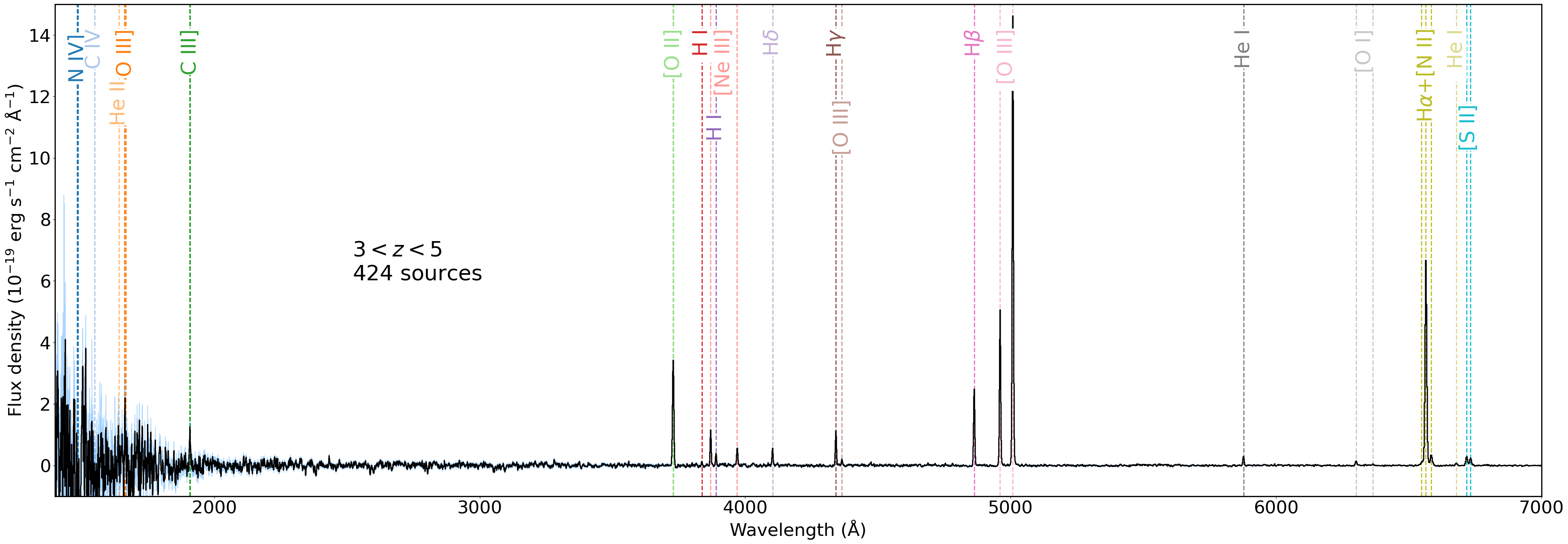}
    \end{subfigure}
    \hfill
    \begin{subfigure}[b]{\textwidth}
        \includegraphics[width=\textwidth]{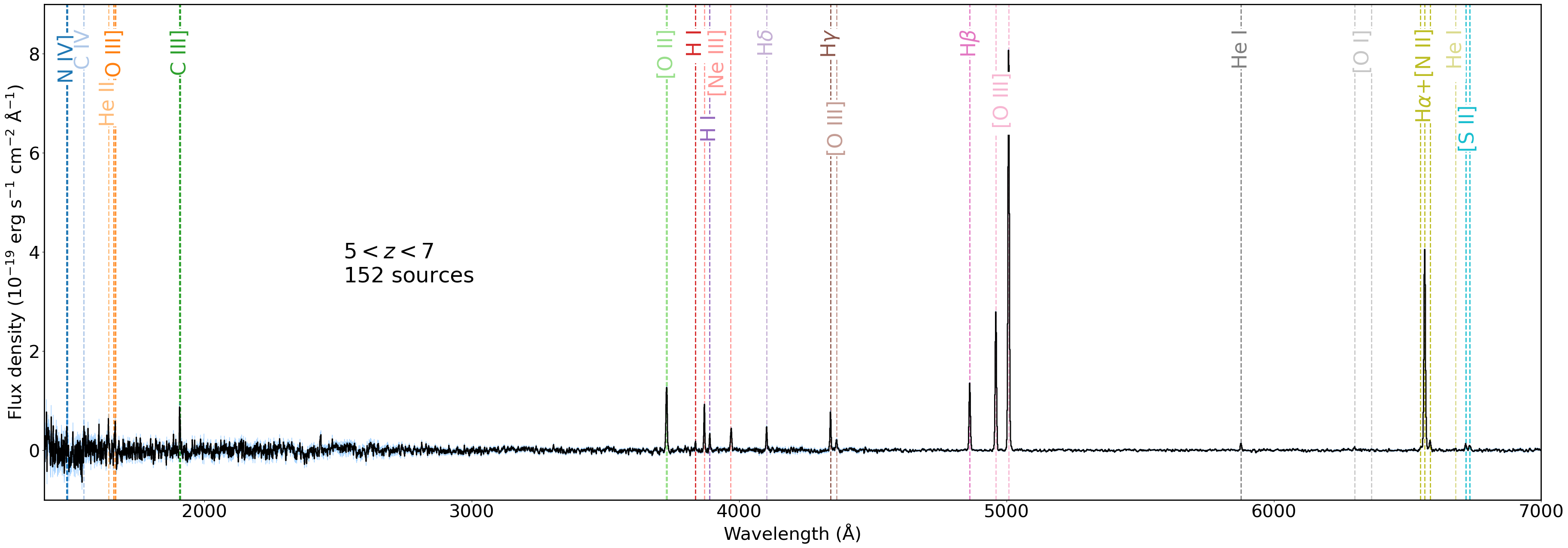}
    \end{subfigure}
    \caption{Stacked spectra of our sample of medium resolution JADES galaxies that have not been previously identified as AGN at $3<z<5$ (top) and $5<z<7$ (bottom). These are mean stacks with no weighting or normalisation. The $3<z<5$ stack contains 424 sources and the $5<z<7$ stack contains 152 sources.}
    \label{fig:combined}
\end{figure*}

\section{Evidence for small, accreting black holes}
\label{sec:evidence}

\subsection{Broad H\texorpdfstring{$\alpha$}{a} detection}
In  Fig.~\ref{fig:combined} we show the final stacks of our sample, with no normalisation or weighting, and with the continuum subtracted for both redshift bins ($3<z<5$ and $5<z<7$). An array of UV and optical emission lines are detected very clearly, similarly to what is depicted in the prism stacks from \cite{roberts-borsani_between_2024}, \cite{hayes_average_2025} and medium resolution (R1000) stacks from \cite{2024arXiv240611997K}. Both redshift bins show bright \OIII $\lambda$5007 emission as well as H$\alpha$. As described in Section \ref{sec:diagnostics}, the \OIII $\lambda$4363 line is the other important indicator of AGN employed in this work, and this line is clearly detected in both stacks. 

\begin{figure*}
    \centering
    \begin{subfigure}[b]{0.49\textwidth}
        \includegraphics[width=\textwidth]{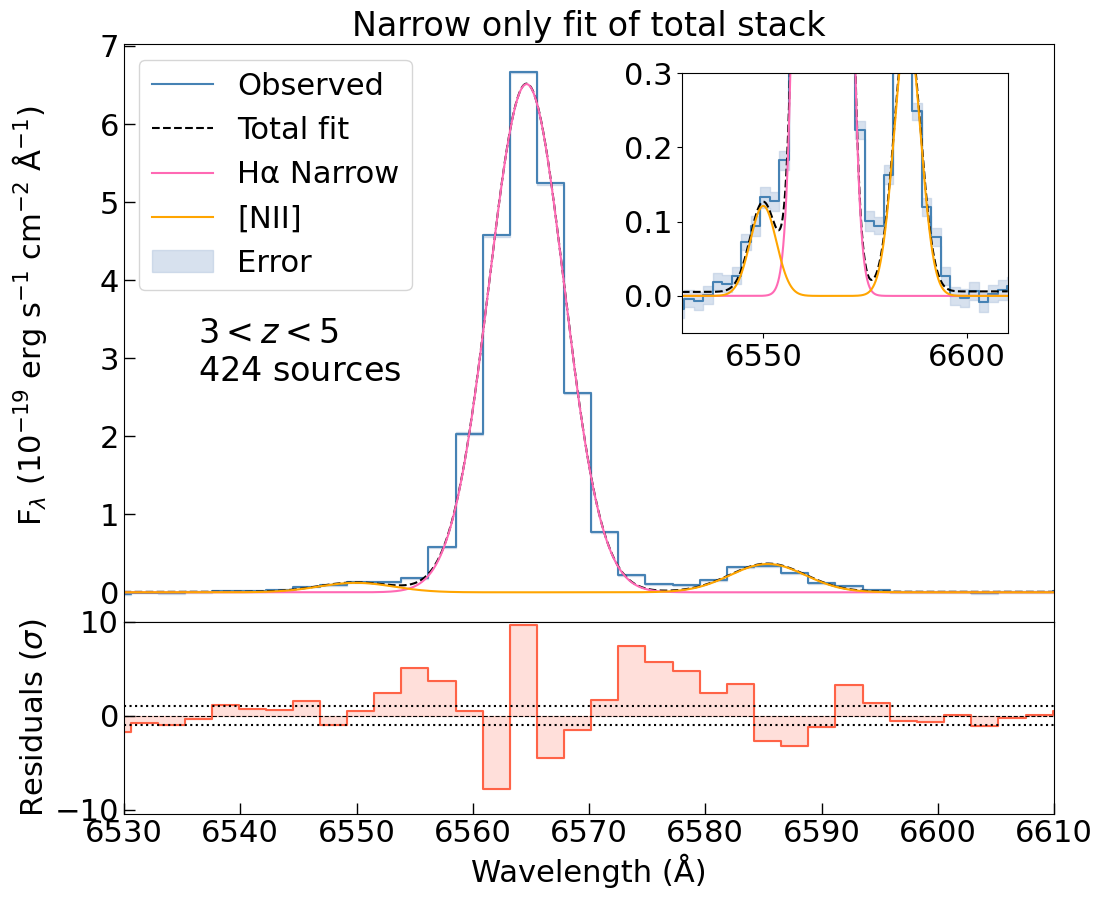}
        \label{fig:halphaz3-5}
    \end{subfigure}
    \hfill
    \begin{subfigure}[b]{0.49\textwidth}
        \includegraphics[width=\textwidth]{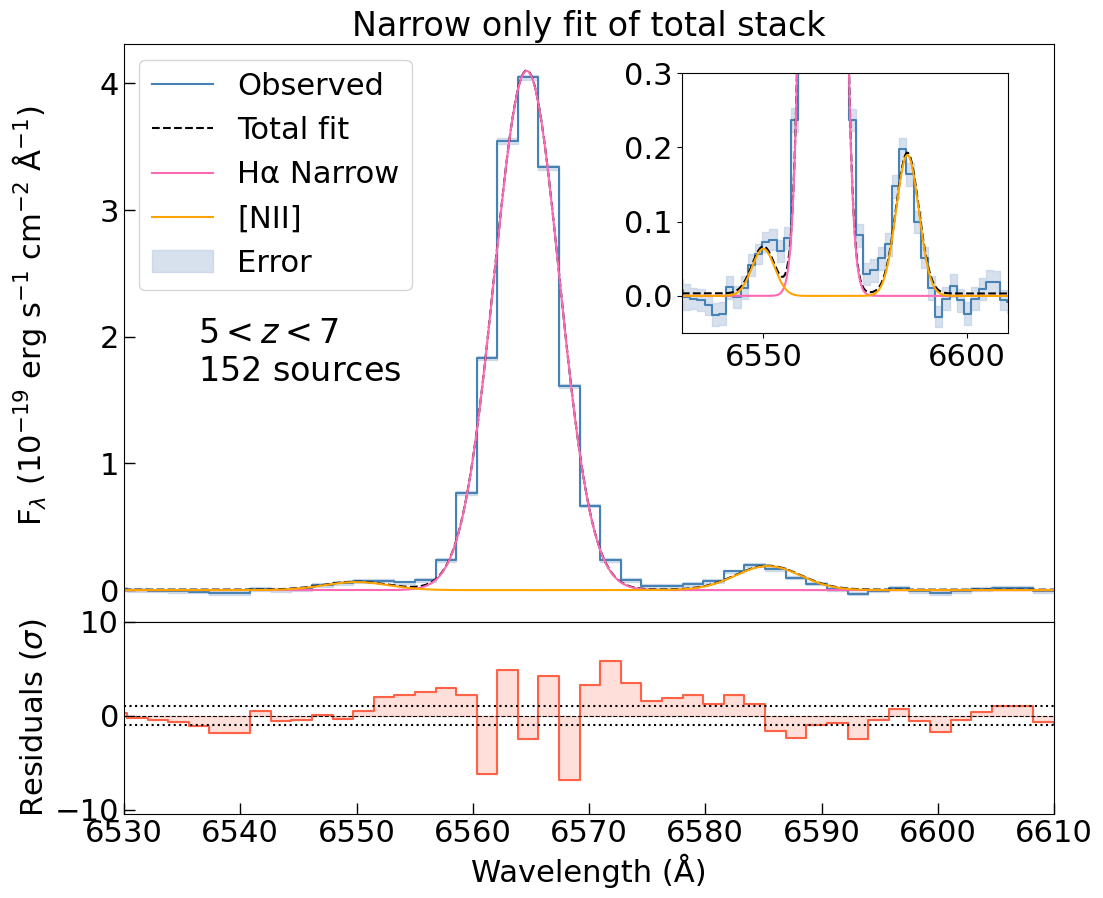}
        \label{fig:halphaz3-5_zoomed}
    \end{subfigure}
    \caption{Zoom of the $3<z<5$ stack (left) and $5<z<7$ stack (right) around the H$\alpha$. The observed spectrum is given by the blue histogram and shaded region is the error. The pink line shows the fit with a narrow component only.
    The orange line is the fit of the [N II] doublet with velocity and width tied to the H$\alpha$ line.
    The dashed line shows the total fit.  The bottom panel shows the residuals of the fit where the dotted line indicates the $\pm 1 \sigma$ levels. The zoom in clearly shows the broad residuals that the narrow-only model fails to reproduce around H$\alpha$ and the excess flux in these regions is also clear from the significant residuals in the bottom panel on both the blue and red sides of the central H$\alpha$ wavelength.}
    \label{fig:halpha}
\end{figure*}

\begin{figure*}
    \centering
    \begin{subfigure}[b]{0.49\textwidth}
        \includegraphics[width=\textwidth]{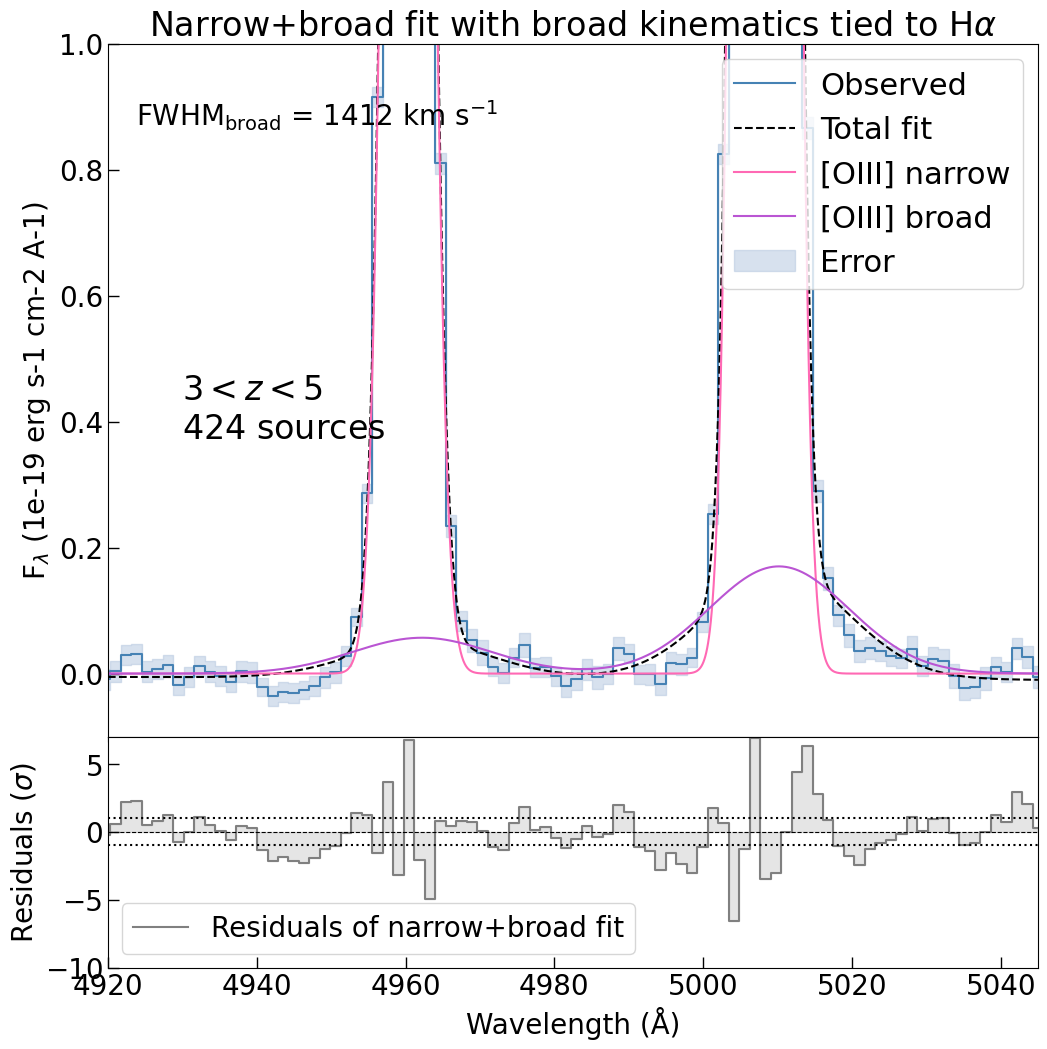}
    \end{subfigure}
    \hfill
    \begin{subfigure}[b]{0.475\textwidth}
        \includegraphics[width=\textwidth]{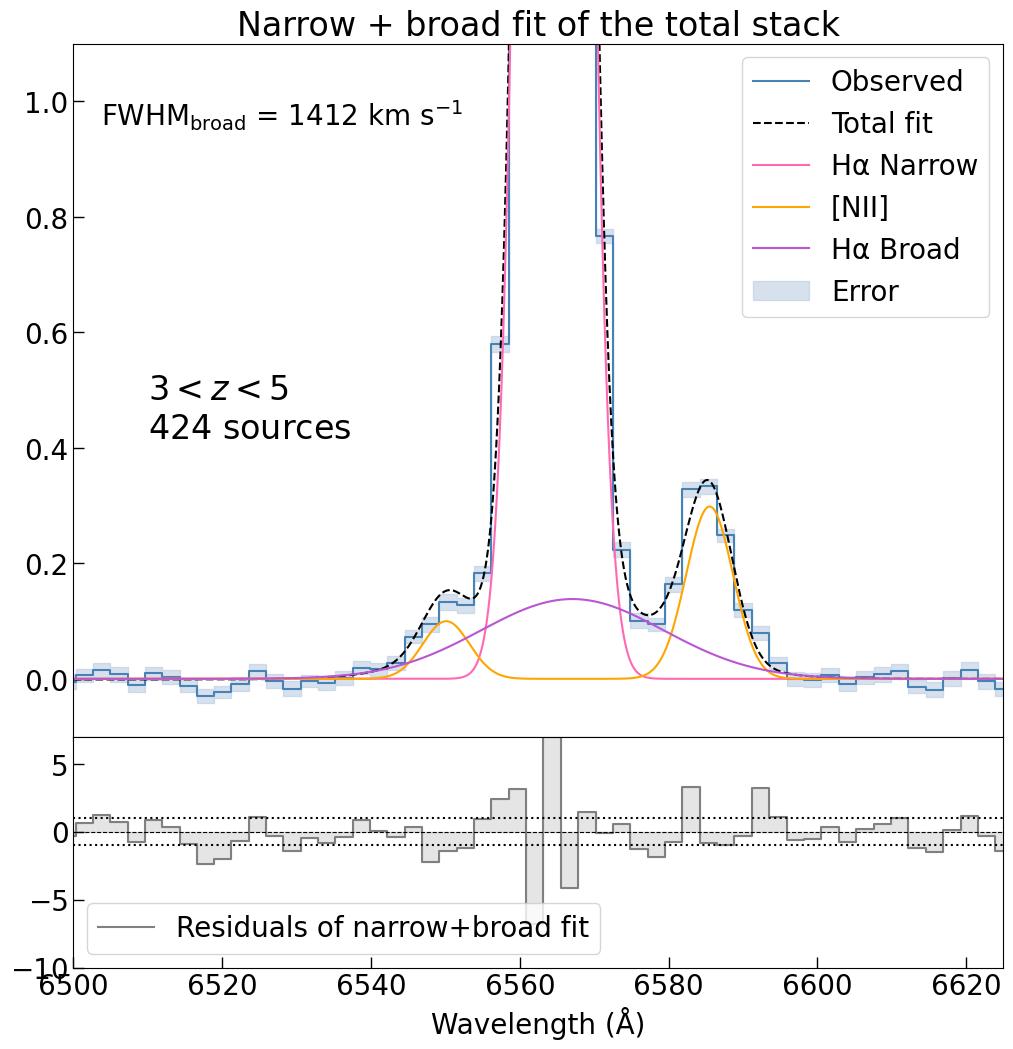}
    \end{subfigure}
    \caption{Zoom around the \OIII and H$\alpha$ lines of the $3<z<5$ total stack. In contrast to Fig.\ref{fig:halpha}, the H$\alpha$ fit (top-right) includes a broad component (violet line), with the residuals shown in the bottom panel.
    On the left the \OIII doublet is fit with narrow and broad components tied to the H$\alpha$ line, with the residuals shown in the bottom panel (the dotted horizontal lines show again the $\pm 1\sigma$ deviation). 
    The H$\alpha$ is well fit, as also indicated by the $\Delta BIC = 221$  in favour of the broad component.
    On the contrary, the [OIII] has clear systemic residuals indicating that the broad component of H$\alpha$ is inadequate to reproduce the [OIII].
     The $\Delta BIC$ for [OIII] strongly favours a freely fitted \OIII broad component (with FWHM = 672), indicating that any broad components in \OIII do not have the same origin as the H$\alpha$ broad component.} 
     \label{fig:halpha_broad1}
\end{figure*}

\begin{figure*}
    \centering
    \begin{subfigure}[b]{0.49\textwidth}
        \includegraphics[width=\textwidth]{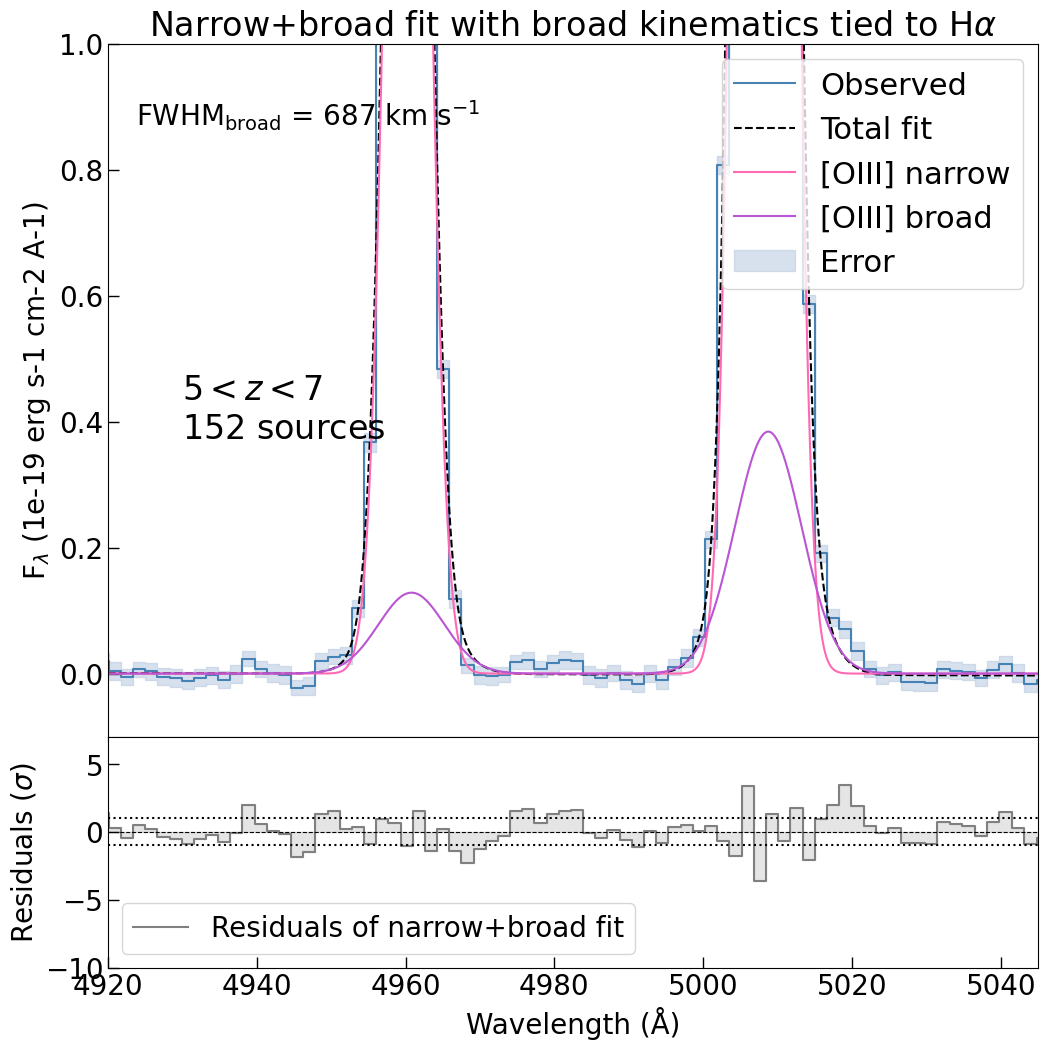}
    \end{subfigure}
    \hfill
    \begin{subfigure}[b]{0.49\textwidth}
        \includegraphics[width=\textwidth]{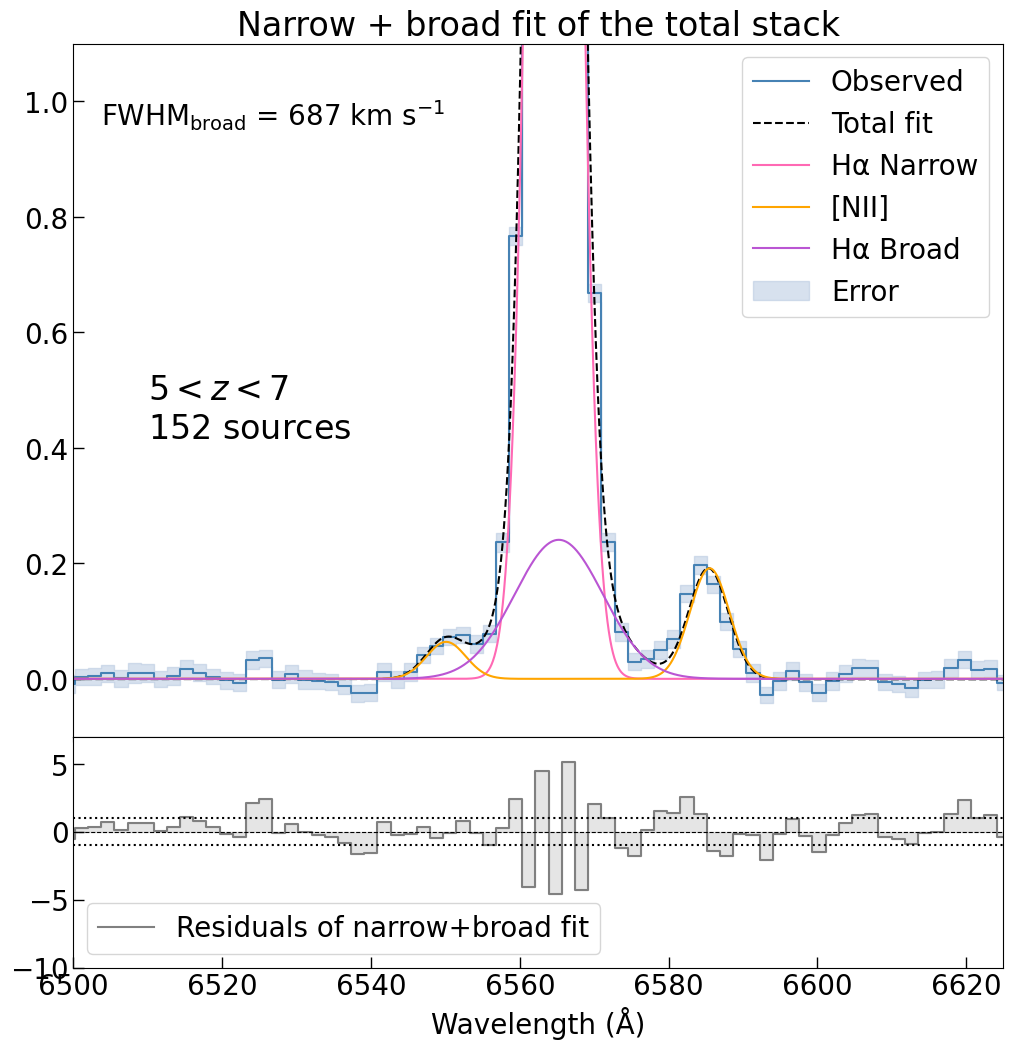}
    \end{subfigure}
    \caption{Same as Fig.\ref{fig:halpha_broad1}, but for the total stack in the $5<z<7$ redshift bin.
    The broad component is clearly a good fit to the H$\alpha$ line. However, in this case, the same broad component added to \OIII is also a good fit and has a similar value of the $BIC$ of a freely fitted \OIII broad component. This could indicate that in this case the broad $H\alpha$ could be due to outflows rather than an AGN BLR.}
    \label{fig:halpha_broad2}
\end{figure*}

Our stacked spectra of the entire sample show hints of broad H$\alpha$ components in both redshift bins. In Fig.~\ref{fig:halpha}, we show the overall stacked H$\alpha$ emission line profile for both redshift bins, with a narrow component fit to both H$\alpha$, and the [NII] doublet, with a zoom in on the broad component. There is a clear excess of flux on both the red and blue side of H$\alpha$, as well as significant residuals which appear as two wings when fitting a single-component model. We add a broad Gaussian on top of the narrow to the model, the results of which are presented in the right panels of Fig.~\ref{fig:halpha_broad1} ($3<z<5$) and Fig.~\ref{fig:halpha_broad2} ($5<z<7$). The FWHM of the fitted broad H$\alpha$ component is $1412^{+113}_{-132}$~\kms  and $687^{+100}_{-67}$~\kms for the $3<z<5$ and $5<z<7$ stack, respectively. To assess the significance of these broad components we calculate $\Delta BIC$ and find that $BIC_\text{narrow only} - BIC_\text{narrow+broad} = 221$ for the 3<z<5 stack and $BIC_{\text{narrow only}} - BIC_{\text{broad+narrow}} = 103$ for the $5<z<7$ stack. Both of these values indicate that a model with a broad component included is preferred.

To assess the presence of outflows, we add a broad component to each line in the \OIII doublet that has the same width and velocity offset as the fitted H$\alpha$ broad component and fit for the peak values, as described in Section \ref{sec:line}. It is important to note that, while we do not know the true intrinsic LSF of the stack, any instrumental broadening would broaden \OIII more than H$\alpha$ since the spectral resolution increases with wavelength. Therefore, we rule out the possibility that broad H$\alpha$ is just an outflow component that has been subject to instrumental broadening not seen in [OIII]. 
We show the results of our \OIII fits in the left panels of  Fig.~\ref{fig:halpha_broad1} (3<z<5) and Fig.~\ref{fig:halpha_broad2} ($5<z<7$). We compute $\Delta BIC$ for the narrow only model compared to the narrow+broad model and find that for the 3<z<5 stack, $\Delta BIC = 493$, which is an extremely strong indication that a model with broad \OIII is preferred. However, it is clear from the residuals in Fig.~\ref{fig:halpha_broad1} that a  narrower broad component would be more suitable. We investigate this further by fitting a broad component to \OIII which can have kinematics that vary freely. The fit produces an \OIII broad component with FWHM = $672$ km/s. We subtract the $BIC$ of this fit, from the fit with the \OIII broad component that has H$\alpha$ broad component kinematics and find a difference of $\Delta BIC =89$ in favour of the narrower component. This indicates that the broad component in H$\alpha$ originates from an AGN BLR, revealing a population of weak AGN that was not detected in the individual spectra.

For the $5<z<7$ stack, the \OIII broad component that has kinematics tied to the H$\alpha$ broad component is clearly a good fit (Figure \ref{fig:halpha_broad2}) with $BIC_{\text{narrow only}} - BIC_{\text{narrow+broad}} = 116$. We also fit a broad component to \OIII and allow the kinematics to vary freely. This results in a broad component of FWHM = 897 km/s. The difference in $BIC$ between this component and the component with kinematics fixed to the H$\alpha$ properties is 8, in favour of the broader component, although with marginal statistical evidence. This indicates that outflows could contribute the broad H$\alpha$ component. The presence of outflow activity in the average of JADES galaxies is in agreement with the discovery of outflows across the redshift range probed by JADES, as reported in \citep{carniani_outflows_2024}. 

We have identified a broad H$\alpha$ component in the $3<z<5$ stack which likely indicates the BLR of AGN due to the absence of a broad counterpart in \OIII. To determine if there is a sub population of these sources that have AGN activity, we next split up the combined stack into bins of \OIII $\lambda$5007 luminosity, equivalent width (EW), $M_{UV}$ and stellar mass of the host galaxy. We also do this for the $5<z<7$ stack to investigate whether there could be a contribution from both outflows and AGN by attempting to disentagle sources with these components. These additional stacking experiments are discussed in the next subsection.

In Appendix \ref{appendix:appendix_tests} we present various tests that are aimed at excluding the possibility that the observed broad lines are resulting from artefacts associated with the stacking technique. Additionally, while the type 2 AGN sample from \cite{scholtz_jades_2025} is not included in our stacks, we also perform a separate stack of only the type 2 AGN sample to compare the results to \cite{scholtz_jades_2025} and \cite{mazzolari_new_2024} who also investigated this. Our results are shown in Appendix \ref{appendix:appendix_type2}.

\begin{figure*}
    \centering
    \begin{subfigure}[b]{0.49\textwidth}
        \includegraphics[width=\textwidth]{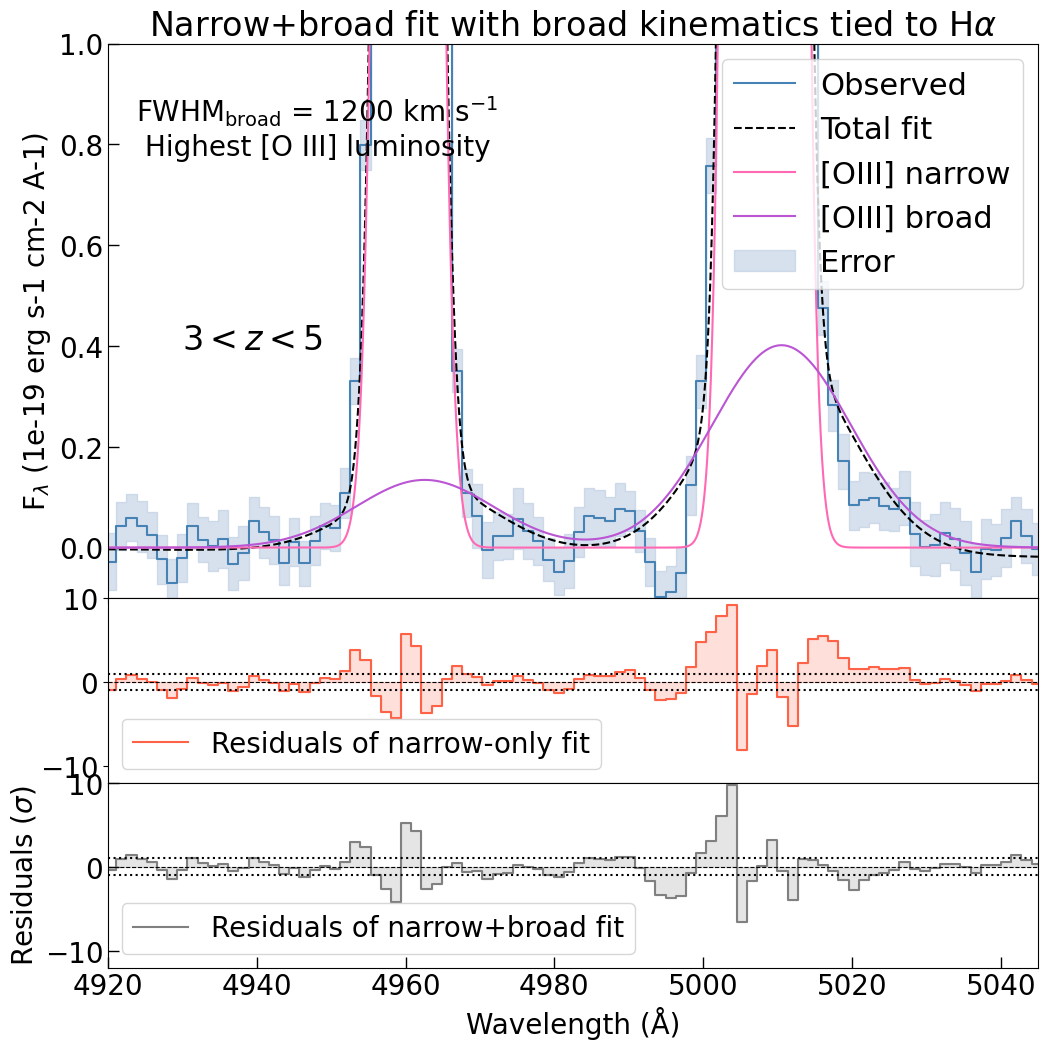}
    \end{subfigure}
    \begin{subfigure}[b]{0.47\textwidth}
        \includegraphics[width=\textwidth]{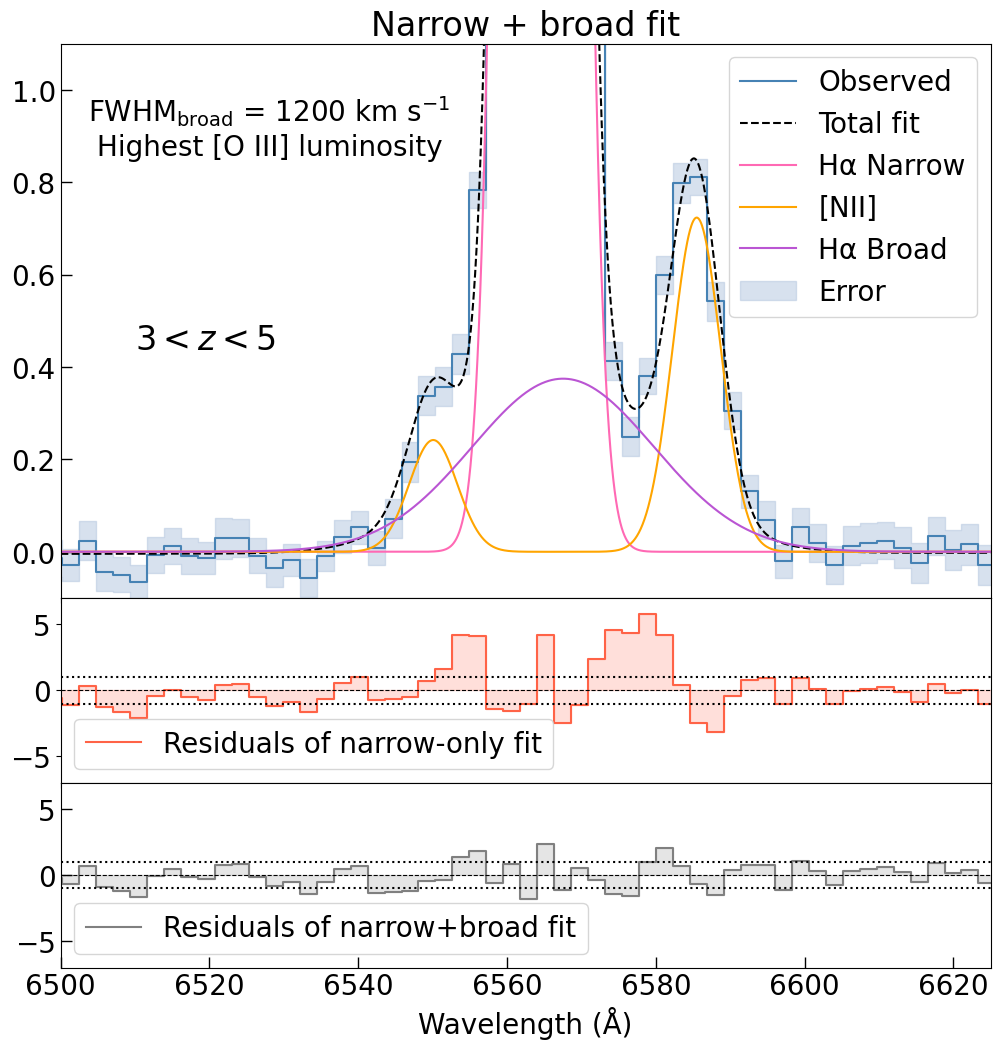}
    \end{subfigure}
    
    \caption{
    Same as Fig.\ref{fig:halpha_broad1}, but for the stack in the $3<z<5$ redshift range and in the highest [OIII] luminosity bin. In this case the middle panel show the residuals without the broad component, while the bottom panel shows the residuals with the inclusion of the broad component (in the case of [OIII] tied to the H$\alpha$ line). Note that the scale on the residuals for the \OIII line and the H$\alpha$ lines are different, but we plot the $\pm 1\sigma$ levels as a dotted line for clarity. 
    The FWHM of the H$\alpha$ broad component, which provides a good fit, is $\gtrsim1000$ \kms, supporting the BLR hypothesis. In the case of [OIII] the same broad component leaves strong residuals.}
    \label{fig:highlum}
\end{figure*}

\begin{figure*}
    \centering
    \begin{subfigure}[b]{0.49\textwidth}
        \includegraphics[width=\textwidth]{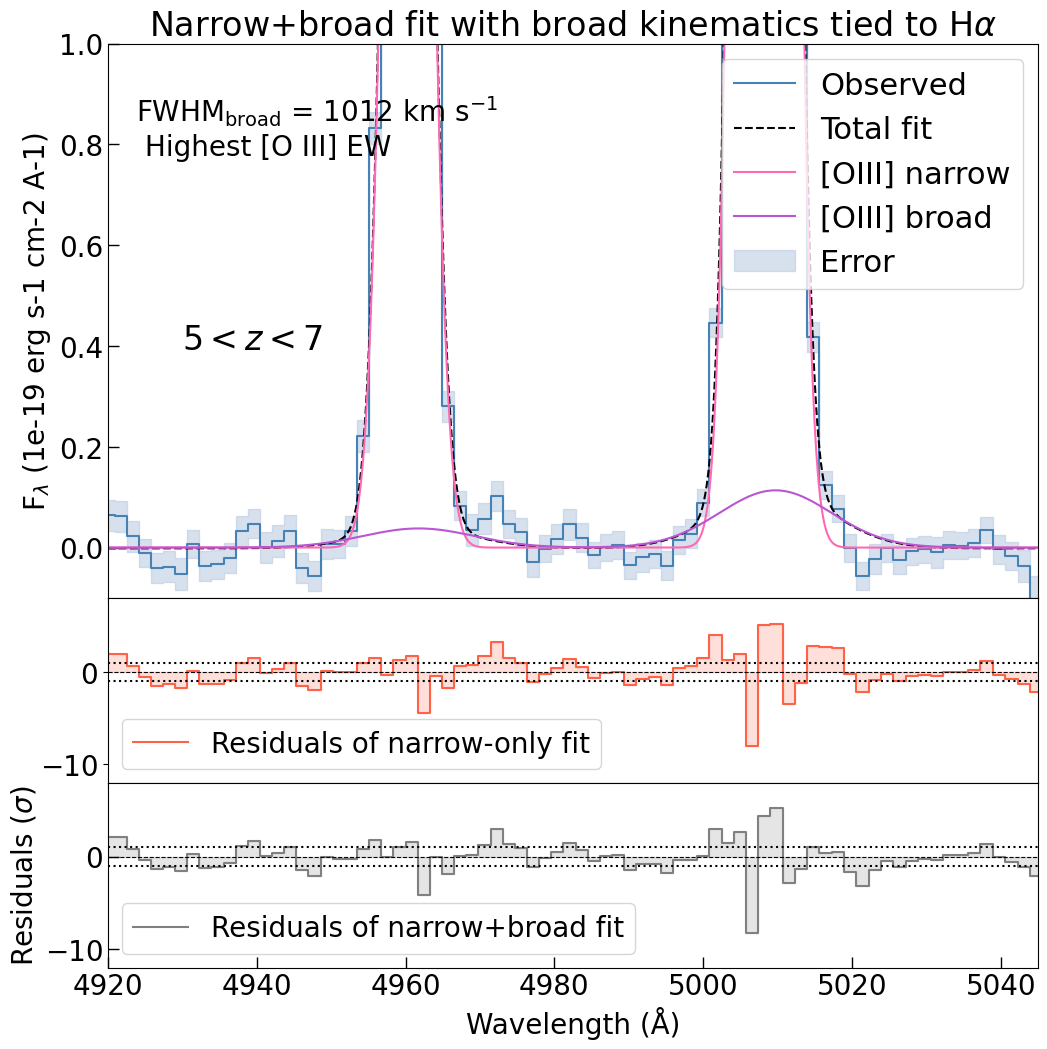}
    \end{subfigure}
    \begin{subfigure}[b]{0.47\textwidth}
        \includegraphics[width=\textwidth]{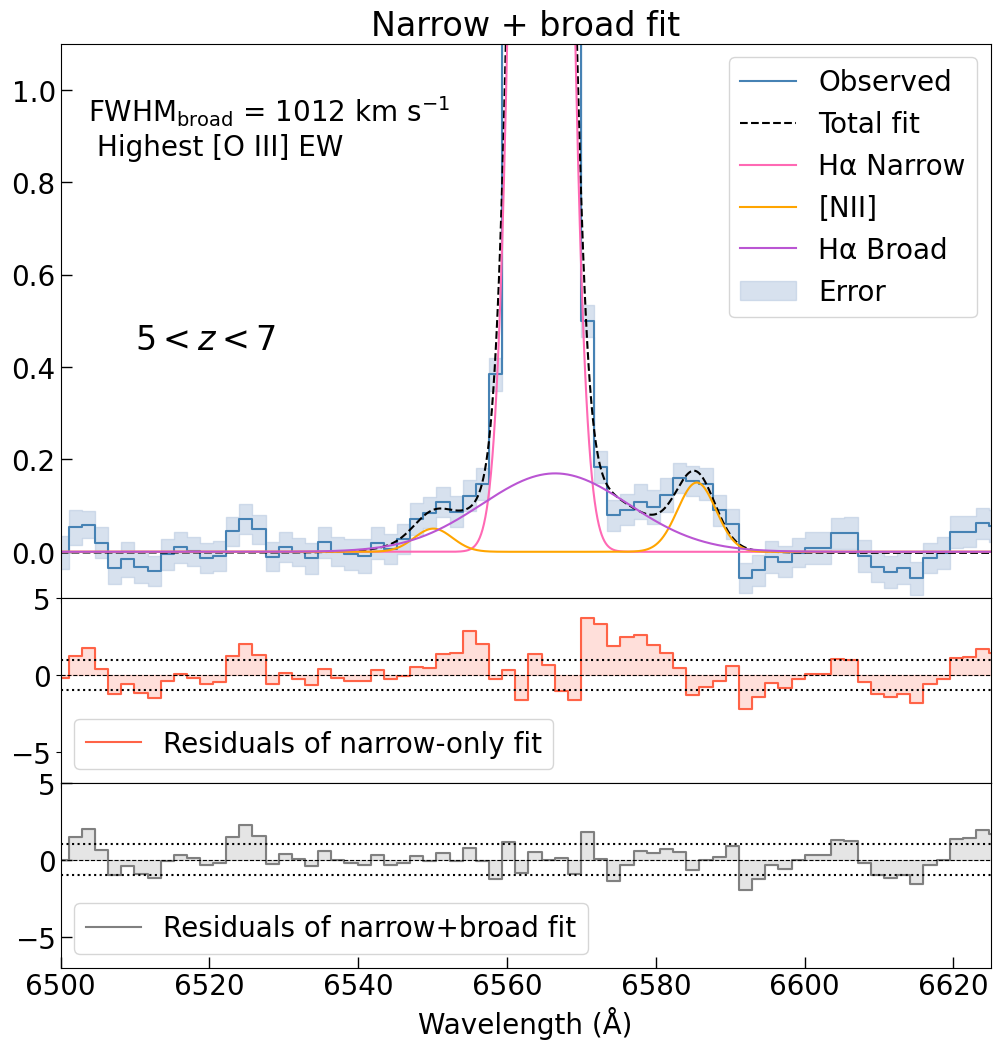}

    \end{subfigure}
    
    \caption{
    Same as Fig.\ref{fig:highlum}, but for the stack in the redshift range $5<z<7$ and highest EW([OIII]) bin. Also in this case the FWHM of the H$\alpha$ broad component that provides a good fit is $\gtrsim1000$ \kms, supporting the BLR hypothesis. The same broad component is not required by the [OIII] as indicated by the residuals and, most importantly, as indicated by the $\Delta BIC$ which favours a much narrower broad component for \OIII (see Table \ref{tab:agn_properties}).}
    \label{fig:highEW}
\end{figure*}

\subsection{Binning by \OIII luminosity, \OIII EW, stellar mass and UV magnitude}
\label{sec:props}

In order to understand which galaxy populations contribute to the broad H$\alpha$ detected in the lower redshift bin stack, and whether there are subpopulations of galaxies which are characterized by a broad H$\alpha$ also in the higher redshift bin,
we create stacks, in each redshift interval, in bins of \OIII $\lambda$5007 luminosity and EW, $M_{UV}$ and host galaxy stellar mass. We bin the population of galaxies in quartiles of each of these quantities. These quantities are chosen for the binning process for the following reasons: the \OIII $\lambda$5007 luminosity is expected to be enhanced under the influence of an AGN due to the high energy photons produced by the accretion disc that are able to reach the narrow line region and increase the fraction of highly ionised oxygen. The presence of AGN also results in a high ionisation parameter which further enhances the \OIII $\lambda$5007 emission \citep{maiolino_re_2019}. Therefore, we expect that by splitting the stack into four luminosity bins we will isolate the AGN sources, and distinguish between broad H$\alpha$ emission arising from an AGN BLR or from outflows. The presence of an AGN is also expected to increase the \OIII EW \citep{Nakajima2022,CidFernandes2011}; on the other hand massive galaxies hosting an AGN typically have a low EW of the nebular lines \citep{Carnall2023,Baker2025_passive,DEugenio2024_passive,Pascalau2025}
; in either case, binning in EW may provide additional valuable insights. Stellar mass binning will be a useful probe as to whether any AGN that are identified are less over-massive relative to their host galaxy than previously identified high-redshift AGN (e.g. \cite{juodzbalis_2025}), as these are the sources we are ultimately searching for. The absolute UV magnitude can also potentially correlate with the presence of an AGN, either because the AGN continuum is directly seen in the UV or, in the case that the UV luminosity is dominated by star formation, because AGN accretion may correlate with the SFR, as both are fueled by high gas content. Additionally, stacking in $M_{UV}$ may help in deriving the black hole mass function by leveraging the UV luminosity function. Therefore, we also explore stacks in bins of $M_{UV}$.

The binning results in 16 stacks for each redshift bin, which we use to search for H$\alpha$ broad components. Table \ref{tab:combined_properties} lists the mean properties of each bin.

The tests described in Section \ref{sec:line} are carried out and we rule out AGN candidates if an \OIII doublet broad component is found with the same width and velocity offset as the H$\alpha$ broad component, if $\Delta BIC$ does not prefer a model with a second broad Gaussian added to H$\alpha$ in addition to the narrow one, or if the SNR of the fitted broad H$\alpha$ component is less than 3$\sigma$.

From the 16 $3<z<5$ stack, we identify five that show significant broad components indicative of AGN, and of the 16 $5<z<7$ stacks, we identify only one. These are the highest (1.5--7$\times 10^{42}$ erg/s) and second-highest (0.8--1.5$\times 10^{41}$ erg/s) [OIII] luminosity bins, the lowest EW([OIII]) bin (12 - 200 \AA), the highest stellar mass ($\log(M_{*}/M_{\odot})=$ 9.34  - 10.55) bins and the highest $M_{UV}$ bin ($-19.29$--$-21.61$) for the $3<z<5$ stacks, and the highest EW bin (1236 - 3060 \AA) for the $5<z<7$ stacks. The highest luminosity stack (3--9.4$\times 10^{42}$ erg/s) at $5<z<7$ also potentially has a broad H$\alpha$ component. However, when we perform our test for outflows, we find that the H$\alpha$ broad component of FWHM =$1000$~\kms is a fairly good fit to the \OIII doublet. We compare the $BIC$ of this fit to the $BIC$ using a broad component with a FWHM of $724$~\kms (found by freely fitting for \OIII broad FWHM rather than forcing the H$\alpha$ width) and find $\Delta BIC = 6$ in favour of the narrower component, but this only just reaches the usual threshold for $\Delta BIC$ criteria, and does not pass our more conservative threshold (equation \ref{bic10}). Therefore, we mark this stack as having a tentative signature of AGN. 

The detection of H$\alpha$ broad components without a counterpart in \OIII in the higher redshift bin suggest that we have been able to separate out AGN sources from outflow sources from the total stack. We present an example in each redshift bin of the broad line detections in Figure \ref{fig:highlum} ($3<z<5$ highest \OIII luminosity) and Figure \ref{fig:highEW} ($5<z<7$ highest \OIII EW). Note that in these figures, the scale on the residuals for the \OIII line and the H$\alpha$ lines are slightly different, but we plot the $\pm 1\sigma$ levels for clarity. The remaining stacks with detected broad components are shown in Appendix \ref{appendix:broadHa}.
The FWHM of the H$\alpha$ broad components that are statistically significant, and without an [OIII] counterpart, are all $\gtrsim 1,000$ \kms; this further supports the BLR scenario. In table \ref{tab:agn_properties} we list $\Delta BIC$ for the H$\alpha$ narrow only minus the broad+narrow model, and the broad H$\alpha$ FWHM for each stack with confirmed AGN signatures. We also list the velocity offset of the fitted broad H$\alpha$ component. Crucially, even though the [O~III] doublet has a potential broad component in some of these stacks, its velocity and FWHM are different from the broad H$\alpha$, as highlighted by the large residuals when we force a broad [O~III] component with the same kinematics as for H$\alpha$. This is also reflected in the $\Delta BIC$ between the \OIII broad component that is restricted to have the same kinematics as the H$\alpha$ broad component, and the \OIII broad component that is fit freely. The FWHM of the freely fitted \OIII components are given in Table \ref{tab:agn_properties} and these are all narrower than the FWHM of the H$\alpha$ broad component which is also given in this table. Table \ref{tab:agn_properties} also lists the values of $BIC_{H\alpha, \text{kinematics}} - BIC_{\text{free fit}}$ which are all $\geq6$. Therefore, the broad H$\alpha$ has no matching kinematic component in [O~III], strongly supporting the BLR interpretation. The absence of a broad counterpart in \OIII also indicates that the broad features in H$\alpha$ are not an artifact of our stacking. For example, if this was caused by small errors in the redshifts when converting the individual spectra to the rest frame, we would expect the same feature to appear in all of the strong emission lines.

We also investigated the possibility that the detected broad components in the stacks are due to individual bright sources dominating the stack. We implemented a Jackknifing test where we re-stack each stack that has a broad line detection N times (where N is the number of sources in the stack), and each time we remove one of the sources from the stack. If any of the resampled stacks do not show a broad component, this means the source that was removed is the one that causes the broad component and there is not an average of AGN activity but rather one galaxy with an AGN. Our testing ruled out this possibility, since for each stack with a broad line detection, the resampled stacks all maintained their broad components. This shows that we have detected an average of AGN activity within the JADES stacks between z=3-7.

We do not detect any broad components in H$\beta$, as shown by the fits of the double Gaussian models presented in Appendix \ref{appendix:appendix_a}. However, as described in Section \ref{sec:line}, the H$\beta$ emission line is at least three times fainter than H$\alpha$, so we expect that if H$\alpha$ broad lines are not detected in individual galaxies but appear in the stacks, then the H$\beta$ could be too faint to even detect in the stacks.

Our results show that there is an average population of AGN within the JADES galaxies at both $3<z<5$ and $5<z<7$, which are below the detection limit for individual objects. These AGN have the potential to be the missing population that is expected to have BHs that lie along the local M$_{BH}$-M$_{*}$ relation. 
To confirm this, we will need to assess whether their BHs are over-massive relative to their host galaxies. This is addressed in Section \ref{sec:agn_props}.

Within the lower redshift bin, the stack of the highest \OIII $\lambda$5007 luminosity sources show evidence for a broad H$\alpha$ associated with BLRs. This is expected, due to the enhancement of \OIII emissions from high energy photons from the accretion disk. There is a tentative detection of broad H$\alpha$ associated with BLRs in the highest \OIII luminosity sources also within the higher redshift bin.

The highest stellar mass sources at $3<z<5$  and the highest EW sources at $5<z<7$ also show evidence in their stacks for a broad H$\alpha$ associated with BLRs. 
As mentioned, higher EW nebular lines are often associated with AGN, and more massive galaxies generally host more massive BHs, which can therefore easier to detect. So the findings of a broad H$\alpha$ associated in these bins is not totally surprising.
However, we must also note that a non-detection does not imply that there are no AGN in these bins - it could mean that there are Type-1 AGN that are still too faint to be detected, or type 2 AGN that are not detectable by their BLR.

In Appendix \ref{appendix:appendix_b}, we describe the reasons why we excluded the remaining stacks as having a broad H$\alpha$ associated with an AGN BLR.

In Appendix \ref{appendix:appendix_a} we also show the H$\alpha$ narrow and broad fits of the AGN candidates that we determined from the normalised stacks. For the highest stellar mass stack, the broad H$\alpha$ emission is weakened by the normalisation, which the narrower value of FWHM$_{\text{broadH}\alpha}$ reflects. The other stacks do not exhibit significant changes to their broad H$\alpha$ profiles. This indicates that it is the sources with the highest $F_{\text{[O III]}}$ that contribute the broad component to the stack because they are weighted down from the normalisation, reducing the broad component in the highest stellar mass stacks, but causing no change in the other stacks with the brightest \OIII emission. This is in agreement with our results that the most luminous \OIII sources are those with the broad components.

We also present the results for the stacks weighted by rms$^{-2}$ in Appendix \ref{appendix:appendix_a}.

 \section{Narrow line AGN diagnostics}
\label{sec:diagnostics}

\begin{figure*}
    \centering
    \begin{subfigure}[b]{0.9\textwidth}
        \includegraphics[width=\textwidth]{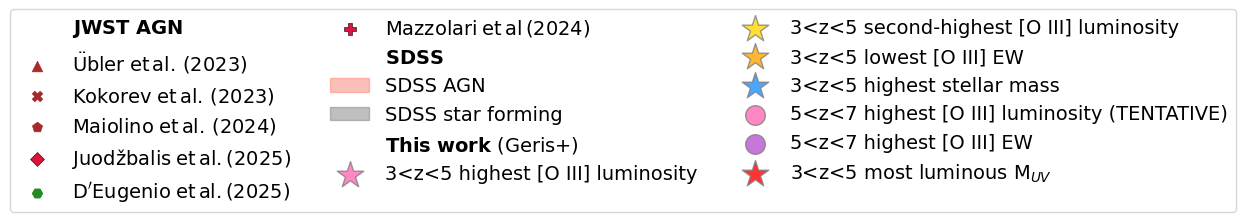}
    \end{subfigure}
    \begin{subfigure}[b]{0.48\textwidth}
        \includegraphics[width=\textwidth]{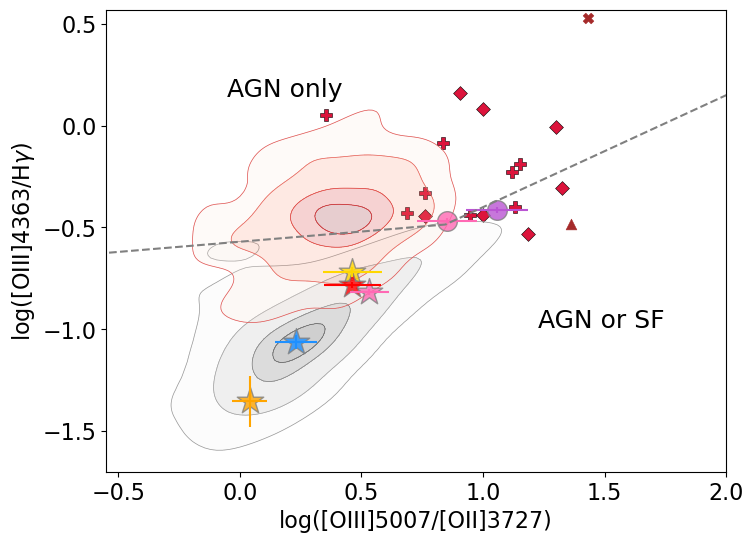}
    \end{subfigure}
    \hfill
    \begin{subfigure}[b]{0.50\textwidth}
        \includegraphics[width=\textwidth]{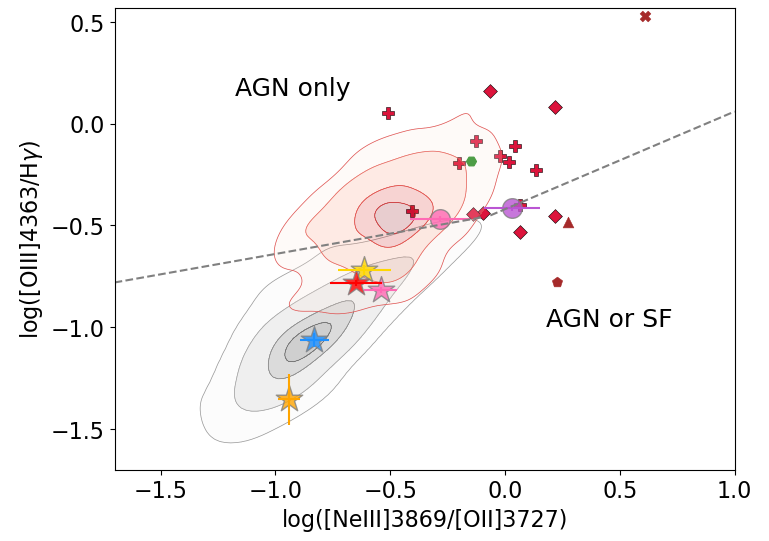}
    \end{subfigure}
    \caption{Location of our stacks on the narrow-line diagnostic diagrams involving the auroral line [OIII]4363, proposed by \citet{mazzolari_new_2024}. The dashed segments provide the separation between the region populated only by AGN (top) from the region populated by AGN and Star Forming galaxies (bottom), as identified by \citet{mazzolari_new_2024}. 
    We also show the locations of AGN identified in \citet{mazzolari_ceers} (red crosses), \citet{ubler_ga-nifs_2023} (red triangle), \citet{kokorev_uncover_2023} (red crosses), \citet{maiolino_small_2024} (red pentagon), \citet{juodzbalis_2025} (red diamonds) and \citet{D'Eugenio+25} (green point).
    Interestingly, the only stack we have labelled as tentative ($5<z<7$, highest [OIII] luminosity), shown as a pink circle, is the only one to fall in the AGN-only region of both diagnostic plots. This provides further evidence that this stack might be a solid detection of AGN, despite the tentative BLR detection. The other stacks fall in the region that can be populated both by AGN and SF glaxies, hence less conclusive,  although the highest EW([OIII] at $5<z<7$ stack is very close to the demarcation line.}
    \label{fig:oiii_diagnostic}
\end{figure*}

In the previous section we have shown that the galaxy population is hosting a population of AGN identified through the broad H$\alpha$ associated with their BLRs. Here we aim to consider further evidence of these AGN in this galaxy population. We investigate the emission line diagnostics based on the \OIII$\lambda$4363, which has proved to be one of the most promising diagnostic in search for AGN even in low metallicity sources \citep{mazzolari_new_2024}, compared to the more typical emission line diagnostics such as BPT \citep{1981PASP...93....5B} and VO87 \citep{1987ApJS...63..295V}, which fail to identify AGN at lower metallicities \citep{harikane_jwstnirspec_2023,kocevski_hidden_2023,maiolino_small_2024,ubler_ga-nifs_2023}.

These diagrams use the following line ratios:

\begin{enumerate}
\item \OIII $\lambda$ 4363 / H$\gamma$
\item \OIII 5007 $\lambda$ / [O II] $\lambda$ 3726+3729
\item {[Ne III] $\lambda$ 3869 / [O II] $\lambda$ 3726+3729}
\end{enumerate}

The effectiveness of this diagnostic is primarily driven by the auroral \OIII$\lambda$4363 emission line, which is sensitive to the temperature of the ISM \citep{maiolino_re_2019}. \jwst detections of strong \OIII $\lambda$4363 have been proposed to indicate AGN activity \citep{brinchmann_high-z_2023, ubler_ga-nifs_2024}. As described in \cite{mazzolari_new_2024}, it is expected that the \OIII $\lambda$4363 line would be boosted by AGN activity, as the high energy photons produced by AGN increase the temperature of the gas in the narrow line region. 
Typically, this line is faint at $z<2$, however, with the sensitivity of \jwst and the high SNR in our stacks, this line is well detected in all the stacks we have presented. The other lines required are also well detected. We measure the fluxes of the emission lines using the same method as described in section \ref{sec:line}, but without adding a second Gaussian component. To correct the fluxes of the \OIII $\lambda$5007 and [O II] $\lambda$3726+3729 lines for dust extinction, we assume an intrinsic narrow line Balmer decrement of 2.86, consistent with the standard scenario of Case B recombination at an electron temperature $T_\mathrm{e} = 10000$K and electron density $n_\mathrm{e} = 100~\mathrm{cm}^{-3}$ \citep[e.g.,][]{2012MNRAS.419.1402G}. Comparing the observed Balmer decrement to the intrinsic ratio and assuming the SMC extinction curve from \cite{2003ApJ...594..279G}, we derived $A_{V}$ for each stack, and we use this to correct the observed narrow lines for extinction. The observed Balmer decrement and $A_{V}$ are reported  in Table \ref{tab:agn_props1}. 
Note that the effect of dust extinction is negligible for the ratio of \OIII $\lambda$4363 / H$\gamma$ due to the close proximity in wavelength of these lines. Figure \ref{fig:oiii_diagnostic} shows the location of our stacks with detected H$\alpha$ broad lines on two of the new \OIII $\lambda$4363 diagnostics proposed by \citet{mazzolari_new_2024}. All of the lower redshift stacks fall in the  region of the diagram where it is not possible to distinguish between AGN and star forming galaxies. However, as we have already determined that sources within these stacks host Type-1 AGN, this diagram does not discount our results. It is also important to note that a conservative approach has been used by \cite{mazzolari_new_2024} to define their demarcation lines, so while ratios above the demarcation lines are unambiguously associated to AGN, many known AGN (and AGN models) fall within the AGN-SFG region. This has also been seen in other studies including \cite{juodzbalis_2025}, where their stacks of Type-1 JADES AGN also fall into the composite region. Since the AGN we detect are low-mass and faint, it is possible that their emission line ratios are influenced by both AGN activity and star formation, as the AGN likely does not dominate the galaxy as much as in brighter high-redshift AGN. Therefore, it would make sense that our AGN fall in the composite region. The positions of the $5<z<7$ highest EW stack in the AGN-only region on the diagnostic involving [Ne III], and at the boundary for the diagram involving \OIII $\lambda$5007, confirm that this stack displays a population of AGN. An interesting result from these diagnostics is that the stack we have marked as a tentative detection of broad H$\alpha$ (highest luminosity $5<z<7$), is the only one to fall directly in the AGN-only region of both diagnostics. This implies that this stack does in fact show evidence of AGN activity, despite the presence of outflows that we detected and makes the identification of the broad H$\alpha$ from the BLR more problematic.

\section{AGN and BH properties}
\label{sec:agn_props}
\subsection{Black hole masses}
\label{sec:mbh}
We can estimate the average masses of the black holes detected in our stacks using the luminosity of the broad H$\alpha$ line and its FWHM. Using local virial relations that relate these quantities is the standard method to determine the BH masses of type-1 AGN at high redshift \citep{maiolino_jades_2024, juodzbalis_2025, harikane_jwstnirspec_2023}.
It has been suggested that these relations might not apply to high redshift, especially in the super-Eddington regime. However, it is not clear in which direction the systematics would go, with some models claiming that those relations would overestimate the BH masses by up to a factor of 5 for accretion rates as high as 100 times higher than the Eddington limit \citep{Lupi2024,Lambrides2024,King2025}, while other studies suggest that black hole masses using virial relations might be underestimated by more than an order of magnitude already at the Eddington limit, because of the strong radiation pressure that reduces the effective gravitational force felt by the BLR clouds \citep{Marconi2008,Marconi2009}. Additionally, there have been claims that for the specific case of Little Red Dots (LRDs), the virial relations may be overestimating the black hole masses by about two orders of magnitude because of electron or Balmer scattering \citep{2025arXiv250316595R, Naidu2025_BHS}. Yet, \citet{juodzbalis_2025} discuss in detail that the latter scenario is likely untenable on multiple grounds.
Additionally, it is comforting that surveys with the GRAVITY interferometer have directly measured the black hole mass of a quasar at z=2.3, accreting well above the Eddington limit ($L/L_{Edd}\sim 7-20$), and found that it is within a factor of 2.5 of what expected from the virial relations (when using the broad H$\alpha$) \citep{GRAVITY2024}, i.e. well within the scatter of the relations. Direct BH mass measurements have been recently extended with GRAVITY+ to more high-z quasars and finding a similar level of consistency with the local scaling relations (GRAVITY+ collaboration, in prep.). Regarding LRDs, the variability studies in one case at z=7 \citep{Ji2025QSO1} has indicated no deviation in the relation between BLR radius and luminosity, which is underlying the virial relation. Even more compelling, Juodzbalis et al. (in prep.) have obtained the direct measurement of the black hole mass in a lensed LRD at z=7, resolving its sphere of influence, and finding a value fully consistent with the virial relations.

Summarising, there is no compelling evidence that the Black Hole masses of high-z AGN are significantly different from those inferred assuming the local virial relations. Therefore, 
we use the relation from \cite{2015ApJ...813...82R}, given by

\begin{equation}
\begin{split}
\log (M_{\rm BH}/\msun) = 6.60 + 0.47 \log & \left(\frac{L_{\rm H\alpha}}{10^{42}~\mathrm{erg\,s^{-1}}} \right) + \\
2.06 & \log \left(\frac{\text{FWHM}_{\rm H\alpha}}{1000~\kms}\right),
\label{mass}
\end{split}
\end{equation}

where $L_{\rm H\alpha}$ is the luminosity of the broad H$\alpha$ line and ${\text{FWHM}}_{H\alpha}$ is its FWHM.

To obtain the luminosity of the broad line, we first correct its flux for dust attenuation. Estimating the dust-corrected flux for the BLR of AGN is difficult, as the intrinsic ratio of the H$\alpha$ and H$\beta$ fluxes can vary depending on the enhancement of H$\alpha$ relative to H$\beta$ from collisional excitation in the extreme, high density conditions of the BLR. Regardless of this problem, we do not detect broad H$\beta$ in the stacks. Therefore, we use the narrow line Balmer decrement, H$\alpha$/H$\beta$, to obtain the dust-corrected broad H$\alpha$ flux using the values of A$_V$ in Table \ref{tab:agn_props1}. By using the narrow line properties, we are assuming that the dust obscuration of the broad line region has the same origin as the narrow line obscuration, i.e. the ISM of the galaxy \citep{2022A&A...666A..17G}, similarly to \cite{juodzbalis_2025}. We derive the corrected broad H$\alpha$ flux which are given in Table \ref{tab:agn_props1} along with values of the observed broad line flux, observed Balmer decrement, $A_{V}$, and median redshift of the stack. The corrected fluxes are converted to luminosity, using the median redshift from each stack, to calculate the luminosity distance.

We find average black hole masses of $\log(M_{\rm BH}/M_{\odot})\sim 6.4$ (see Table \ref{tab:agn_props2}). The uncertainty was derived by propagating the errors in $L_{H_{\alpha}}$ and ${\text{FWHM}}_{H\alpha}$, but also includes a 0.3 dex uncertainty added to the measured uncertainty in quadrature, that comes from the systematic scatter in equation \ref{mass}. These values are consistent with the lower end of the mass distribution obtained from \jwst observations of individual high-z AGN \citep{juodzbalis_2025,Taylor2024,harikane_jwstnirspec_2023,Kocevski2025} -- including lensed systems \citep[e.g.,][]{furtak+2024,Ji2025QSO1,DEugenio_QSO1}. This is expected from broad lines that are revealed through stacking, since these black holes must be small for their broad H$\alpha$ to go undetected in the individual spectra, but must be massive enough to produce a reasonable broad component in the stack that can be confidently attributed to AGN origin within the R1000 data. Although AGN of similar masses have been discovered in individual galaxies, the AGN discovered in our stacks are some of the smallest black holes ever discovered at these redshifts. For example, the smallest Type-1 AGN identified in \cite{maiolino_jades_2024} has a mass of $\log \left(\frac{M_{\text{BH}}}{M_{\odot}}\right) = 5.65$ and potentially comes from a dual AGN system. Our stacks show that there is an average population of these relatively small objects among galaxies between $3<z<7$, as predicted by various seeding models. For instance, DCBH models predict seed masses of $\log \left(\frac{M_{\text{BH}}}{M_{\odot}}\right) = 4-6$ \citep{greene_intermediate-mass_2020}.

Finally, in Appendix \ref{appendix:masses}  we compare the black hole masses derived from the unweighted and un-normalised stacks, to those derived from the stacks normalised by \OIII flux, and weighted by inverse variance, with the warning that in these cases the information on the intrisic fluxes is partially lost due to the normalization and weighting strategy.

\begin{figure*}
	\includegraphics[width=1.6\columnwidth]{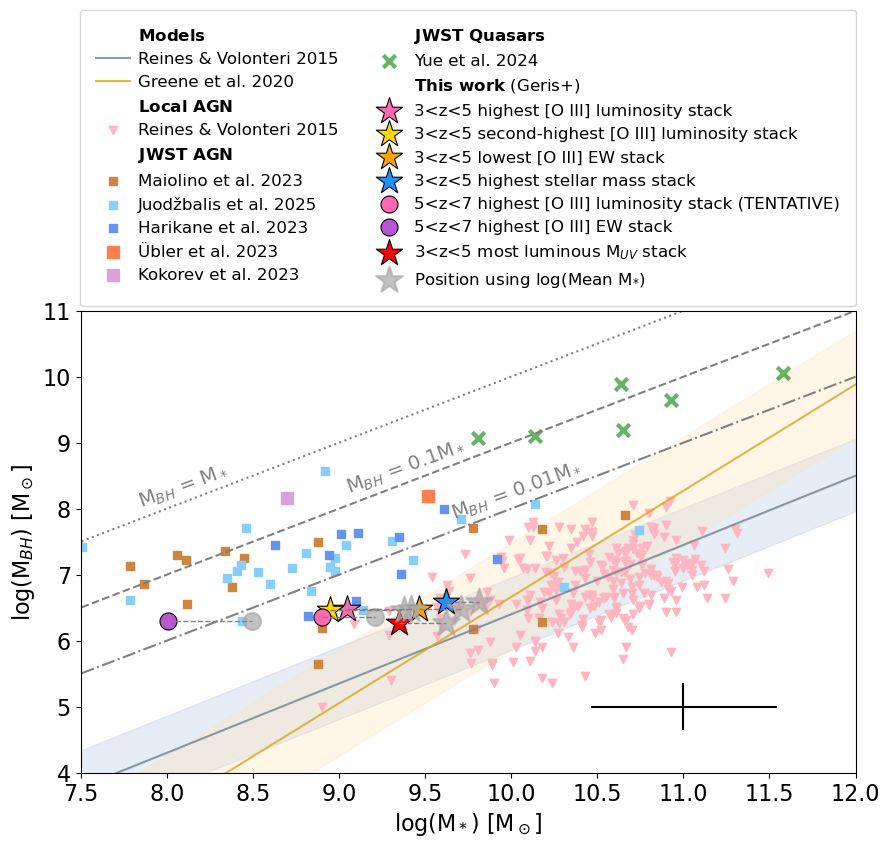}
    \caption{Black hole masses $M_{\text{BH}}$ derived from the stacks with confirmed broad H$\alpha$ components plotted against the median $\log$ stellar mass $M_{*}$ of their host galaxies. The star ($3<z<5$) and circle ($5<z<7$) data points show the positions of our stacks. The squares  show previously identified \jwst broad line AGN from \citet{maiolino_jades_2024} (brown) and \citet{juodzbalis_2025} (light blue), \citet{harikane_jwstnirspec_2023} (dark blue), \citet{ubler_ga-nifs_2023} (orange) and \citet{kokorev_uncover_2023} (purple). The green crosses show the locations of quasars whose galaxies have been studied with \jwst \citep{2024ApJ...966..176Y}. The grey stars and circles show the positions of the stacks if plotted against $\log$(Mean M$_{*}$) instead, with dashed lines connecting to their counterpart - see Section \ref{s.discussion} for a discussion on this. The pink triangle data points show the positions of local AGN from \citet{2015ApJ...813...82R}. The solid lines show local relations derived from \citet{2015ApJ...813...82R} (grey) and \citet{greene_intermediate-mass_2020} (gold). The dashed/dotted lines show the relations for $M_{\text{BH}} = M_{*}$, $M_{\text{BH}} = 0.1M_{*}$ and $M_{\text{BH}} = 0.01M_{*}$. The black cross shows the uncertainty in the estimation of the BH masses (vertical error bars), and the typical range of stellar masses in each bin (horizontal error bars).}
    \label{fig:mbh-mstar_new}
\end{figure*}

\subsection{Black hole versus stellar mass relation}

In Fig.~\ref{fig:mbh-mstar_new} we show the black hole masses from each stacked bin versus the median stellar mass (given in Table \ref{tab:agn_props2}), together with measurements from the literature on other individual objects selected with JWST. It is clear that this population of black holes mostly falls below the masses of previously discovered high-z AGN. Both the highest M$_{\rm UV}$ and highest stellar mass bins at $3<z<5$ lie within the scatter of the local scaling relation. This indicates that within the highest stellar mass galaxies (M$_{*}$/M$_{\odot} \sim 9.7$) and within the galaxies with the brightest UV magnitude (M$_{\rm UV}$ = -19.85), there is an average population of Type-1 AGN that are consistent with the local scaling relation and their BHs are therefore less over massive relative to their hosts than the previously discovered \jwst AGN. Our stacks reveal the population of Type-1 AGN that were missing from previous \jwst observations. 

An interesting point is that higher stellar masses generally indicate more evolved systems. Therefore, the location of the highest stellar mass bin within the scatter of the local scaling relation could imply that BHs in dwarf galaxies could start off as overmassive, but head towards the local scaling relation as they evolve.

Most of the other stacks are also potentially consistent with the local scaling relation within uncertainties, although on average still above the local relation, and with the notable exception of the stack with the highest EW([OIII]), which is  about 2 dex above the relation.

This shows that black holes of mass $\sim 10^{6}$ M$_{\odot}$ exist in both more massive and less massive galaxies resulting in both an overmassive and non-overmassive population.

Our findings are also consistent with other studies suggesting that selection effects alone, although playing an important role, do not fully account for the deviation from the local $M_{\text{BH}}$–$M_{*}$ relation \citep{juodzbalis_2025} because we are still observing over-massive BHs even in stacked spectra, which reveal those that are too difficult to see in the individual spectra. However, we still fail to probe the region beneath the local relation, even when analysing the black holes in the {\it highest stellar mass galaxies}, which we would expect to be the least over massive, therefore lying below the relation. This further indicates that the relation is intrinsically skewed towards overmassive black holes.

Another interesting point, as highlighed above, is that the highest EW stack in the higher redshift bin deviates far from the local scaling relation. This indicates that we have also identified a population of Type-1 AGN that, although having low BH masses, are more similar to the over-massive BHs previously discovered by \jwst than our newly identified less over-massive population.

\begin{figure*}
	\includegraphics[width=2\columnwidth]{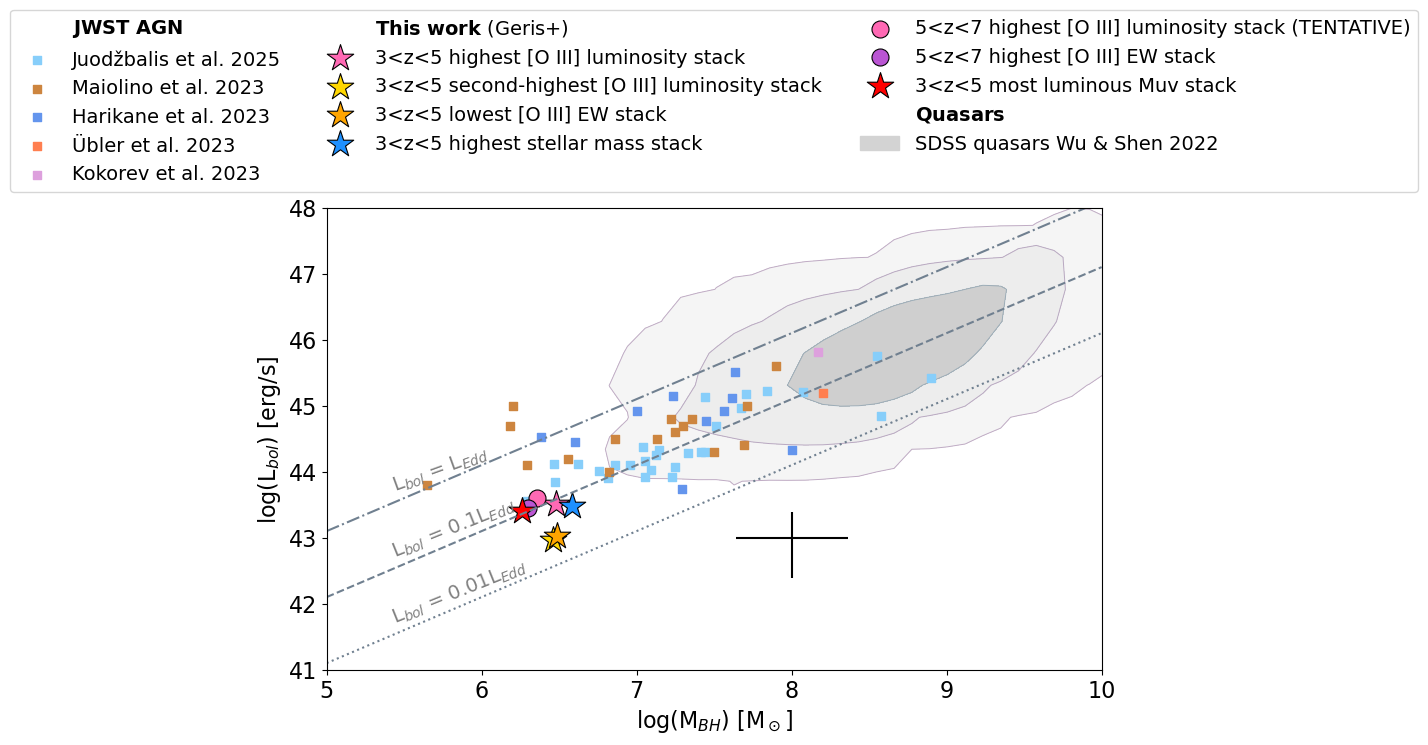}
    \caption{Average black hole bolometric luminosity $L_{\text{Bol}}$ as a function of mass $M_{\text{BH}}$ derived from the stacks with confirmed broad H$\alpha$ components associated with BLRs. The star and circle data points show the positions of our stacks. The squares show previously identified \jwst broad line AGN from \citet{maiolino_jades_2024} (brown) and \citet{juodzbalis_2025} (light blue), \citet{harikane_jwstnirspec_2023} (dark blue), \citet{ubler_ga-nifs_2023} (orange) and \citet{kokorev_uncover_2023} (purple). The dashed/dotted lines show the relations for $L_{\text{Bol}} = L_{Edd}$, $L_{\text{Bol}} = 0.1L_{Edd}$ and $L_{\text{Bol}} = 0.01L_{Edd}$. The contours show the SDSS quasars from \citet{2022ApJS..263...42W}. The black cross shows the average uncertainties of our stacks.}
    \label{fig:Lbol-mbh_new}
\end{figure*}

\subsection{Luminosity and accretion rates}

The key properties of an AGN are its luminosity and Eddington ratio ($L_{\text{bol}}/L_{\text{Edd}}$). In order to estimate the bolometric luminosity we use the well known scaling relation between the H$\alpha$ luminosity and bolometric luminosity (L$_{\rm bol}$) from \cite{2012MNRAS.423..600S}: 

\begin{equation}
L_{Bol} = 130 L_{H\alpha}
\end{equation}

where $L_{H\alpha}$ is the luminosity of the broad H$\alpha$ component. Similarly to Equation \ref{mass}, this relation has a scatter of 0.3 dex, which is taken into account in the uncertainties in $L_{\text{bol}}$, presented in Table \ref{tab:agn_props2}. We also calculate the Eddington ratio (see table \ref{tab:agn_props2}), to assess how efficiently this population of high-z AGN is growing. In Fig.~\ref{fig:Lbol-mbh_new} we show the average bolometric luminosity of the AGN in our stacks, against black hole mass, together with results on individual targets from the literature. In the same figure we show the luminosity corresponding to 1\%, 10\%, and 100\% the Eddington luminosity. All  AGN detected in our stacks are on average less luminous than the individual Type-1 AGN detected in JADES.
The bolometric luminosities between the different bins are consistent within uncertainties with an average bolometric luminosity of $\log(L_{\text{bol}}/\text{ergs}^{-1}) \approx 43.3$. In terms of black hole masses, we are probing the low mass end of the distribution of previous JWST-identified AGN.
Not surprisingly, the accretion rates inferred from our stacks are typically of the order of $L/L_{Edd}\sim 0.1$, slightly lower than the individual AGN, which is not surprising given that in the stacks we probe fainter AGN. Interestingly, two of the stack probe accretion rates as low as $L/L_{Edd}\sim 0.02-0.03$, i.e. nearly dormant black holes. It is likely that if there are AGN in the the other stacks in which we have not identified a broad H$\alpha$ from a BLR, they are accreting at even lower rates, suggesting a large population of dormant black holes at these redshifts, consistent with previous findings \citep{juodzbalis_dormant_2024}.

\begin{figure*}
	\includegraphics[width=1.5\columnwidth]{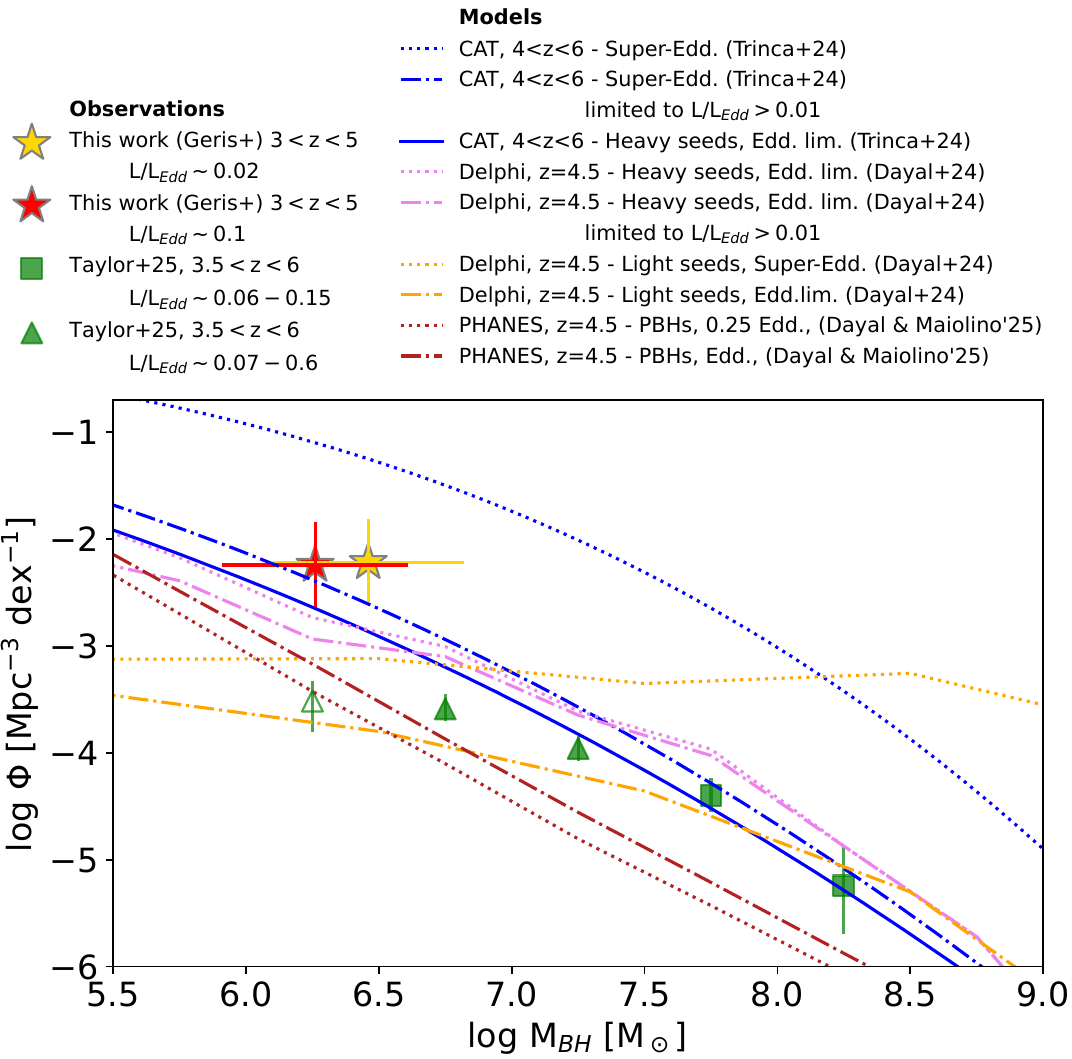}
    \caption{Black Holes Mass Function inferred from two of our stacks at $3<z<5$ (red and gold stars), together with the previous estimate by \citet{Taylor2024} at higher masses and at $3.5<z<6$ (green symbols, where the hollow symbol indicates the point flagged by them as affected by low completeness). It should be noted that each of these points probe only active black holes, with characteristic accretion rates given in the legend. The various lines show predictions by semi-analytical models with various black hole seeding and growth assumptions, as indicated in the legend and as discussed in the text. In nearly all cases, the lines associated with models provide the predicted BHMF for the whole population of BHs, including dormant ones and those accreting at rates below the detection limits. The only exceptions are: the dot-dashed blue line, which shows the  CAT SAMs in the case where BHs are allowed to accrete in super-Eddington bursts (dotted blue lines), but where only active black holes with $L/L_{Edd}>0.01$ are selected; the dot-dashed violet line, which shows the Delphi SAM with heavy seeds and Eddington limited (dotted violet line), but where only active black holes with $L/L_{Edd}>0.01$ asre selected. These two dashed lines are therefore more directly comparable with the observations.
    }
    \label{fig:bhmf}
\end{figure*}

\section{Black Hole Mass Function}
\label{sec:bhmf}

Our stacking is probing low black hole masses, $M_{BH}\sim 10^6~M_\odot$, that were hardly probed by previous studies. It is worth attempting to estimate the density of these low mass black holes, i.e. their associated BH mass function (BHMF), as it may be providing important constraints on the BH seeding and growth scenarios. Previous studies have attempted to derive the BHMF at high redshift 
\citep{Taylor2024,matthee_little_2024,He2024_BHMF,Wu2022_BHMF,Shen_Kelly_2012_BHMF}, with partially consistent results, especially when considering caveats of the studies associated with incompleteness in the various BH mass bins, given by the capability of identifying smaller black holes (which are associated with narrower and fainter broad lines). One additional issue to consider is that these studies only take into account the black hole mass function of accreting black holes. The complete BHMF should take into account the duty cycle and/or the distribution of accretion rates. We further discuss this aspect below.

We estimate the BH mass function associated with our stacking as follows. We consider the stacks in bins of M$_{UV}$. Only the most luminous bin (or 'highest' as referred to in other parts of this paper), centered at $M_{UV} = -19.7$ shows evidence for a broad component of H$\alpha$, from which we have inferred a BH mass of $1.7\times 10^6~M_\odot$.
The inferred Eddigton ratio is $L/L_{Edd}\sim 0.1$.
We can make the assumption that each galaxy in this $M_{UV}$ bin contributes to the stack with a BH of this mass. In reality, there will be a mixture of BHs accreting with different rates around $L/L_{Edd}\sim 0.1$ contributing to this UV luminosity bin, and also somewhat different BH masses. Yet the BH masses cannot change by a large factor, both because they scale quadratically with the width of the lines (hence would result in different line profiles) and because more massive black holes would be detectable in individual Type-1 AGN, hence have already been revealed and presented in the JADES Type-1 sample paper \citep{juodzbalis_2025}
and excluded from the stack. Then, although the JADES spectroscopic selection function is complex, we assume that the galaxies spectroscopically observed by JADES in this UV luminosity bin are representative of the broader population of galaxies in the same redshift interval and in the same luminosity bin. We then take the volume density of galaxies in this bin via the analytical equation for the UV luminosity function as a function of redshift provided by \citet{Bouwens2021_UVLF}, at the redshift corresponding to the median redshift in our bin. Since the UV luminosity function is per unit magnitude, we multiply the value given by \citet{Bouwens2021_UVLF}'s equation by the size of our $M_{UV}$ bin ($\Delta M_{UV}=2.3$). The BHMF must be per unit BH mass interval, so we need to assess the BH mass interval spanned by our stack, which we do not know as we do not have the BH mass measurements of each individual source going into the stack. However, for the purpose of having an estimate of this interval, we can make the approximation that the BH mass in this interval is (to a first order) proportional to the stellar mass, hence we can approximatively derive the BH mass interval from the stellar mass interval in ths bin (i.e. $\Delta \log{M_{BH}} \sim \Delta \log{M_*} \approx 1.1 ~dex$, obtained by the 16\%-84\% of the mass distribution in the bin). Hence we obtain the BH mass density per dex by dividing the volume density by this value. The resulting estimated volume density of BHs with $\log{(M_{BH}/M_\odot)}=6.26$ in the redshift interval $3<z<5$ is $\Phi = 5.7\times 10^{-3}~Mpc^{-3}~dex^{-1}$. The uncertainty on this value is probably dominated by the cosmic variance, which \cite{juodzbalis_2025} estimated to be about 0.4~dex around this UV luminosity. One caveat of our method, is that BHs with similar masses may be present in the other $M_{UV}$ bins, but there are not enough (or they are not accreting enough) to be detected when diluted in the stack. Therefore, our estimation might be considered a lower limit, althogh the non-detection of broad H$\alpha$ in other $M_{UV}$ bins indicates that the real density cannot be higher than a factor of about 2, or else those BHs would be detected \citep{juodzbalis_dormant_2024}.

In Fig.\ref{fig:bhmf} we report our value for the BHMF at $3<z<5$ at $\log{(M_{BH}/M_\odot)}=6.26$ (red star). We recall that the median Eddington ratio in this bin is $L/L_{Edd}\sim 0.1$. Therefore, although we reveal very low black hole masses, we are still probing fairly active black holes, although on average less active than those identified through individual detections of the broad lines.

In order to probe black holes with even lower accretion rates, and therefore possibly more representative of the global population, we repeat the same calculation by taking our stacking bin which has the lowest average Eddington ratio. This is the bin with second highest [OIII] luminosities at $3<z<5$, which has $L/L_{Edd}\sim 0.025$. Because of detectability limits, the lower Eddington ratio is paid with a higher average BH mass of $M_{BH}\sim 3\times 10^6~M_\odot$. We estimate the black hole density in this case with the same method as for the highest $M_{UV}$. However, for this stack, we do not have a sharp UV luminosity bin out of which to extract the volume density from the UV luminosity function. Therefore, in this case we take the median UV luminosity of the galaxies in the bin, and the 16\%--84\% percentiles of their distribution, and then derive the volume density as in the previous case. We derive a volume density for black holes with 
$\log{(M_{BH}/M_\odot)}=6.46$ in the redshift interval $3<z<5$ of $\Phi = 6.0\times 10^{-3}~Mpc^{-3}~dex^{-1}$. This value is reported with an gold star in Fig.\ref{fig:bhmf}.
We remind that in this case we are probing black holes that are accreting at $L_{Edd}\sim 0.02$.

We compared our derived values with the BHMF inferred by \cite{Taylor2024} (green symbols) at higher masses, although at somewhat higher redshift ($3.5<z<6$). The hollow symbol is for their lowest mass point where they warn about very low incompleteness. We note that in the two highest BH mass bins (green squares) their sample probes accretion rates in the range $L/L_{Edd}\sim 0.06-0.15$ (10\%--90\% percentiles), while their lower mass sample ($M_{BH}<3\times 10^7~M_{\odot}$, green triangles) probes Eddington ratios of $L/L_{Edd}\sim 0.07-0.6$, so significantly higher than our stack and probably missing a significant fraction of the mildly accreting or dormant black holes. This confirms the increasing incompleteness of individual BH detections at low accretion rates when exploring lower masses, as already pointed out by \citet{juodzbalis_dormant_2024}. This is highly relevant for the comparison with the predictions of models, as discussed in the following.

In Fig.\ref{fig:bhmf} we show the predictions of some semi-analytical models (SAMs) and analytical models. The CAT SAMs \citep{Trinca2022,Trinca2023,Trinca2024} that include heavy seeds and assume Eddington limited accretion are indicated with a blue solid curve. This, in principle, reproduces fairly well our $M_{UV}$ stack (red star) and the (more complete) high mass bins of Tylor et al. However, one must consider that the model BHMF includes all BHs, including those that are totally dormant. Therefore, it is likely that this model actually underpredicts the BHMF. The dotted blue line shows the  CAT SAMs where seeds, both light and heavy, are allowed to accrete in short super-Eddington bursts. Although short, these super-Eddington bursts can be very effective in rapidly boosting the black hole mass in the simulation. The simulation is clearly above all observed points of the BHMF. However, also in this case the theoretical mass function is for all BHs, including the dormant ones, while observations are for active BHs. In this super-Eddington version of the CAT model the issue is even more relevant; indeed, in this case BHs spend most of their time in a dormant phase \citep[see e.g.][]{juodzbalis_dormant_2024}. Therefore, for this accretion scenario, we have also derived the SAM BHMF for those black holes that are accreting at $L/L_{Edd}>0.01$, more comparable with our observational points, and which is shown with a blue dot-dashed line. In this case the super-Eddington CAT model passes well through the value inferred from our $M_{UV}$ stacked bin (which has $L/L_{Edd}\sim 0.1$, red star), it is also consistent (within 1$\sigma$) with the second highest [OIII] luminosity stacked bin (which has $L/L_{Edd}\sim 0.02$, gold star),
and also reproduces well the high mass (more complete) points by Taylor et al. ($L/L_{Edd}\sim 0.06-0.15$). Given the several uncertainties and assumptions both in the observational points and in the models, the agreement across a broad range of BH masses is remarkable and is probably indicating that super-Eddington accretion is an important aspect of the early growth of black holes.
As a byproduct, this comparison may confirm that 
the largest population of BHs at these epochs is mostly dormant, corroborating other independent findings \citep{juodzbalis_dormant_2024}.

We also show, with a violet dotted line, the Delphi SAM fiducial model, which assumes heavy seeds and Eddington-limited accretion
\citep{Dayal2019,Dayal2025}. This model is also reasonably close to the observational points. However, as discussed, for comparing with the observations one should only consider the active population in the model. This is shown with the violet dot-dashed line, which illustrate the Delphi SAMs when only BHs accreting at $L/L_{Edd}>0.01$ are considered. In this case the deviation is not major (indicating that in this simulation most black holes are in the relatively active phase), although there is an increasing discrepancy with the lowest black hole mass points probed by our stacks.

The DELPHI SAM that include only light seeds fall short of our BHMF at $M_{BH}\sim 2-3\times 10^6~M_\odot$, regardless of the assumption that they accrete Eddington limited (dot-dashed orange line) or at super-Eddington (dotted orange line). The discrepancy is certainly large because the DELPHI mass function also includes dormant black holes which do not contribute to our stack. This suggests that light seeds alone are probably not a viable scenario for explaining the early population of BHs.

The dotted and dot-dashed red lines lines show the BHMF predicted by the PHANES analytical models \citep{Dayal2024_PBH,Dayal2025_PBH}, which provide a prescription for the cosmic evolution of the population of Primordial Black Holes (PBHs), which are predicted to form shortly after the Big Bang according to some theories
\citep[e.g.][]{Escriva2024,Carr2024}. The dot dashed blue line is for PBHs accreting on average at the Eddington limit, while the dotted blue line is for PBHs accreting at 0.25 their Eddington limit. These models have a slope similar to the observed BHMF, but underpredict the observed values by about 1 dex. At face value this result may indicate that PHB are unlikely to make the bulk of the BH population at z$\sim$3--5. Yet, it may be possible that by varying the accretion prescriptions or the initial PBH mass spectrum, these PBH models might reproduce better the observed points. It is also true that the possible imprint from PBHs is likely washed away at this late epoch. A more proper constraint might come from BHMFs at higher redshifts.

Overall, the comparison of our BHMF at low BH masses ($M_{BH}\sim 10^6~M_\odot$) with models seems to favour super-Eddington accretion and probably also heavy seed models.

We conclude by warning that a more accurate comparison between models and observations, which properly takes into account the full distribution of accretion rates and their detectability, should be done via forward modelling of each simulation through the sensitivity and selection criteria of each survey. This a complex approach that is beyond the scope of this paper and will be presented in a separate work.

\begin{figure}
	\includegraphics[width=\columnwidth]{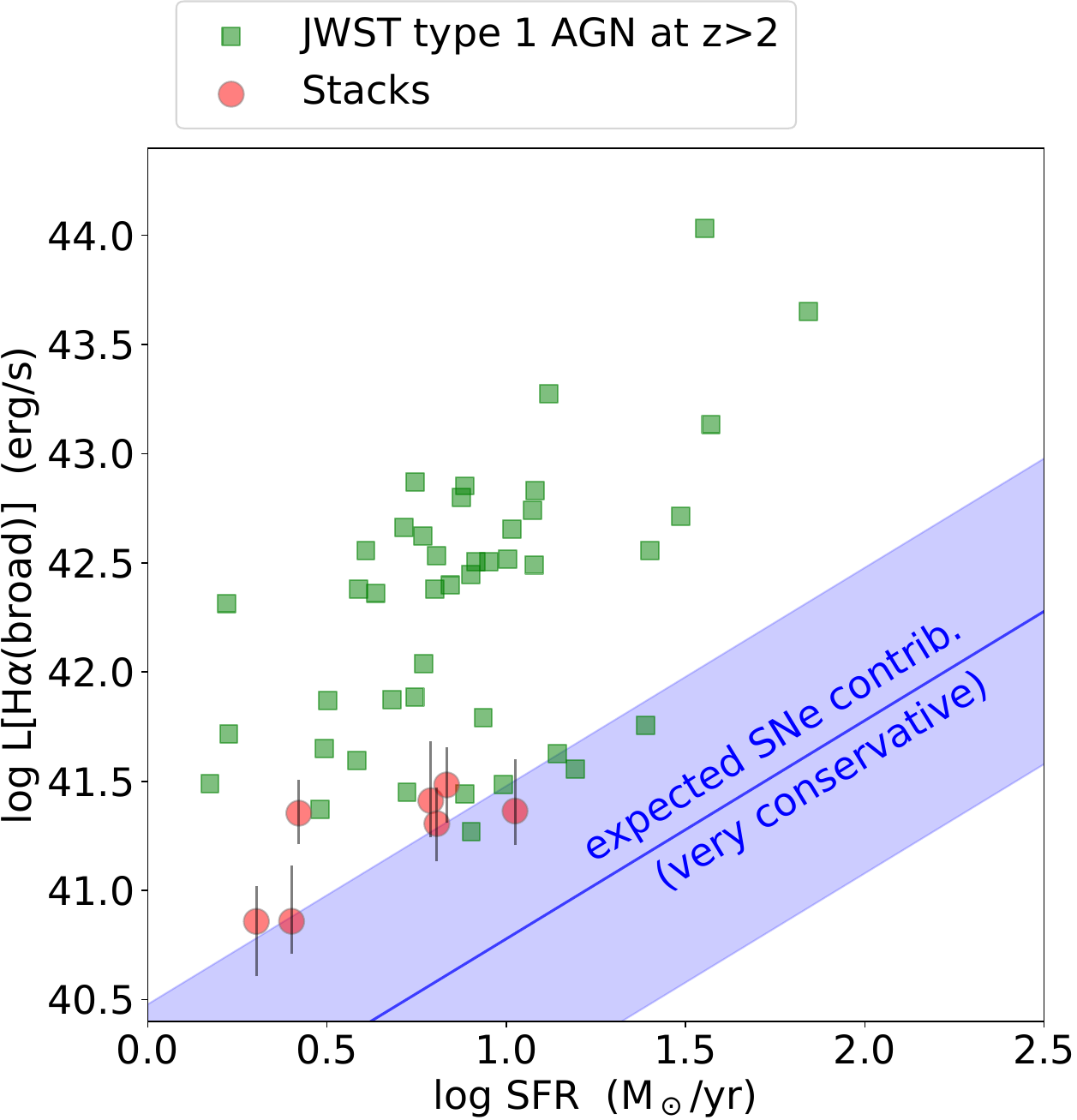}
    \caption{Luminosity of the broad H$\alpha$ in the stacked spectra versus SFR, in the scenario where the broad H$\alpha$ is produced by core collapse SNe. The blue line and shaded region is the expected relation according to the very conservative simple model presented in the text. Green squares are Type-1 AGN for which the broad H$\alpha$ is detected individually \citep{maiolino_jades_2024,juodzbalis_2025,matthee_little_2024,Maiolino2025_Xrays}. The red circles are for the stacked spectrum with broad H$\alpha$ detections presented in this paper.
    }
    \label{fig:SNe_stack}
\end{figure}

\section{Exploring the case of supernovae}
\label{sec:SNEs}

At the faint activity levels that we are probing with the stacking the broad H$\alpha$ line may potentially be contributed also by phenomena not related to AGN.
We have ruled out outflows as possible origin of the broad H$\alpha$ via the non-detection of the same component in [OIII]. However, a broad component of H$\alpha$ could also originate from core-collapse supernovae (SNe). This possibility was discussed by \citet{Maiolino2025_Xrays} in the context of individual galaxies with broad H$\alpha$ and lacking X-ray emission. They concluded that SNe could not be responsible for the broad H$\alpha$ in the vast majority of cases, based on the luminosity of this component and lack of variability. Yet, they did not exclude that SNe might have contributed to the objects displaying the faintest broad H$\alpha$ emission.

In the case of the stacked spectra, we are reaching such faint emission levels in the detection of broad H$\alpha$ that it might be possible that the cumulative contribution of core-collapse SNe in the galaxies included in the stack might be significant, possibly detectable and contributing to the signal that we are observing. We investigate this scenario in this section.

Core collapse SNe have a very diverse distribution of H$\alpha$ lumionsities and profiles, which vary in different ways with time after explosion.
We estimate the expected luminosity of H$\alpha$(broad) contributed by SNe by making a very simple, but conservative toy model. Specifically, we assume that core-collapse SNe have a $L(H\alpha_{broad})=10^{41}$ erg s$^{-1}$ for 3 years. This is a very conservative assumption as most core collapse SNe do not reach such high luminosities in H$\alpha$ and decline on shorter timescales \citep{Taddia2013_SNe,Pastorello2002_SNe,Kokubo2019_SNe}. There are rare superluminous SNe that barely reach H$\alpha$ luminosities of $10^{42}$ erg s$^{-1}$, but they decline even faster (probably because they are releasing energy more quickly). We then assume a SN rate of one SN every 50 years for a SFR of 1 $M_\odot~yr^{-1}$. With these assumptions we derive that the overlap of the SN signatures in the different galaxies in the stack should result into a broad H$\alpha$ luminosity  as a function of SFR given by $L(H\alpha _{broad})\approx 6\times 10^{39} \left( \frac{SFR}{M_\odot ~yr^{-1}}\right)$ erg s$^{-1}$, which is shown with a solid blue line in Fig.\ref{fig:SNe_stack}. We assume an uncertainty on this conservative relation of a factor of 5, to take into account the possible average variation of the luminosity of the individual SNe and light curves.

The green squares are for the individual AGN with broad lines reported in \citet{maiolino_jades_2024}, \citet{juodzbalis_2025} and \citet{matthee_little_2024}, as already reported in \citet{Maiolino2025_Xrays}. The red circles show the location of the points from our stacks. They are all well above the conservative line, and some of them also above the generous uncertainties. It is therefore possible that some of the broad \Halpha emission can potentially have some contribution from  SNe. 

\begin{figure}
	\includegraphics[width=\columnwidth]{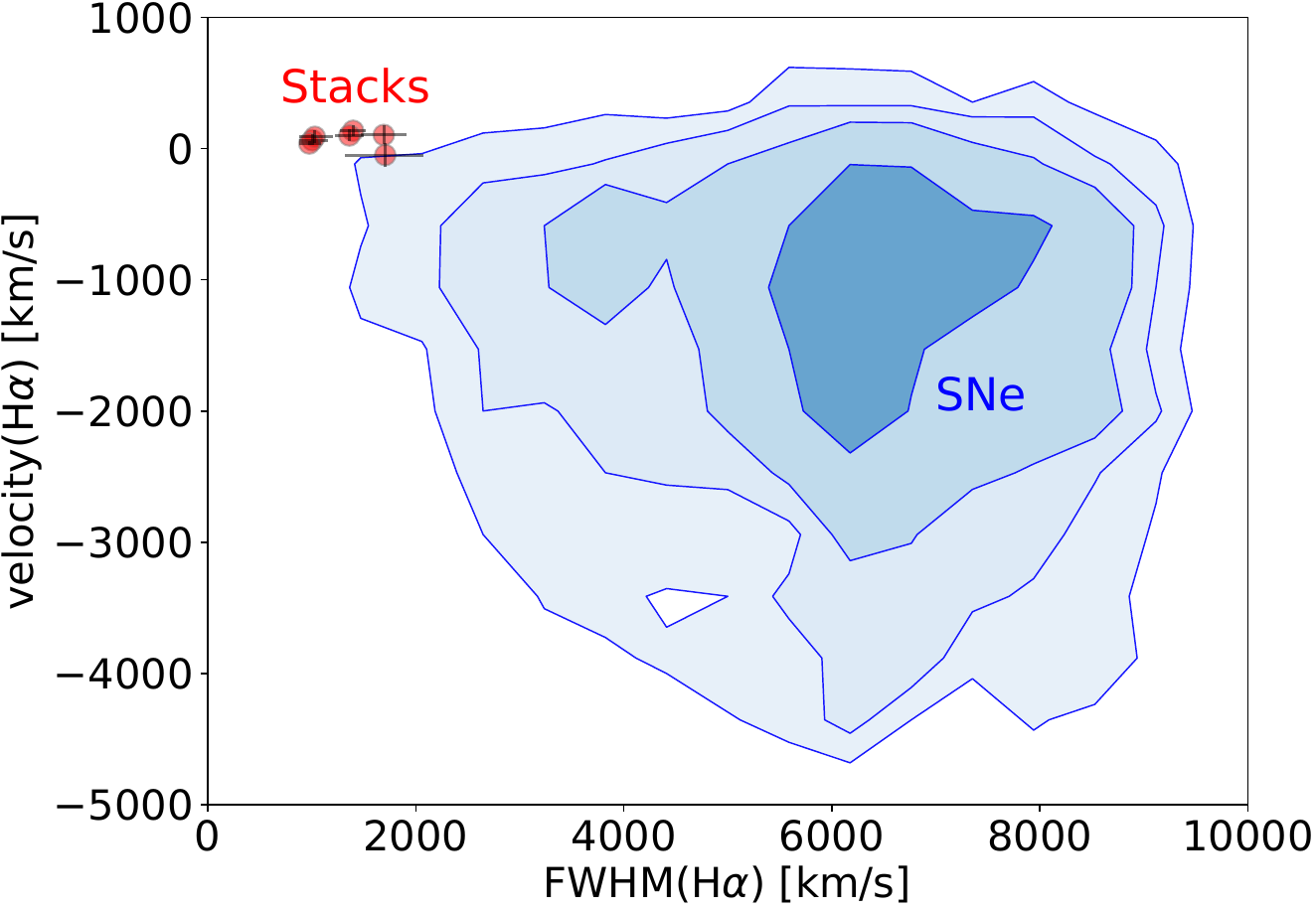}
    \caption{Distribution of line shift and FWHM for the broad H$\alpha$ lines. The profile of the broad H$\alpha$ lines in our stacks are shown with red circles and are compared with the distribution observed in core collapse SNe \citep[blue contours, ][] {Anderson2014_SNe, Guiterrez2017_SNe}.
    }
    \label{fig:SNe_FWHM_shift}
\end{figure}

However, the SN scenario has issues in terms of line profile. Indeed, most core collapse SNe show a P-Cygni profile of the broad H$\alpha$, which we do not detect in the stack. However, it is possible that the P-Cygni profile is not seen because of the low spectral resolution and limited SNR. Yet, an additional issue is that all core collapse SNe typically show a much broader H$\alpha$ line (typically with FWHM between 2,000 and 10,000 \kms), much wider than the H$\alpha$ broad component in our stacks. It is still possible that in the stacks we are seeing only the narrower core of the broad component H$\alpha$ and that the broader wings are hidden in the noise. However, the SN scenario has an additional problem; indeed, the broad H$\alpha$ in SNe quickly becomes blueshifted (typically by a few 1000~\kms) because of dust formation in the ejecta. There is no evidence for blueshift in the broad component of H$\alpha$ of the JADES stacked spectra; actually, all of them are slightly redshifted. The comparison of the broad H$\alpha$ profile, in terms of FWHM and shift, between the lines detected in our stack (red circles) and the distribution for core collapse SNe \citep[blue contours, inferred from][]{Anderson2014_SNe,Guiterrez2017_SNe}, is shown in Fig.\ref{fig:SNe_FWHM_shift}. Clearly, there is little overlap between the two, suggesting that the bulk of the observed broad H$\alpha$ is unlikely to come from SNe.

Finally, we note that core collapse SNe are often accompanied also by other features, such as OI and CaII emission at $\sim$8500 \AA \citep{Nicholl2019_SNe,Pessi2023_SNe,Kokubo2019_SNe} which should be visible in our lower redshift bin stacks, but are not seen.

In summary, most of the evidence indicates that the broad H$\alpha$ lines seen in our stacks are very unlikely due to SNe. However, we cannot exclude some contribution by SNe.

Our analysis in this section also reveals that it might be difficult to search for smaller BHs or BHs that are accreting at lower rates, as we may hit the limit where they are confused with the cumulative contribution of SNe.

\begin{figure}
	\includegraphics[width=\columnwidth]{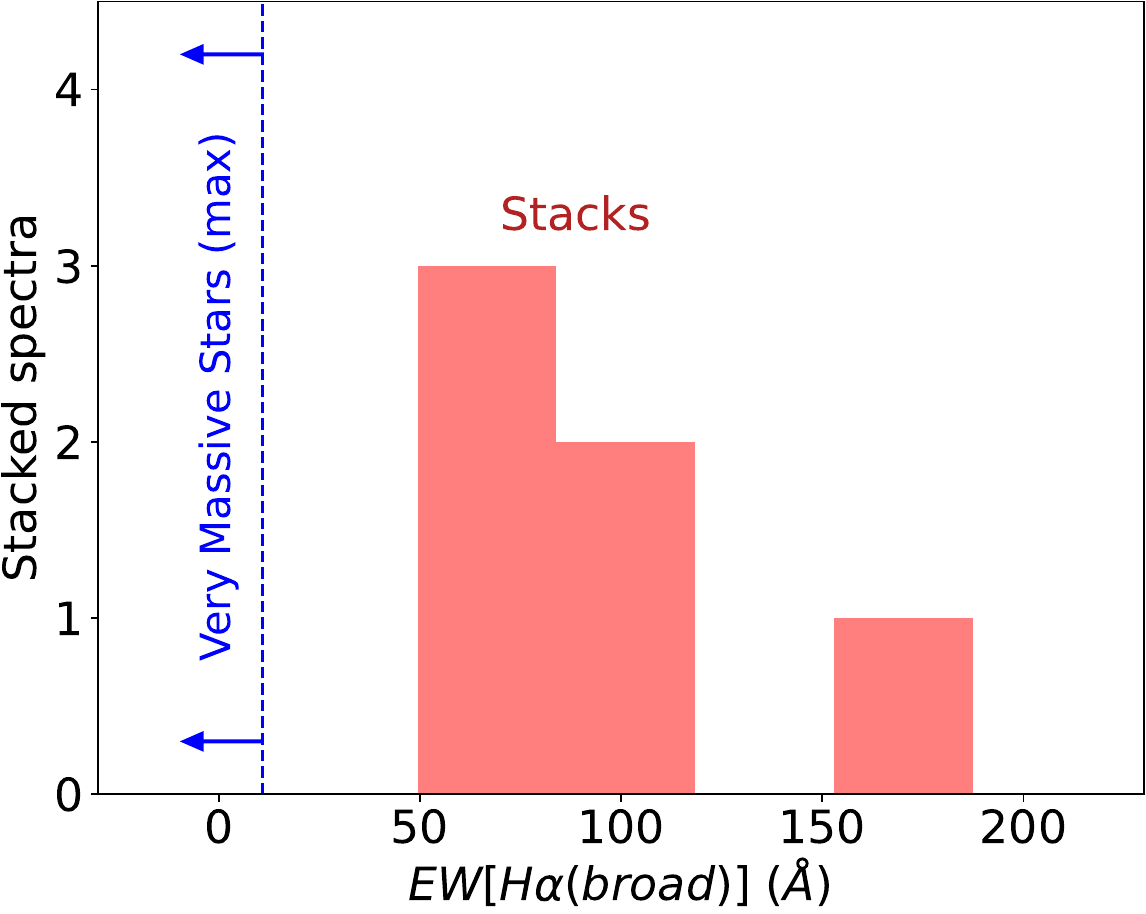}
    \caption{Distribution of the broad H$\alpha$ EW in our stacked spectra, compared with the EW expected from Very Massive Stars (dashed vertical line), which should be considered as a tight upper limit because of contribution to the continuum from less massive and older stars.
    }
    \label{fig:VMS_EW}
\end{figure}

\section{Exploring the case of very massive stars}
\label{sec:stars}

One additional possible scenario for the origin of the broad lines is that part of them are assocaited with Very Massive Stars (VMS), i.e. stars with masses in excess of 100~M$_\odot$. Indeed, these are charcterized by broad wings of H$\alpha$ and H$\beta$
\citep{Martins2020A&A,Martins2022,Martins2023}. However, they also present prominent, broad HeII4686 emission, which is not seen in our stacks. Additionally, the broad Balmer lines associated with VMS have very low Equivalent Widths, of less than $\sim$ 10\AA ~\citep{Martins2020A&A}, and actually the observed EW is expected to be even lower because of contribution to the continuum by lower mass and older stars. On the contrary, as illustrated in Fig.\ref{fig:VMS_EW}, the EW of the broad lines observed in our spectra are of the of the order of several tens or hundreds of \AA. This further indicates that the  broad H$\alpha$ observed in the stacks cannot be coming from the contribution of VMS.

\section{Discussion}\label{s.discussion}

The first two years of \jwst observations have uncovered a ubiquitous population of low-mass SMBHs at redshifts $z\sim4-7$, identified primarily via their broad H$\alpha$ emission \citep{greene_uncover_2023,harikane_jwstnirspec_2023,
maiolino_jades_2024,matthee_little_2024}.
When compared to the stellar mass of their host galaxies, these SMBHs are found to be
`over-massive', i.e. they are more massive than what we would infer from the local $M_\mathrm{BH}-M_*$ relation \citep[e.g ][]{harikane_jwstnirspec_2023,maiolino_jades_2024,juodzbalis_2025}.
At the same time, more massive galaxies at lower redshifts $z=2-3$ are found to lie already on the $M_\mathrm{BH}-M_*$ relation \citep[e.g.,][]{sun+2025}. A comprehensive census of broad-line AGN demonstrates a clear redshift-dependent shift in the average $M_\mathrm{BH}$ away from the local $M_\mathrm{BH}-M_*$ relation \citep{juodzbalis_2025}. However, it was unclear whether the SMBH population is overmassive overall, or whether there are selection effects in play.

Clearly, lower-mass black holes ($\lesssim 10^{6} M_{\odot}$) are likely to exist at high-redshift since they are known to exist in the local Universe (e.g. \citealt{2009ApJ...699L.113L}). Our stacked spectra show strong evidence of type 1 AGN, suggesting that they are common in high-redshift galaxies, yet too faint to be detected individually. Indeed, even though none of our galaxies show 
individual detections of broad H$\alpha$, this is consistent with the sensitivity of current \jwst surveys such as JADES \citep[see ][ for completeness simulations of \jwst observations of AGN]{juodzbalis_dormant_2024}.

In Section~\ref{sec:agn_props} we adopted the single-epoch virial estimator from \citet{2015ApJ...813...82R} to estimate the BH mass.
While this relation has been derived from local AGN, \cite{maiolino_jades_2024} showed that any differences in the high-z AGN (including decreased metallicity and associated reduction of dust content) would likely have a small effect on the BH masses derived using this relation due to the weak (square root) dependence of $M_\mathrm{BH}$ on the H$\alpha$ luminosity. 
Interestingly, 
\cite{GRAVITY2024} directly measured the BH mass in a luminous quasar at z=2.6 and finding that it is consistent, within a factor of 2.5, with that derived adopting the locally calibrated virial relations when using the broad H$\alpha$.
A similar finding has recently being obtained for more high redshift quasar by the GRAVITY+ collaboration (in prep.).

In the specific case of LRDs,
 recent works including \cite{2025A&A...696A..30S} and \cite{2025arXiv250316595R} have suggested that the broad H$\alpha$ components caused by the BLR of an AGN are exponential in shape rather than Gaussian. They suggest that this would indicate that BH masses of known AGN have been significantly overestimated, by about two orders of magnitude, and accretion rates consequently underestimated. However, \cite{juodzbalis_2025} investigated this on their sample of type 1 AGN from JADES and found that there is no evidence that the black hole masses have been overestimated by orders of magnitude based on various lines of evidence. Additionally, \citet{Ji2025QSO1} found for a lensed LRD at z=7, based on variability arguments, that there is no evidence that the relation between size of the BLR and luminosity (which is at the based of the virial relations) is different from the local trend.
 Most importantly, recent studies of the same lensed LRD at z=7 have resolved the BH's sphere of influence, directly measuring the BH mass, and finding consistency with the mass derived based on the local virial relations (Juodzbalis et al., in prep.).

For what concerns the stellar masses of the host galaxies,
a distinctive advantage of our stacking analysis is that the SED of individual galaxies is dominated by the stars, so we can measure $M_{*}$ without the additional uncertainties present in galaxies dominated by AGN continuum emission
\citep[e.g.,][]{maiolino_jades_2024,2025ApJ...981...19L,juodzbalis_2025}. By comparing our black-hole masses to the typical
$M_*$ of our galaxies, we find that our stacked sources lie within $\sim 1-2$~dex above the local $M_\mathrm{BH}-M_*$ 
relation -much closer than typical broad-line AGN from individually detected sources which lie at up to $\sim3$ dex above the local relation (Fig.~\ref{fig:mbh-mstar_new}).
This proximity supports the hypothesis that current over-massive black holes are the high-$M_\mathrm{BH}$ tail of
the distribution at fixed $M_*$ \citep[e.g.,][]{li+2024,juodzbalis_dormant_2024}, although there are still substantial uncertainties.
On the other hand, even our highest-$M_*$ stack, which should be the most ''biased'' in terms of stellar mass, does not reach below the local scaling relation; if the average galaxy
population were to lie perfectly on the local relation, we would expect the high-mass tail of the sample to scatter below the relation, contrary to our results. However, this could be due to at least three reasons, as we discuss below.

The main source of uncertainty is the complex dependence of $M_\mathrm{BH}$ on the properties of the stack.
The stacked $M_\mathrm{BH}$ may still be biased to the most luminous black holes. In principle, this
should not be a dominant effect, due to the aforementioned square-root dependence of $M_\mathrm{BH}$ on
$L_\mathrm{H\alpha}$. At the same time, at fixed $L_\mathrm{H\alpha}$, the most massive black holes would be harder to detect, since their emission is spread over more spectral pixels than for lower-mass black holes. This
latter effect applies until the broad H$\alpha$ is as narrow as the broad [OIII] outflows; at this point we can no
longer securely assign a BLR origin to the H$\alpha$ line. Due to these complex effects, we consider two
limiting scenarios, where $M_\mathrm{BH}$ from the stack is representative either of the median $\log M_\mathrm{BH}$
or of the mean $M_\mathrm{BH}$ from the underlying population.
In our comparison with $M_*$, we implicitly assumed the former, since we compare with the median $\log M_*$.
Assuming instead the second scenario, a fair comparison between our stack and individual sources on the $M_\mathrm{BH}-M_*$ plane would require using the mean $M_*$. Since the latter can be 0.3-1dex higher than the median (as shown by the grey points in Figure \ref{fig:mbh-mstar_new}), this would bring our sample closer to the local relation, thus making our results stronger, while still failing to probe the region below the local relation. Additionally, the mean of the (linear) stellar mass may be dominated by a few very massive galaxies, and therefore not very representative of the population in that bin and giving a stellar-mass biased result.

Still, several questions remain open. On one hand, we need to reconcile the $M_\mathrm{BH}-M_*$ relation with the
$M_\mathrm{BH}-\sigma$ relation. This is outside the scope of this paper since the LSF of the stack is not easily obtained. \citet{maiolino_jades_2024} and \citet{juodzbalis_2025} have shown that even over-massive black holes at high
redshift still follow the local $M_\mathrm{BH}-\sigma$ relation. This result can be potentially explained in terms of these galaxies following the local $M_{BH}-M_{dyn}$ relation, but being undermassive in terms of stellar mass. This is expected in simulations, where at high redshift (at least low mass galaxies) are expected to be gas and dark-matter dominated, and therefore this could explain both findings \citep{McClymont2025_BHs}.
Yet, the picture becomes potentially more complicated with our results from stacking. Indeed,
given that our implied $M_\mathrm{BH}$ is a factor
of 10 lower than the typical value from \citet{maiolino_jades_2024} or \citet{harikane_jwstnirspec_2023}, this
implies either that our AGN do not follow the $M_\mathrm{BH}-\sigma$ relation, or that our host galaxies have
systematically different structure (e.g., size, gas or dark-matter fraction) compared to the previously studied
samples. Exploring these scenarios would require stacking spectra at high spectral resolution to determine the average velocity dispersion of the narrow components, which is not possible with the current data. 

The black hole mass function (BHMF) inferred from our stacks is also very intriguing. These are close to a power-law extrapolation of the higher mass points inferred by \citet{Taylor2024}, at masses where their BHMF is more complete in terms of identification and in terms of accretion rates. It is remarkable that, despite all assumptions and uncertainties, some Semi-Analytical Models reproduce reasonably well the observed BHMF, especially when considering only the active BHs ($L/L_{Edd}>0.01$), which are those probed by the observations. In particular, the CAT SAM which envisage short bursts of super-Eddington accretion \citep{Trinca2024} can account well for most of the observational points in the BHMF. This finding suggests that eposodic super-Eddington accretion is possibly a key aspect of the early evolution of black holes.

It is important to note that the same model predicts a much larger population of dormant black holes, with $L/L_{Edd}<0.01$. 
As found by \citet{juodzbalis_dormant_2024}, these may be detected with \jwst  at high masses ($M_{BH}$ a few times $10^8~M_\odot$), but the bulk of them, at lower masses, are probably out of reach even for \jwst because we are approaching the limit where they may be confused by the contribution of SNe. Detecting such a large population of dormant black holes may only be achievable with the next generation of Gravitational Waves observatories, such as LISA.

It is also interesting to note that the Delphi SAM with heavy seeds \citep{Dayal2019} is reasonably close to the observed points, although they fall short by a factor of about 6 ($\sim 2\sigma$) the low mass points obtained by our stacks. The same models involving only light seeds are instead quite inconsistent with the BHMF results. These findings suggest that light seeds alone are unlikely to reproduce the observed population of BHs at z$\sim$3--5.

Future observations with the \jwst high-resolution gratings (R$\sim$2700) are essential, as they
would constrain the dynamical mass of the host galaxies, and enable exploring the $M_\mathrm{BH}-\sigma$ relation.
Additionally, high-resolution would reduce the contamination between the broad and narrow components, enabling a
more accurate measurement of the broad-line profile and would possibly probe even smaller black holes ($M_{BH}<10^6~M_\odot$).

Clearly, it is also important to confirm our findings independently by detecting SMBH in the individual galaxies that lie on the local $M_\mathrm{BH}-M_*$ relation.
Since our stacks have 10--20 higher SNR than typical spectra in JADES, individual
detections can only be achieved for SMBHs with very high accretion rates, above the Eddington limit. Alternatively,
exploiting gravitational lensing will enable us to probe low-mass SMBHs with regular accretion rates.

\section{Summary and conclusions}
\label{sec:conclusions}
We have presented stacks of JADES R1000 spectra in redshift bins of $3<z<5$ and $5<z<7$ to search for faint broad line AGN, possibly associated with low mass black holes. The increased signal to noise due to stacking allows us to detect signatures of AGN in both redshift bins, with $\log(M_{\rm BH}/M_{\odot}) \approx 6.4$. These AGN are low luminosity, but are still accreting at a fairly significant fraction of the Eddington limit. While possible black holes of these masses have been discovered at high redshift \citep[e.g., ][]{maiolino_jades_2024, maiolino_small_2024}, these results present the first indication that an average population of small black holes exist at high redshifts, confirming that previous observations were primarily probing massive black holes due to selection effects.
Our main findings are summarised below:

\begin{itemize}

    \item We find potential broad components of H$\alpha$ in the stacks in both the $3<z<5$ and $5<z<7$ samples. The broad component in the highest redshift stack has a potential counterpart in \OIII, suggesting a potential contribution from outflows to the broad component.  Determining whether the broad H$\alpha$ is driven by AGN or outflows is difficult in the combined stacks because sources with AGN and outflows are intertwined. Therefore, the stacks are split into bins of stellar mass, UV luminosity, \OIII luminosity and \OIII equivalent width, with the goal of separating different types of activity.
    
    \item At $3<z<5$, the two highest \OIII luminosity bins, the lowest EW bin, the highest stellar mass bin and the highest $M_{UV}$ bin have a statistically significant broad H$\alpha$ component of FWHM$\gtrsim 1000~\kms$ that does not have a counterpart in \OIII. Therefore, the populations of galaxies within these stacks host, on average, a population of broad line AGN. The properties of the stacks that include AGN-associated broad H$\alpha$, compared to those without, tell us that it is the galaxies with highest \OIII luminosity that are hosting these AGN, and that a population of these galaxies can have high or low \OIII EW. Also galaxies with the highest stellar masses and M$_{\rm UV}$ tend to preferentially host the new population of previously undetected AGN. In the $5<z<7$ stacks the highest EW stack has a broad component in H$\alpha$ with no \OIII counterpart and the highest [OIII] luminosity stack has a tentative detection of broad H$\alpha$.
    
    \item We locate our stacks on the new \OIII $\lambda$ 4363-based narrow line diagnostic diagrams for high redshift AGN. The highest [OIII] luminosity stack  in the $5<z<7$ bins, which has tentative detection of broad H$\alpha$ associated with BLR, is in the AGN-only locus of the diagram, supporting the AGN origin of the broad H$\alpha$. The location of our other stacks on these diagrams are less conclusive as they are in the region where both AGN and SF galaxies can be located. 
    
    \item The average mass of these black holes in both redshift bins is $\log(M_{\rm BH}/M_{\odot}) \approx 6.2-6.6$, which is at the lowest end of the mass range of previously discovered high redshift AGN, and approaching the predicted range of black hole seed masses, according to some models. The stacks are also revealing black holes that are accreting at rates somewhat lower than previous studies ($L/L_{Edd}\sim 0.1$), and with some stacks showing accretion rates as low as $L/L_{Edd}\sim 0.02$. These results indicate that studies of individual objects have been mostly probing the higher mass and more active population of high redshift black holes.

    \item In terms of $M_{BH}$-$M_{*}$ distribution, the average black holes inferred from the stacks are still above the local relation, but much closer than found by previous studies and, in many cases, consistent with the local relations within the scatter. This result indicates that previous studies on individual sources probed the high-mass tail of a very scattered intrinsic relation. Yet, the finding that all stacks are systematically above the local relation, even when probing such faint AGN, indicates that there is an intrinsic offset of the global relation. In particular, the fact that the stack of the most massive galaxies, which should be highly biased for high stellar masses (hence would be expected to lie below the relation) is still above the relation (although consistent with it within the scatter). It is also interesting that some stacks, despite probing very weak AGN, are also well above the relation, further confirming the finding of previous studies that, although in a tail, the Universe manages to produce extremely massive black holes in small galaxies. 
    
    \item We use our stacks to derive the Black Hole Mass Function (BHMF) at BH masses of a few times $10^6~M_\odot$. The volume density of low mass BHs at $3<z<5$, with accretion rates of $L/L_{Edd}\sim 0.02-0.1$, is quite high, about $6\times 10^{-3}~Mpc^{-3}~dex^{-1}$. The resulting BHMF obtained by combining with previous results at high masses, is well reproduced by models assuming an evolution characterized by short bursts of super-Eddington accretion. The same model expects the existence of a much larger population of dormant BHs ($L/L_{Edd}< 0.01$) which are not detectable even with our stacking technique.

    \item We have explored the possibility that the faint broad H$\alpha$ detected in our stacks is due to the cumulative signature of SNe. We find that, even making very conservative assumptions, SNe are very unlikely to be responsible for the observed signal, based on luminosity and line profile arguments. However, we cannot exclude some contribution by SNe to some of the stacks.

    \item We have also explored the possibility that the broad H$\alpha$ is contributed by populations of Very Massive Stars and find that also this scenario is very unlikely, due the the very low EW(H$\alpha _{broad}$) expected in this scenario, inconsistent with that found in our stacks.

\end{itemize}

It is important in the future to expand our findings to individual sources via deeper observations or lensed AGN. Additionally, it will be important to find and characterize even lower mass BHs, possibly approaching the seeding regime, via higher spectral resolution observations.

\begin{table*}
    \centering
    \caption{Mean properties of stacks in the redshift bins $3<z<5$ and $5<z<7$. Each redshift bin contains 12 stacks, 4 with different \OIII luminosity ranges, 4 with different \OIII EW ranges, and 4 with different stellar mass ranges. The first column shows the redshift bin, the second column describes each of the four bins (highest, second-highest, third-highest, lowest) of the four properties of the stacks (\OIII $\lambda$5007 Luminosity, \OIII $\lambda$5007 EW and $\log(M_*)$ ($M_{\odot}$)), M$_{\rm UV}$. The third column gives the mean \OIII luminosity within each bin, the fourth column gives the mean \OIII EW, the fifth column gives the mean $\log(M_{*})$ and the sixth column gives the mean M$_{\rm UV}$.}
    \label{tab:combined_properties}
    \begin{tabular}{lccccc} 
        \hline
        Redshift Bin & Bin (highest to lowest) & \OIII $\lambda$5007 Luminosity (erg/s) & \OIII $\lambda$5007 EW (Å) & $\log(M_*)$ ($M_{\odot}$) & M$_{\rm UV}$\\
        \hline
        \multirow{4}{*}{$3<z<5$} & 1 & $2.83 \times 10^{42}$ & 1254 & 9.7 & -19.85\\
               & 2 & $1.09 \times 10^{42}$ & 602  & 9.1 & -19.03\\
               & 3 & $6.03 \times 10^{41}$ & 307  & 8.7 & -18.51\\
               & 4 & $2.81 \times 10^{41}$ & 125  & 8.1 & -17.64\\
        \hline
        \multirow{4}{*}{$5<z<7$} & 1 & $4.59 \times 10^{42}$ & 1661 & 9.4 & -20.00\\
               & 2 & $2.05 \times 10^{42}$ & 990  & 8.7 & -19.38\\
               & 3 & $1.20 \times 10^{42}$ & 610  & 8.3 & -18.82\\
               & 4 & $4.97 \times 10^{41}$ & 335  & 7.7 & -17.81\\
        \hline
    \end{tabular}
\end{table*}

\begin{table*}
    \centering
    \caption{Properties of the broad H$\alpha$ line for the stacks (unweighted and not normalised) with confirmed H$\alpha$ broad components. Column 1 is the specific stack, column 2 is the  value of $\Delta BIC = BIC_{\text{narrow only}} - BIC_{\text{broad+narrow}}$ for H$\alpha$, column 3 is the FWHM of the detected H$\alpha$ broad component, column 4 is the velocity offset of the detected H$\alpha$ broad component, column 5 is the  FWHM of the \OIII broad component that is fit freely (without kinematics constrained to those of H$\alpha$), and column 6 is $\Delta BIC_{\text{O III}} = BIC_{H\alpha \text{kinematics}} - BIC_{\text{free fit}}$.}    
    \label{tab:agn_properties}
    \begin{tabular}{lcccccccc}
        \hline
        Stack & $\Delta BIC$ & H$\alpha$ broad FWHM (km/s) & H$\alpha$ broad velocity offset (km/s) & FWHM$_{\text{O III}}$ (km/s) &  $\Delta BIC_{\text{O III}}$\\
        \hline
        3<z<5 highest \OIII luminosity & 143 & $1200^{+102}_{-98}$ & $135^{+41}_{38}$ & $442^{+20}_{-17}$ & 192\\[0.15cm] 
        3<z<5 second highest \OIII luminosity & 48 & $1568^{+197}_{-232}$ &$103^{+71}_{68}$ & $561^{+80}_{-67}$& 39\\[0.15cm] 
        3<z<5 lowest \OIII EW & 26 & $1515^{+186}_{-203}$ & $313^{+89}_{97}$ & $540^{+52}_{-40}$ & 29\\[0.15cm] 
        3<z<5 highest stellar mass & 127 & $1365^{+125}_{-141}$ & $101^{+44}_{43}$ & $592^{+37}_{-35}$ & 64\\[0.15cm] 
        3<z<5 highest M$_{\rm UV}$ & 127 & $998^{+97}_{-94}$ & $38^{+15}_{12}$ & $474^{+22}_{-18}$ & 196\\[0.15cm] 
        5<z<7 highest \OIII luminosity (tentative) & 52 & $1000^{+162}_{-135}$ & $63^{+47}_{36}$ & $722^{+62}_{-58}$ & 6\\[0.15cm] 
        5<z<7 highest \OIII EW & 52 & $1012^{+164}_{-145}$ & $89^{+53}_{45}$ & $360^{+63}_{-10}$ & 92\\[0.15cm] 
        \hline
    \end{tabular}%
\end{table*}

\begin{table*}
    \centering
    \caption{Properties used to correct for dust extinction and compute $M_{\text{BH}}$ for the stacks with confirmed H$\alpha$ broad components. Column 1 gives the stack, column 2 gives the observed flux of the H$\alpha$ broad component, column 3 gives the observed Balmer decrement, column 4 gives the $A_{V}$, column 5 gives the corrected flux of the broad component, column 6 gives the median redshift of the stack.}
    \label{tab:agn_props1}
    \small
    \begin{tabular}{lcccccccc}
        \hline
        Stack & \makecell{$F_{H\alpha}$ broad (observed) \\ ($10^{-20}$ erg\,s$^{-1}$\,cm$^{-2}$)} & \makecell{$\frac{F_{H\alpha}}{F_{H\beta}}$ \\ observed} & $A_{V}$ &  \makecell{$F_{H\alpha}$ broad (corrected) \\ ($10^{-20}$ erg\,s$^{-1}$\,cm$^{-2}$)} & Median z\\
        \hline
        3<z<5 highest \OIII luminosity & $117.1 \pm 17.2$ & $3.55 \pm 0.06$ & $0.58 \pm 0.20$ & $178.1 \pm 43.8$ & $3.80^{+0.92}_{-0.36}$\\[0.15cm] 
        3<z<5 second-highest \OIII luminosity & $38.2 \pm 8.6$ & $3.42 \pm 0.06$ & $0.48 \pm 0.24$ & $54.3 \pm 14.2$ & $3.68^{+0.80}_{-0.36}$\\[0.15cm] 
        3<z<5 lowest \OIII EW & $35.5 \pm 7.7$ & $4.19 \pm 0.08$ & $1.02 \pm 0.50$ & $74.7 \pm 17.0$ & $3.57^{+0.51}_{0.39}$\\[0.15cm] 
        3<z<5 highest stellar mass & $94.9 \pm 15.0$ & $3.98 \pm 0.05$ & $0.89 \pm 0.44$ & $180.8 \pm 42.4$ & $3.66^{+0.74}_{-0.36}$\\[0.15cm] 
        3<z<5 highest M$_{\rm UV}$ &  $94.6 \pm 20.0$&  $3.44 \pm 0.05$ & $0.50 \pm 0.25$& $135.8 \pm 34.9$ & $3.79^{+0.64}_{-0.46}$\\[0.15cm]
        5<z<7 highest \OIII luminosity  &  $61.9 \pm 18.1$ & $3.17 \pm 0.07$ & $0.28 \pm 0.15$ & $75.8 \pm 21.5$& $5.80^{+0.51}_{-0.41}$\\[0.15cm] 
        5<z<7 highest \OIII EW &  $40.9 \pm 11.8$&  $3.39 \pm 0.09$& $0.46 \pm 0.24$& $57.0 \pm 16.3$& $5.90^{+0.70}_{-0.71}$\\[0.15cm]  
        \hline
    \end{tabular}%
\end{table*}

\begin{table*}
    \centering
    \caption{Average properties of the black holes and host galaxies in the stacks with H$\alpha$ broad components. Column 1 gives the stack, column 2 gives the black hole mass that has been corrected for dust extinction, column 3 gives the bolometric luminosity, column 4 gives the Eddington ratio, column 5 gives the Median of $\log(M_{*})$ in the stack.}
    \label{tab:agn_props2}
    \small
    \begin{tabular}{lcccccccc}
        \hline
        Stack & $\log M_{\rm BH}/M_{\odot}$ & $\log L_{ Bol}/\text{erg\,s}^{-1}$ & $\lambda_{\rm Edd}$ & Median $\log M_{*}/M_{\odot}$\\[0.15cm] 
        \hline
        3<z<5 highest \OIII luminosity & $6.48^{+0.36}_{-0.34}$& $43.52^{\textnormal{+}0.27}_{-0.17}$& $0.09^{+0.08}_{-0.09}$ & $9.05^{+0.40}_{-0.54}$\\[0.15cm] 
        3<z<5 second-highest \OIII luminosity & $6.46^{+0.37}_{-0.36}$ & $42.97^{+0.26}_{-0.15}$& $0.03^{+0.03}_{-0.02}$ &$8.95^{+0.67}_{-0.56}$\\[0.15cm] 
        3<z<5 lowest \OIII EW & $6.48^{+0.36}_{-0.35}$ & $43.08^{+0.15}_{-0.18}$& $0.03^{+0.03}_{-0.03}$ &$9.47^{+0.49}_{-0.48}$\\[0.15cm] 
        3<z<5 highest stellar mass & $6.58^{+0.36}_{-0.35}$ & $43.48^{+0.24}_{-0.16}$& $0.06^{+0.06}_{-0.06}$ &$9.62^{+0.20}_{-0.50}$\\[0.15cm] 
        3<z<5 highest M$_{\rm UV}$ & $6.26^{+0.35}_{-0.35}$ & $43.40^{+0.16}_{-0.17}$ & $0.11^{+0.1}_{-0.1}$&$0.35^{+0.61}_{-0.57}$\\[0.15cm] 
        5<z<7 highest \OIII luminosity & $6.35^{+0.37}_{-0.36}$ &$43.60^{+0.17}_{-0.17}$ & $0.14^{+0.13}_{-0.13}$&$8.9^{+0.83}_{-0.50}$\\[0.15cm] 
        5<z<7 highest \OIII EW & $6.30^{+0.37}_{-0.36}$ & $43.46^{+0.15}_{-0.14}$ & $0.11^{+0.10}_{-0.10}$&$8.01^{+0.59}_{-0.71}$\\[0.15cm] 
        \hline
    \end{tabular}%
\end{table*}

\section*{Acknowledgements}
The authors gratefully acknowledge Will McClymont for insightful discussions, and Giovanni Mazzolari for his assistance in creating the diagnostic diagrams presented in this paper.
YI is supported by JSPS KAKENHI Grant No. 24KJ0202.
FDE, RM, XJ, JS, IJ and CS acknowledge support by the Science and Technology Facilities Council (STFC), by the ERC through Advanced Grant 695671 ``QUENCH'', and by the
UKRI Frontier Research grant RISEandFALL. RM also acknowledges funding from a research professorship from the Royal Society. IJ also acknowledges support by the Huo Family Foundation through a P.C. Ho PhD Studentship.
SA acknowledges grant PID2021-127718NB-I00 funded by the Spanish Ministry of Science and Innovation/State Agency of Research (MICIN/AEI/ 10.13039/501100011033).
AJB and JC acknowledge funding from the "FirstGalaxies" Advanced Grant from the European Research Council (ERC) under the European Union’s Horizon 2020 research and innovation programme (Grant agreement No. 789056)
SC and GV acknowledge support by European Union's HE ERC Starting Grant No. 101040227 - WINGS.
ECL acknowledges support of an STFC Webb Fellowship (ST/W001438/1)
BDJ and BER acknowledge support from the NIRCam Science Team contract to the University of Arizona, NAS5-02015. BER also acknowledges support from JWST Program 3215.
ST acknowledges support by the Royal Society Research Grant G125142.
H\"U acknowledges funding by the European Union (ERC APEX, 101164796). Views and opinions expressed are however those of the authors only and do not necessarily reflect those of the European Union or the European Research Council Executive Agency. Neither the European Union nor the granting authority can be held responsible for them.
The research of CCW is supported by NOIRLab, which is managed by the Association of Universities for Research in Astronomy (AURA) under a cooperative agreement with the National Science Foundation.
JW gratefully acknowledges support from the Cosmic Dawn Center through the DAWN Fellowship. The Cosmic Dawn Center (DAWN) is funded by the Danish National Research Foundation under grant No. 140.

The authors acknowledge use of the lux supercomputer at UC Santa Cruz, funded by NSF MRI grant AST 1828315.

\section*{Data Availability}

We use publicly available data from the JADES Survey, obtained through the \textit{JWST} program IDs 1180, 1181, 1210, 1286, 1287, and 3215. All spectra are available through the Mikulski Archive for Space Telescopes (MAST) at \url{https://dx.doi.org/10.17909/8tdj-8n28}.


\bibliographystyle{mnras}
\bibliography{JADES_BLR_stacking} 


\appendix
\section{Example of the pPXF fit to the continuum for the $3<z<3$ stack}
\label{appendix:appendix_ppxf}
\begin{figure*}
    \includegraphics[width=1\textwidth]{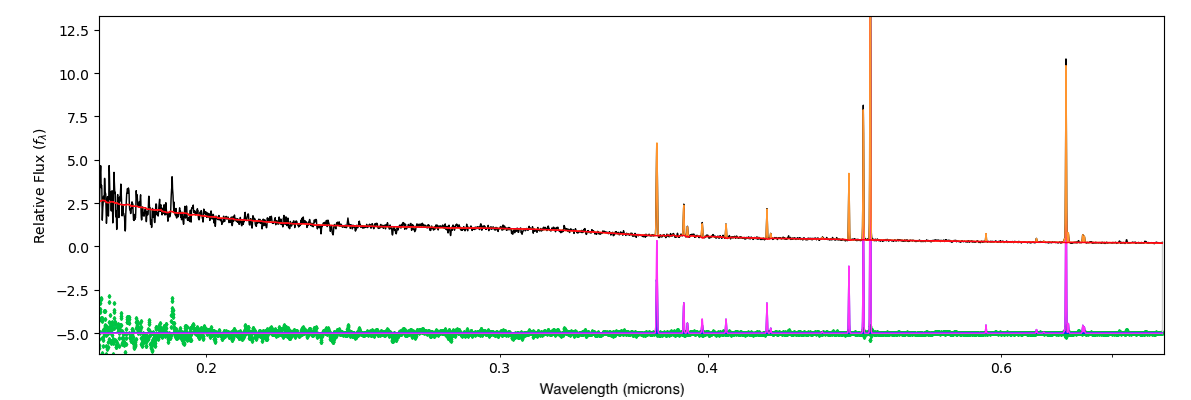}
    \caption{pPXF fit to the $3<z<$ stack. The red line shows the best fit to the continuum, the orange lines show the best fit of the gas components (emission lines) and the green points show the residuals between the data and the model. This figure was produced with the pPXF fitting code \citep{cappellari_full_2023}.}
    \label{fig:ppxf}
\end{figure*}

See Figure \ref{fig:ppxf}.

\section{Testing if the broad-line detection is an artefact of stacking}
\label{appendix:appendix_tests}
\begin{figure}
	\includegraphics[width=1\columnwidth]{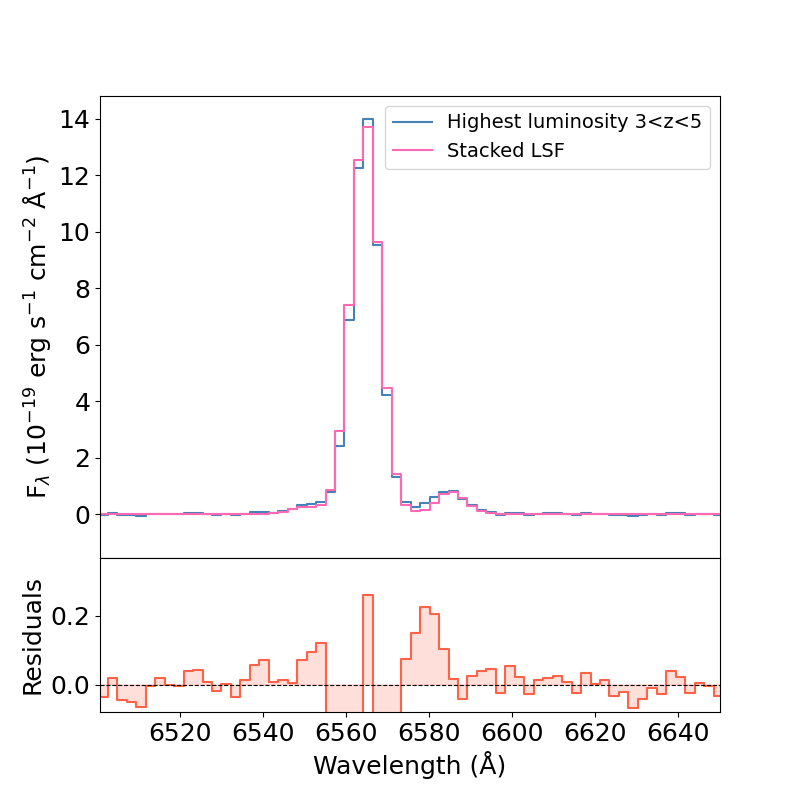}
    \caption{The true stack (blue) of the highest luminosity sources in our sample at z=3-5 compared to the stack of the LSFs (pink). The bottom panel shows the residuals between the true stack and the stacked LSF.}
    \label{fig:lsf_stack}
\end{figure}

We tested possible scenarios in which  the H$\alpha$ broad component results from  issues associated with the stacking technique rather than an AGN BLR. Firstly, the stacked line profiles are not only due to the average of the flux from the sources, but also the average of the line spread functions of the sources. Each galaxy in the stack is at a different redshift and therefore the LSF of each of them will be slightly different. Therefore, we investigate the possibility that the stacked LSFs exhibit broadening and could therefore be the cause of our detected broad H$\alpha$ component. For each galaxy in the stack, we compute the LSF at H$\alpha$ and [N II] by defining a Gaussian of FWHM equal to the known spread of the observed H$\alpha$/[N II] NIRSpec R1000 LSF at that redshift. We use the LSFs for extended sources as this would be the worst case scenario since extended source LSFs are wider, but some of the sources in our stacks are likely point sources which would have a narrower LSF. We then stack each Gaussian using the same method as for the stacks of the spectra. This stacked LSF serves as the mock LSF of the stacked spectra. In Figure \ref{fig:lsf_stack} we show the stacked LSF of the sources in the highest luminosity stack at $3<z<5$ (which is one of the stacks that has a detected H$\alpha$ broad component) overlaid on the true stack. We also show the residuals between the true stack and the stacked LSF. This shows that broad wings remain in the residuals; therefore, we conclude that the H$\alpha$ broadening is not due to the stacking of the LSF. 

\begin{figure}
	\includegraphics[width=1\columnwidth]{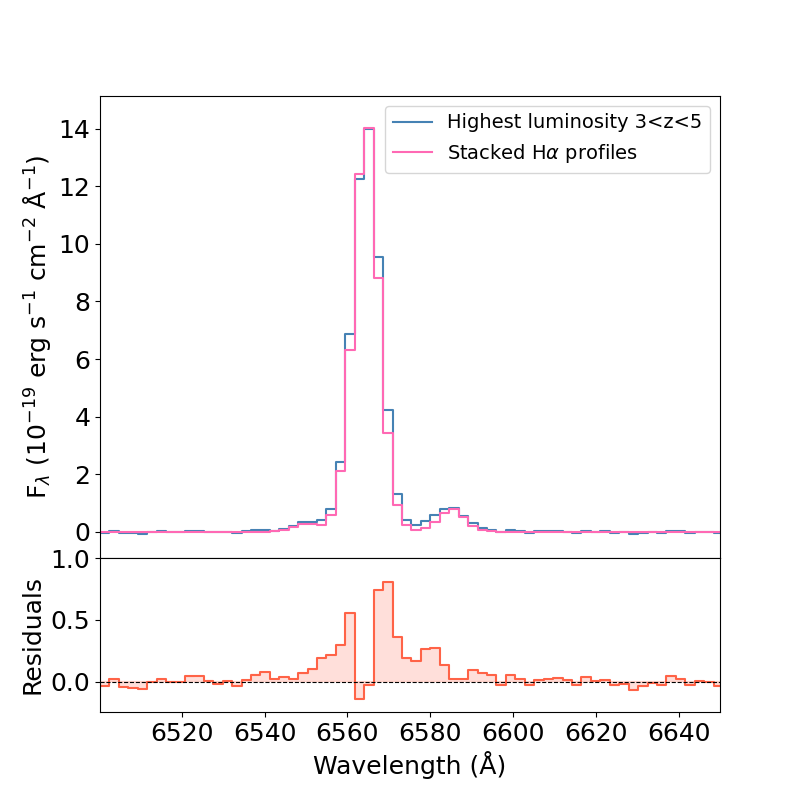}
    \caption{The true stack (blue) of the highest luminosity sources in our sample at z=3-5 compared to the stack of the observed H$\alpha$ profiles (pink). The bottom panel shows the residuals between the true stack and the stacked $H\alpha$ profiles.}
    \label{fig:stacked_halpha}
\end{figure}

The second possible cause of the H$\alpha$ broad component is stacking many H$\alpha$ emission lines with slightly different intrinsic profiles. To rule this out, we repeat the same procedure as for the LSF stacking test, but this time we stack Gaussians that have the observed FWHM of the H$\alpha$ lines. We also use the same FWHM to stack the profiles of the [N II] lines since these lines are close in wavelength and therefore should have very similar LSF. The resulting stack is shown in Figure \ref{fig:stacked_halpha} and once again there are clear flux residuals forming broad wings around H$\alpha$. These tests confirm that stacking the LSFs and observed H$\alpha$ line profiles cannot account for the broadening that we attribute to AGN activity in the stacks.

\section{Stacking type 2 AGN}
\label{appendix:appendix_type2}
\begin{figure*}
    \centering
    \includegraphics[width=0.5\textwidth]{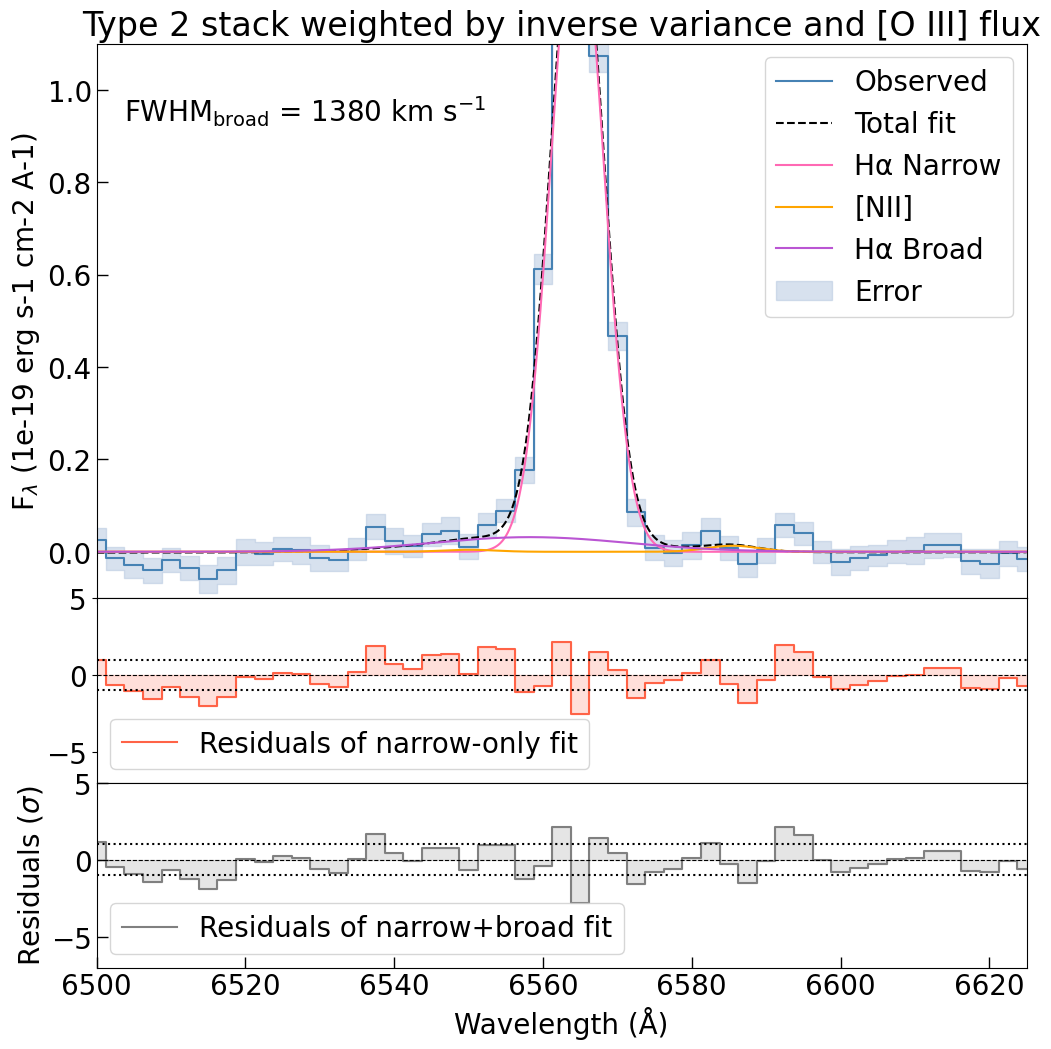}
    \caption{The stack of type 2 AGN weighted by \OIII flux and inverse variance, as was done in \citet{scholtz_jades_2025}. It is clear from the residuals that a broad component is not required for the model to fit the data.} 
    \label{fig:type2_1}
\end{figure*}

\begin{figure*}
    \centering
    \begin{subfigure}[b]{0.5\textwidth}
        \includegraphics[width=\textwidth]{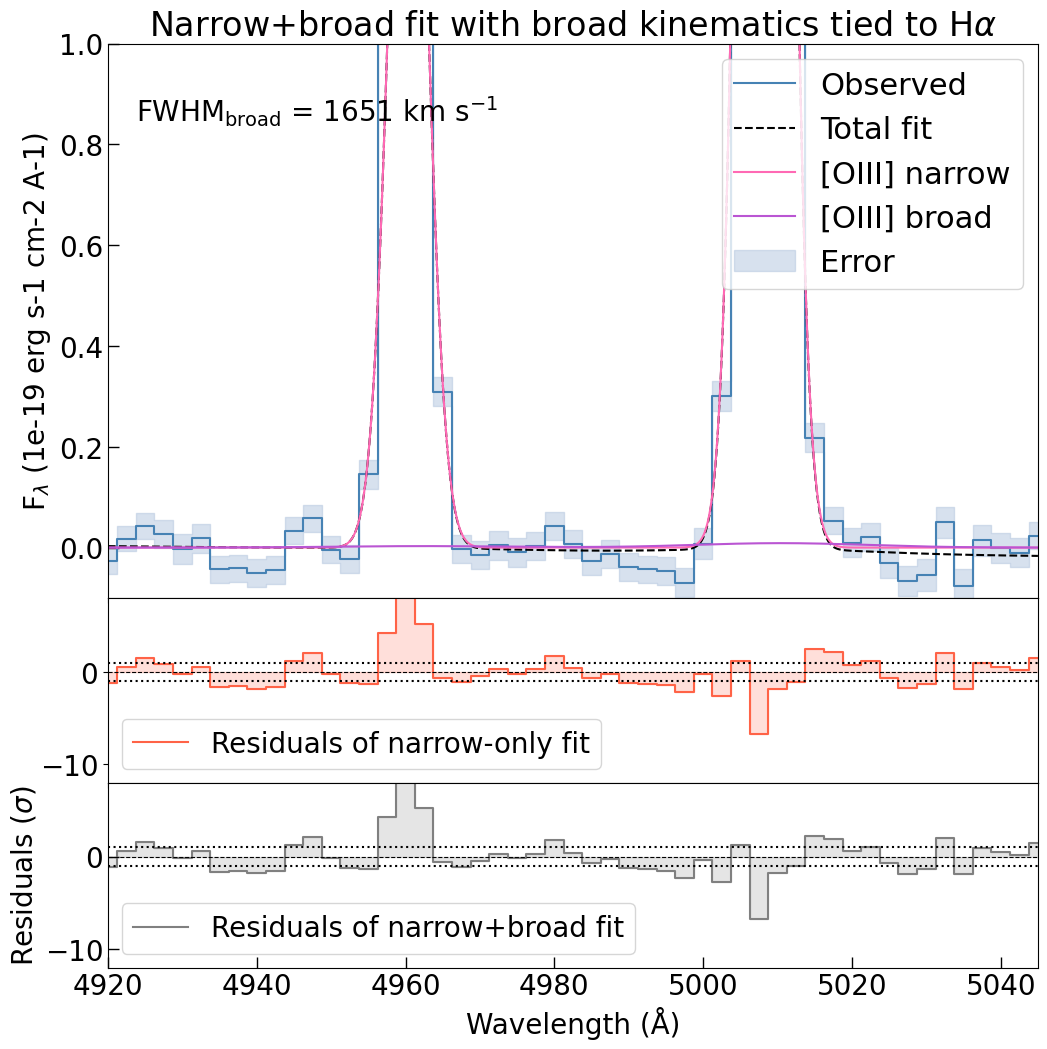}
    \end{subfigure}
    \begin{subfigure}[b]{0.48\textwidth}
        \includegraphics[width=\textwidth]{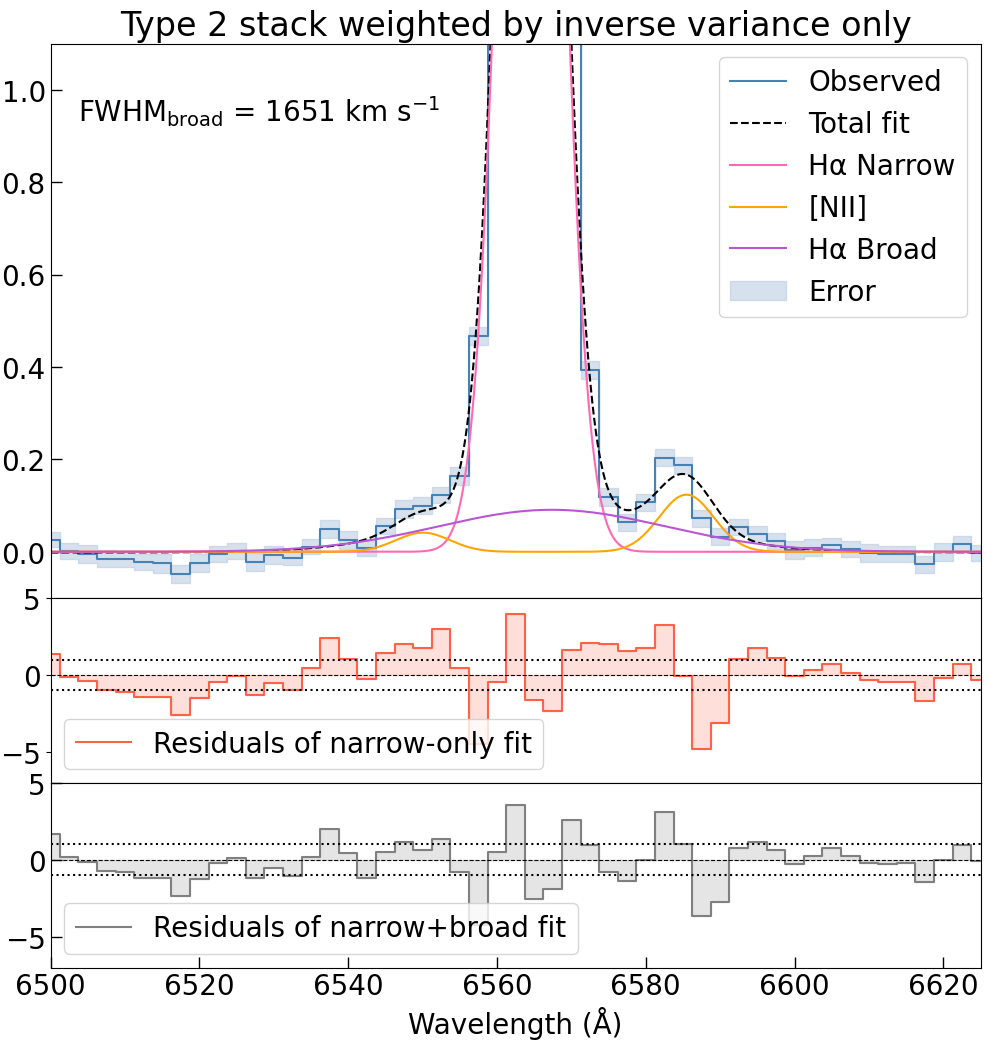}
    \end{subfigure}
    \caption{The stack of type 2 AGN only weighted by inverse variance as was done in \citet{mazzolari_new_2024}. The scale on the residuals for the \OIII line and the H$\alpha$ lines are slightly different, but we plot the $\pm 1\sigma$ levels as a dotted line for clarity. There are clear broad residuals in the narrow only fit of H$\alpha$ and not in \OIII, and we find an H$\alpha$ broad component of FWHM=1651 km/s. $\Delta BIC$ between the narrow only and narrow+broad models is 87 in favour of the broad model.}
    \label{fig:type2_2}
\end{figure*}

Stacks of the JADES type 2 AGN sample from \cite{scholtz_jades_2025} have been investigated by \cite{scholtz_jades_2025} and \cite{mazzolari_new_2024}. Each of these studies showed different results. \cite{scholtz_jades_2025} found that in their stacks weighted by 1/($F_{\rm [O III]}$rms$^{2}$), there was no detection of a broad component while \cite{mazzolari_new_2024} found that in their stacks weighted by 1/rms$^{2}$, there was a broad H$\alpha$ detection. We perform stacks using the methods from both. Figure \ref{fig:type2_1} shows our results of the stack weighted by 1/($F_{\rm [O III]}$rms$^{2}$). It is clear that no broad component appears in the residuals, and fitting a broad line component alongside the narrow model results in $\Delta BIC = -6$ in favour of the narrow model, indicating that the narrow-only model is preferred. In contrast to this are the results from the 1/rms$^{2}$ weighted stacks shown in Figure \ref{fig:type2_2}. In this case, there are clear broad residuals and we retrieve a broad H$\alpha$ component that lowers the BIC ($\Delta BIC = 87$, in favour of the broad model), indicating that the broad model is preferred. The FWHM is also broad enough to be coming from an AGN BLR, and adding this broad component to the \OIII doublet results in a poor fit, indicating that there are no outflows present. 

The different results from the two stacking methods indicates that if the sample is small, as it is for the type 2 AGN included in our stack (17), the chosen stacking method can have a large impact on the result. While we did highlight some differences between the stacking methods in our results presented in the main text, the detection of a broad component was mostly maintained across the different methods, indicating that a large sample (in our case 424 sources at $3<z<5$ and 152 sources at $5<z<7$) may produce more consistent results.

The results from our type 2 AGN stacks indicate that a larger sample of type 2 AGN should be included in the stacks in order to confirm whether there are faint broad H$\alpha$ lines and therefore type 1 AGN hidden within this sample, or if these are true type 2 AGN with no hints of broad H$\alpha$ components.

\section{Broad H$\alpha$ detection}
\label{appendix:broadHa}

\begin{figure*}
    \centering
    \begin{subfigure}[b]{0.49\textwidth}
        \includegraphics[width=\textwidth]{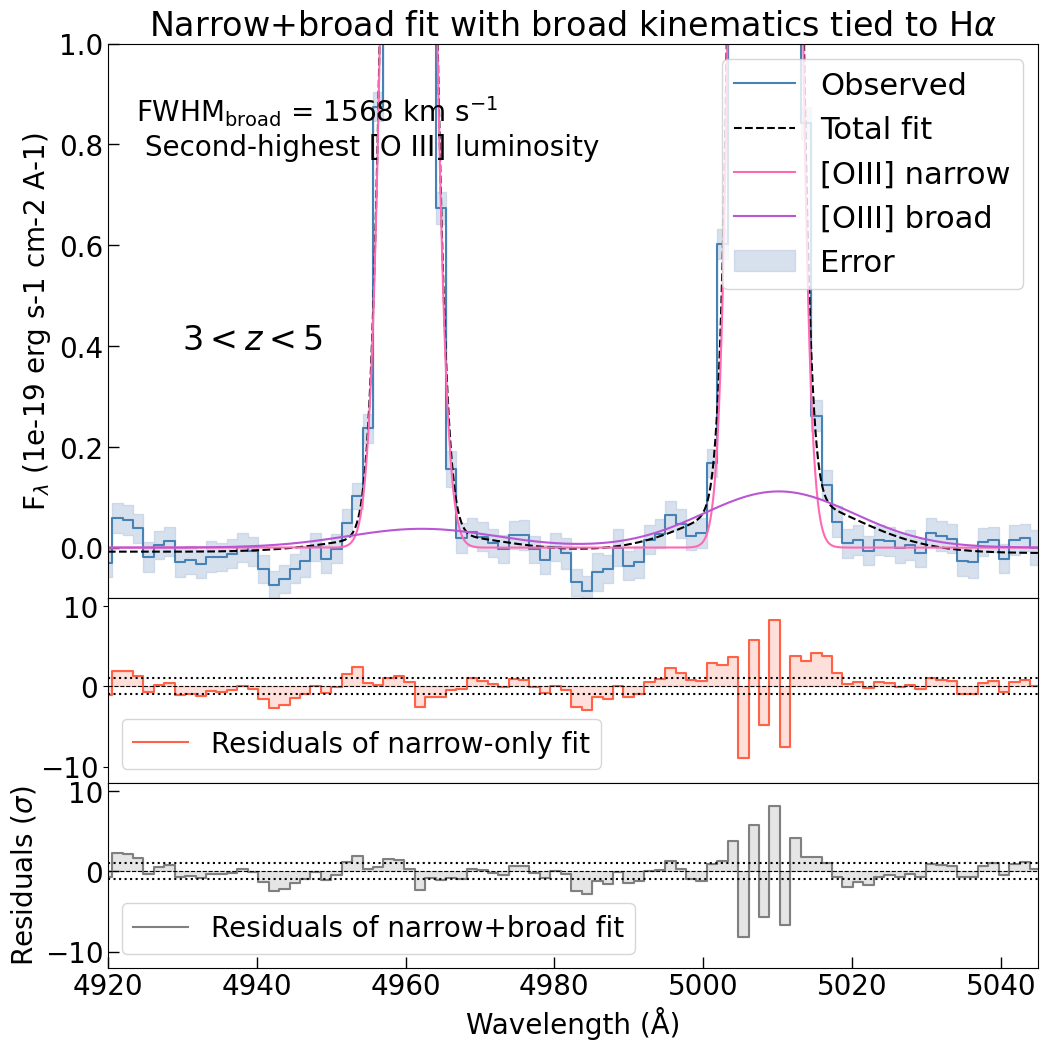}
    \end{subfigure}
    \begin{subfigure}[b]{0.47\textwidth}
        \includegraphics[width=\textwidth]{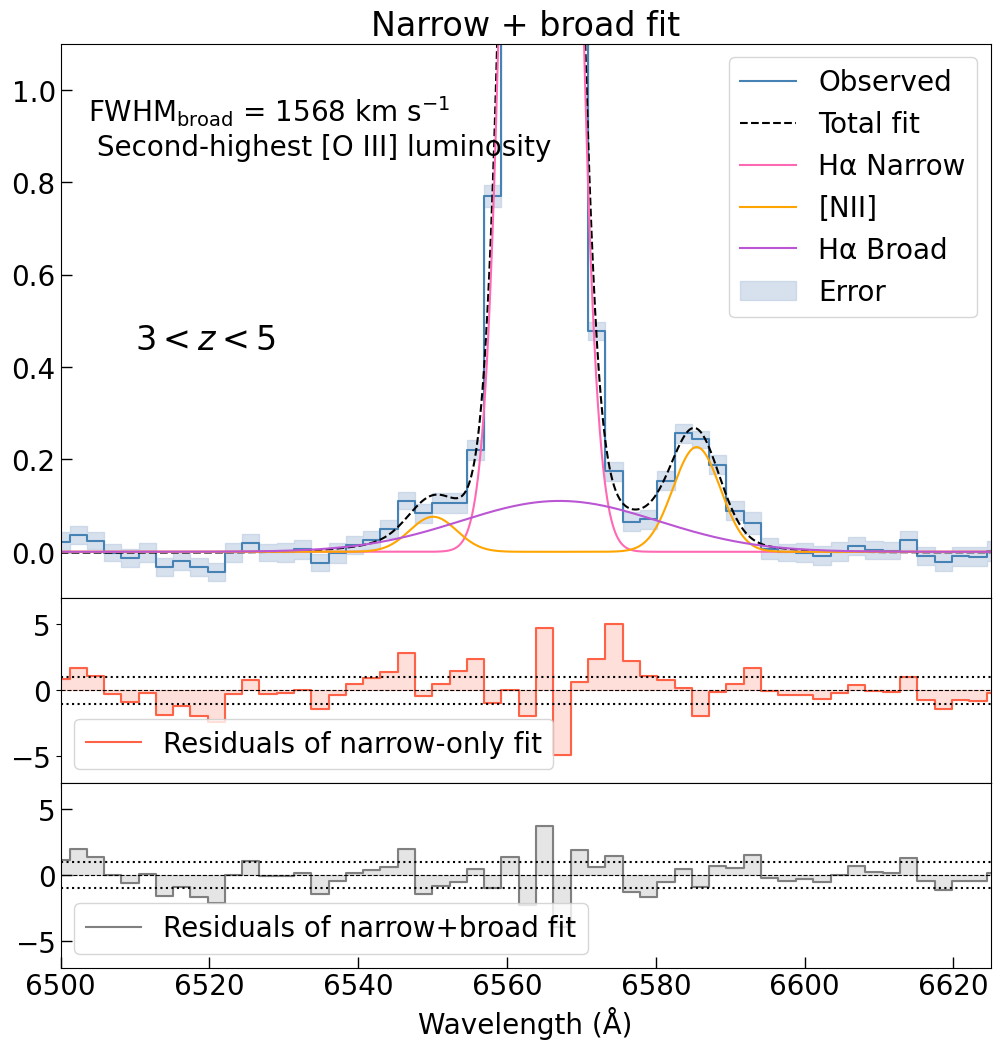}
    \end{subfigure}
    \begin{subfigure}[b]{0.48\textwidth}
        \includegraphics[width=\textwidth]{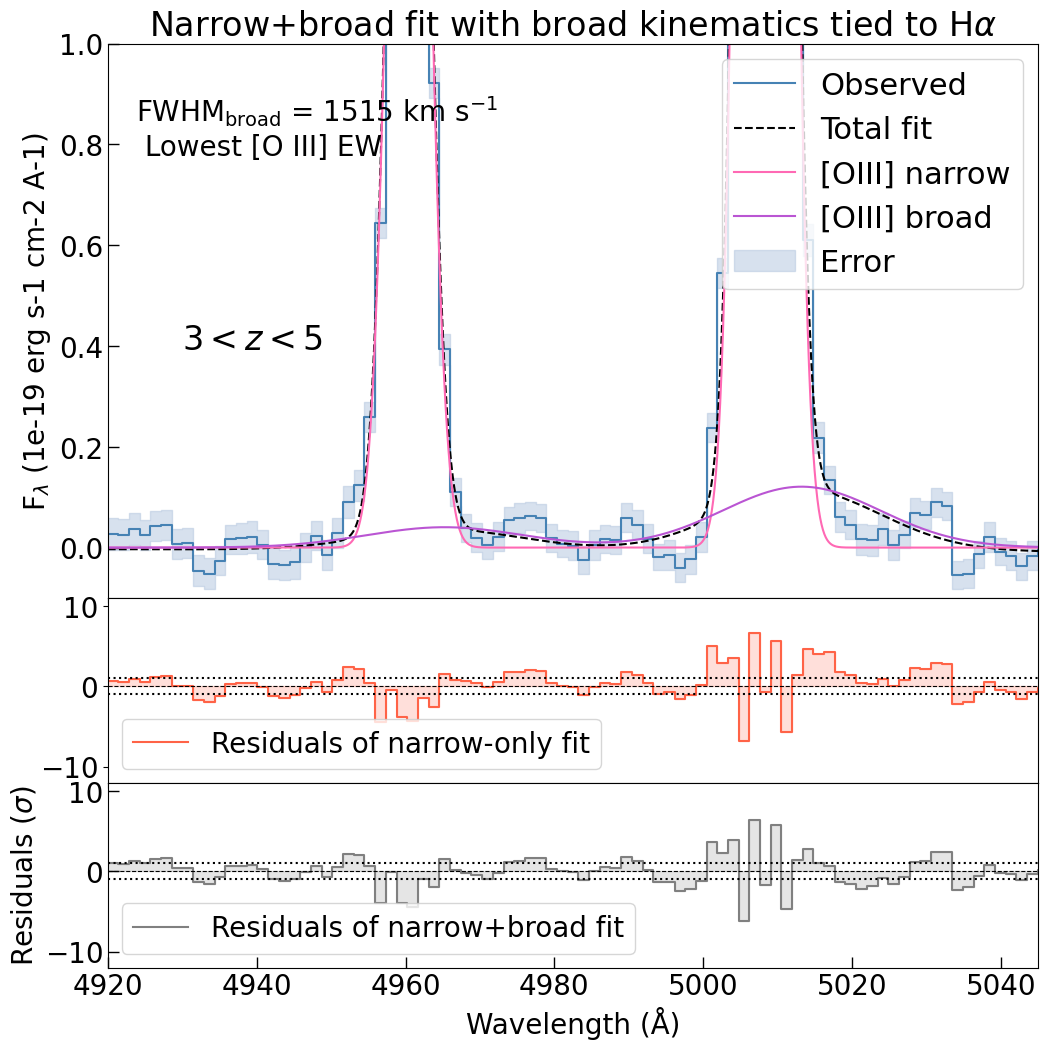}
    \end{subfigure}
    \begin{subfigure}[b]{0.48\textwidth}
        \includegraphics[width=\textwidth]{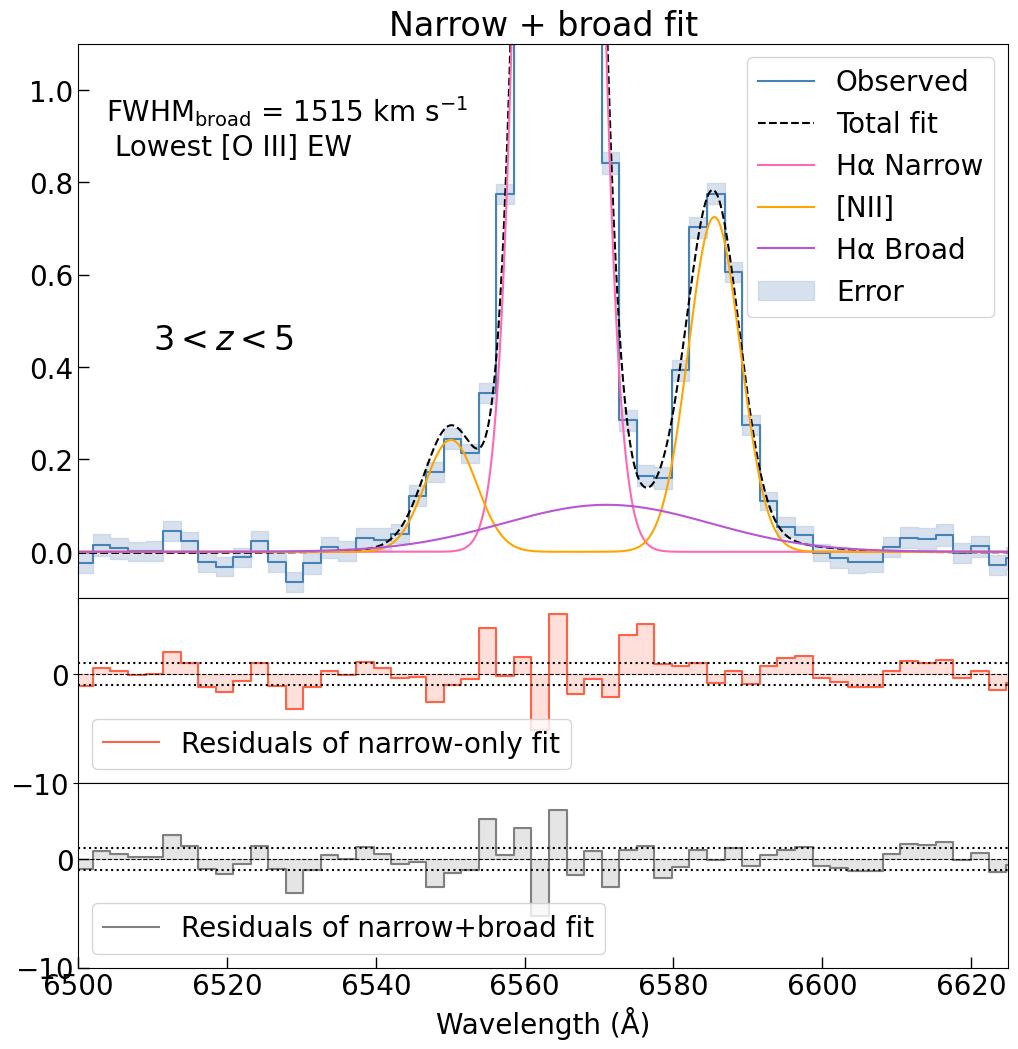}
    \end{subfigure}
    
    \caption{The \OIII and H$\alpha$ lines of the stacks that we have detected AGN activity in. The left panel shows the fit to \OIII with a broad component added, using kinematics tied to the H$\alpha$ broad component, while the right panel shows the broad component detected in H$\alpha$ for the highest \OIII EW stack at $5<z<7$. The pink line shows the fit of the narrow model, the purple line shows the fit of the broad component, the dashed line shows the total fit of the narrow, broad and linear continuum fit and the shaded region is the error. The centre panel in each image shows the residuals of a narrow only fit and the bottom panel shows the residuals of a narrow+broad fit. The scale on the residuals for the \OIII line and the H$\alpha$ lines are slightly different, but we plot the $\pm 1\sigma$ levels as a dotted line for clarity. The FWHM of the H$\alpha$ broad component is $\gtrsim1000$ \kms, supporting the BLR hypothesis. Some of these stacks display a broad component in \OIII but this does not match the broad H$\alpha$ kinematics. Thus, these stacks either only have AGN, or both AGN and outflows.}
    \label{fig:1}
\end{figure*}

\begin{figure*}
    \centering
    \begin{subfigure}[b]{0.48\textwidth}
        \includegraphics[width=\textwidth]{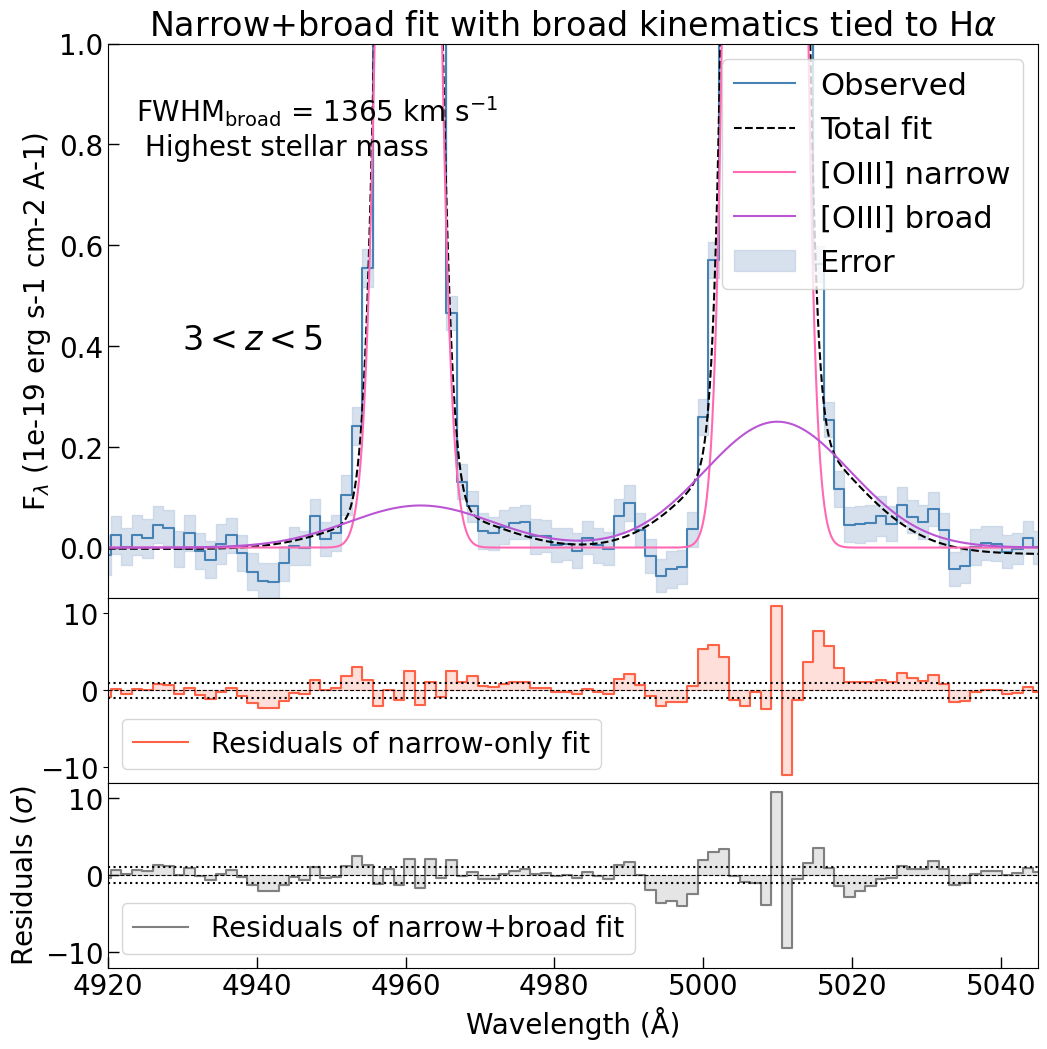}
    \end{subfigure}
    \begin{subfigure}[b]{0.48\textwidth}
        \includegraphics[width=\textwidth]{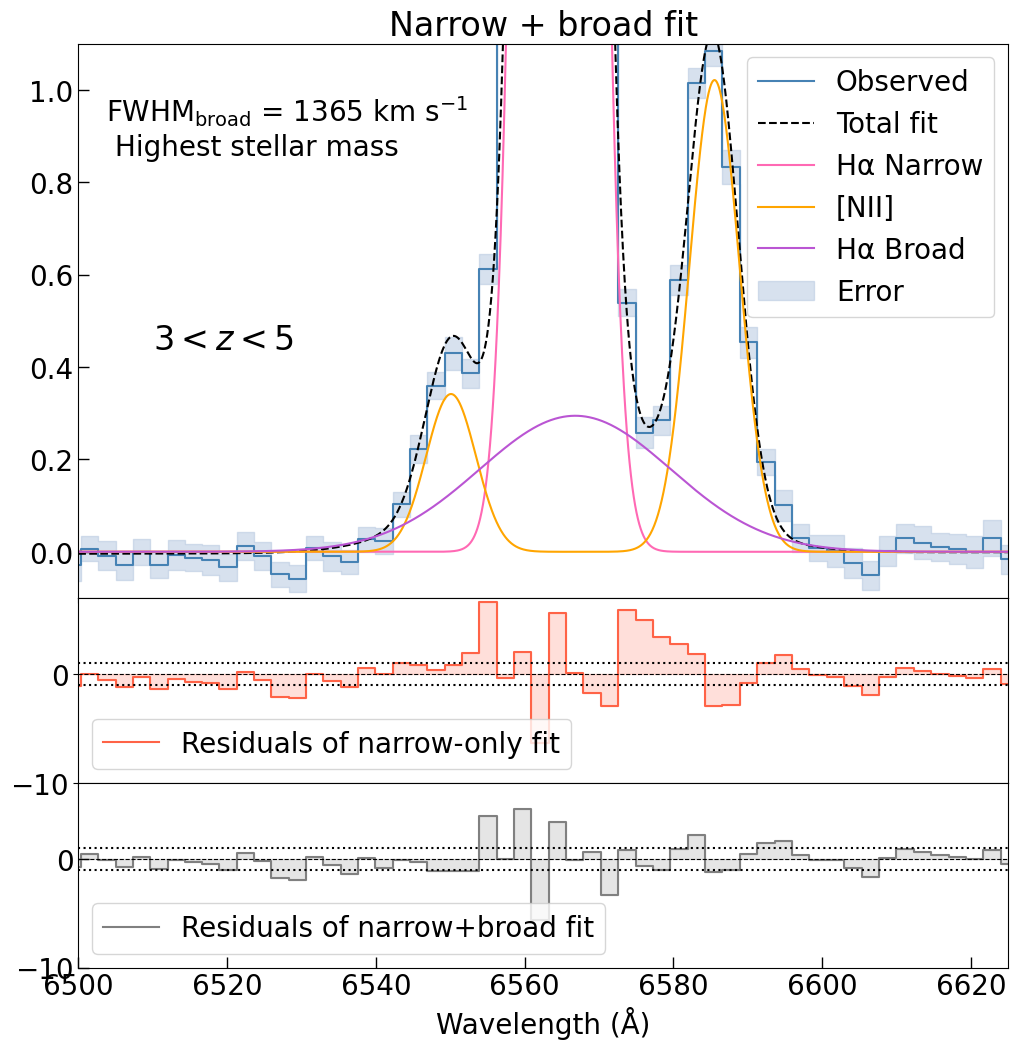}
    \end{subfigure}
    \begin{subfigure}[b]{0.49\textwidth}
        \includegraphics[width=\textwidth]{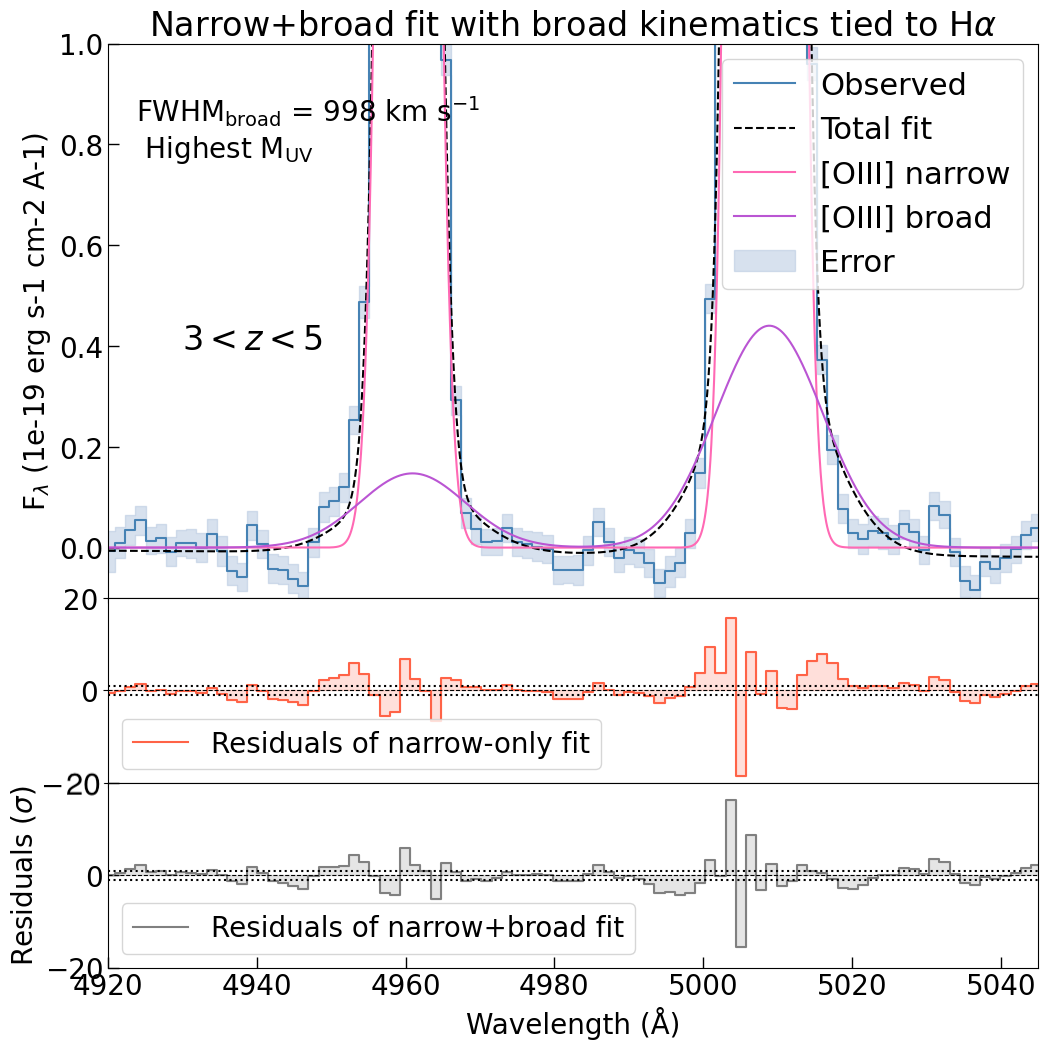}
    \end{subfigure}
    \begin{subfigure}[b]{0.47\textwidth}
        \includegraphics[width=\textwidth]{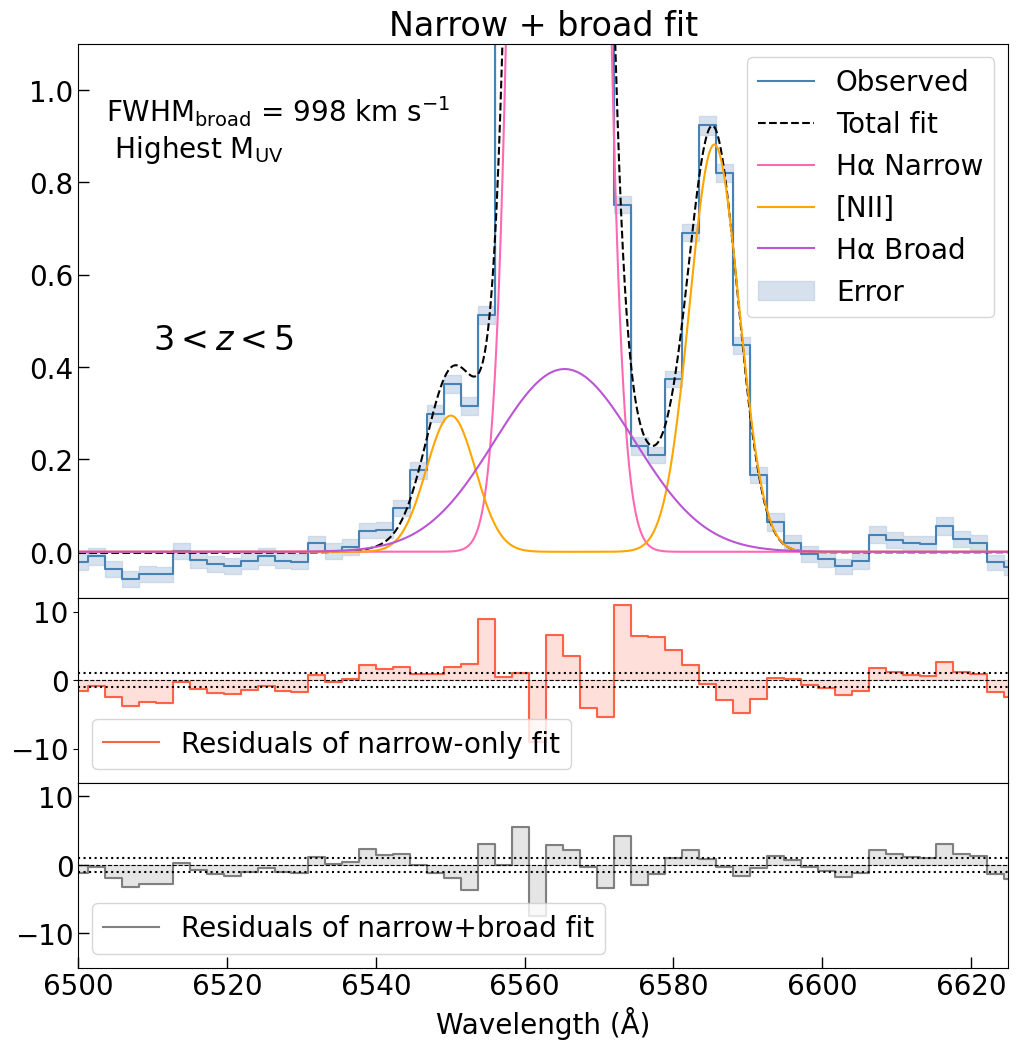}
    \end{subfigure}
    \caption{Extension of Figure \ref{fig:1}.}
    \label{fig:2}
\end{figure*}

\begin{figure*}
    \centering
    \begin{subfigure}[b]{0.49\textwidth}
        \includegraphics[width=\textwidth]{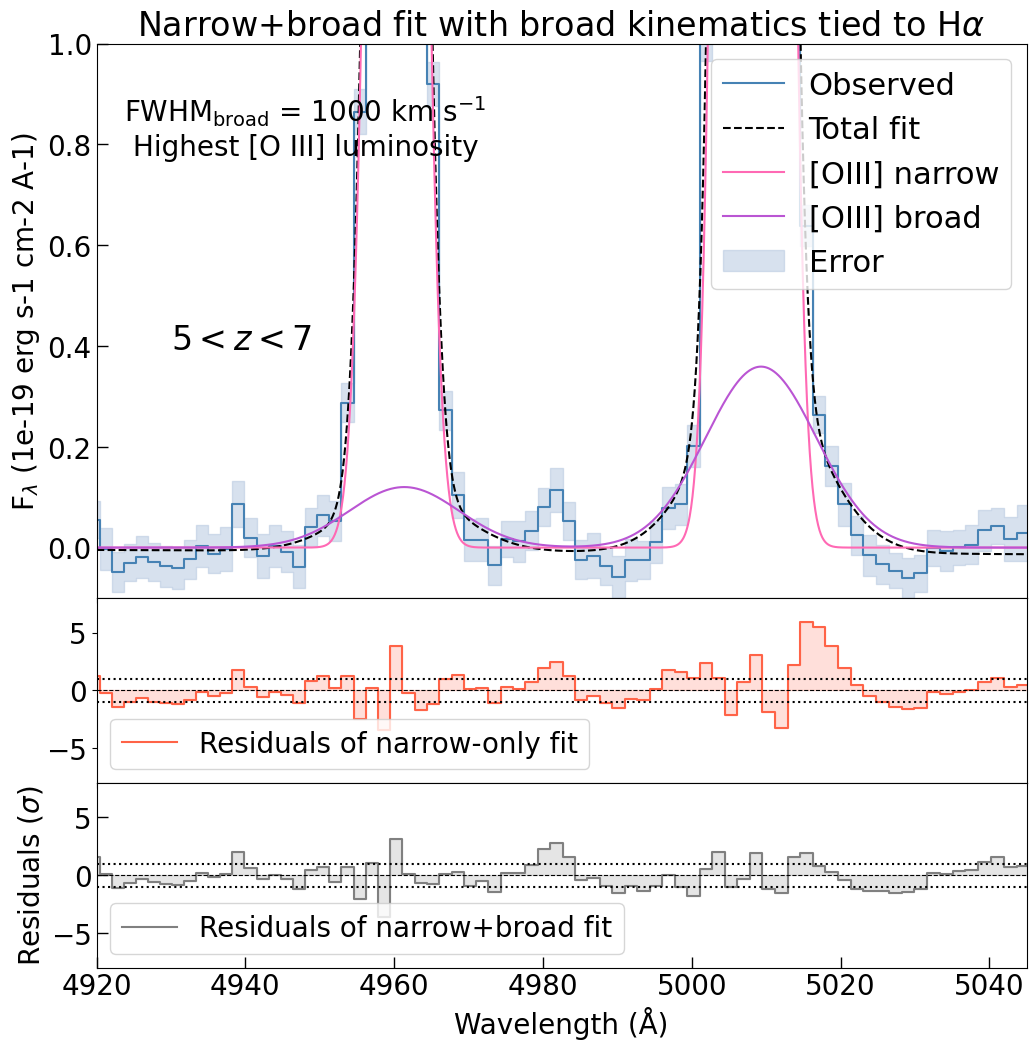}
    \end{subfigure}
    \begin{subfigure}[b]{0.47\textwidth}
        \includegraphics[width=\textwidth]{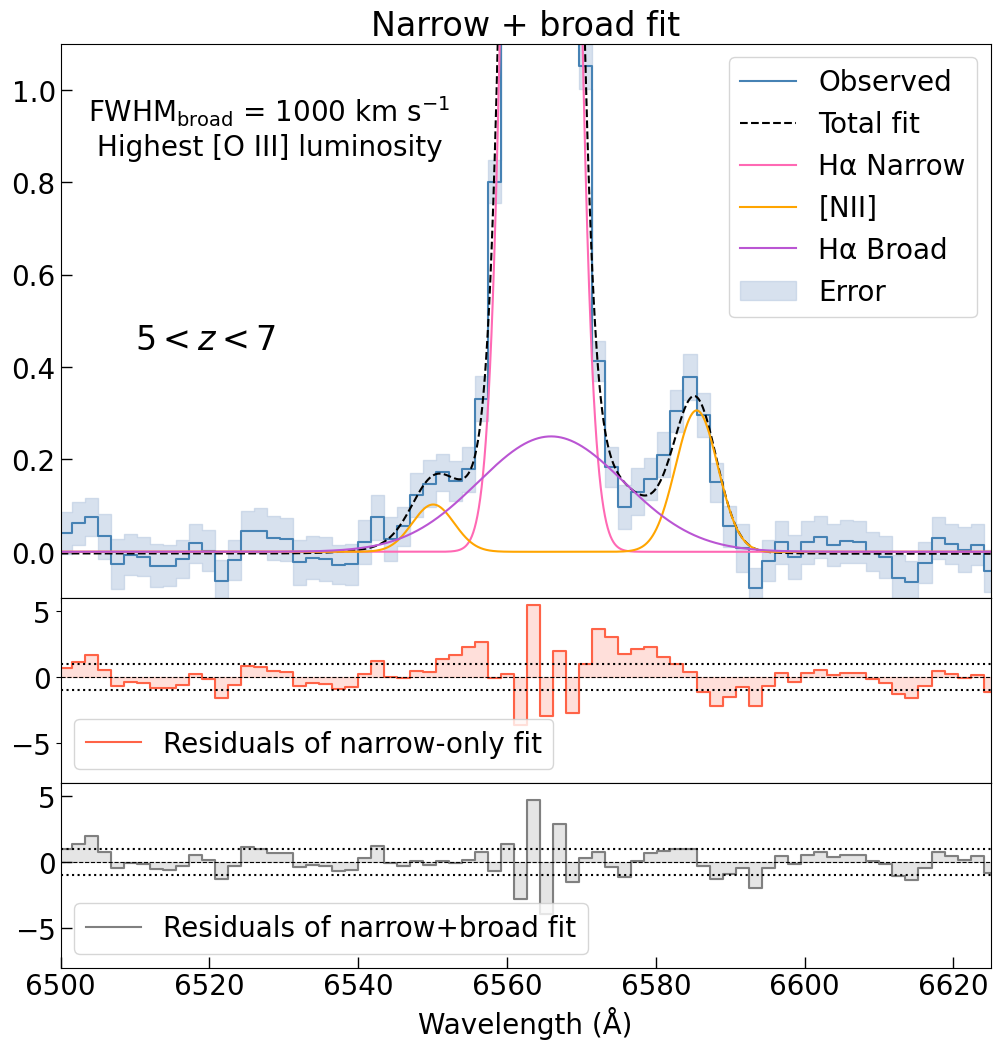}
    \end{subfigure}
    \caption{Extension of Figures \ref{fig:1} and \ref{fig:2}.}
    \label{fig:3}
\end{figure*}

Here we present the fits of the stacks that have a detected H$\alpha$ broad component attributed to AGN BLR, that were not shown in the text. These are shown in Figures \ref{fig:1}, \ref{fig:2} and \ref{fig:3}.

\section{Fitting of normalised/weighted spectra, and H\texorpdfstring{$\beta$}{b}}
\label{appendix:appendix_a}
\begin{figure*}
    \centering
    \begin{subfigure}[b]{0.33\textwidth}
        \includegraphics[width=\textwidth]{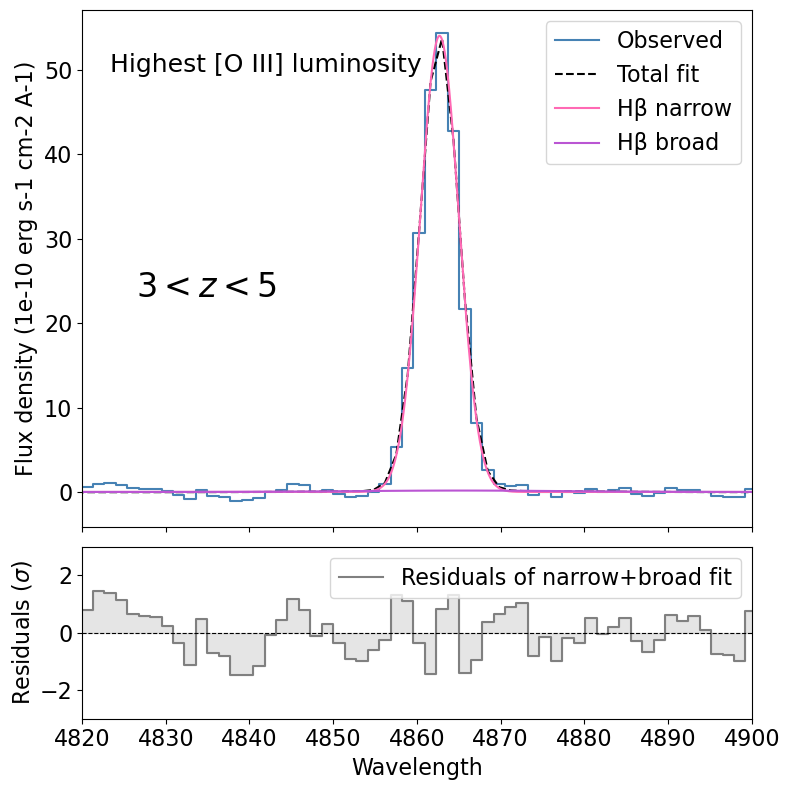}
    \end{subfigure}
    \begin{subfigure}[b]{0.33\textwidth}
        \includegraphics[width=\textwidth]{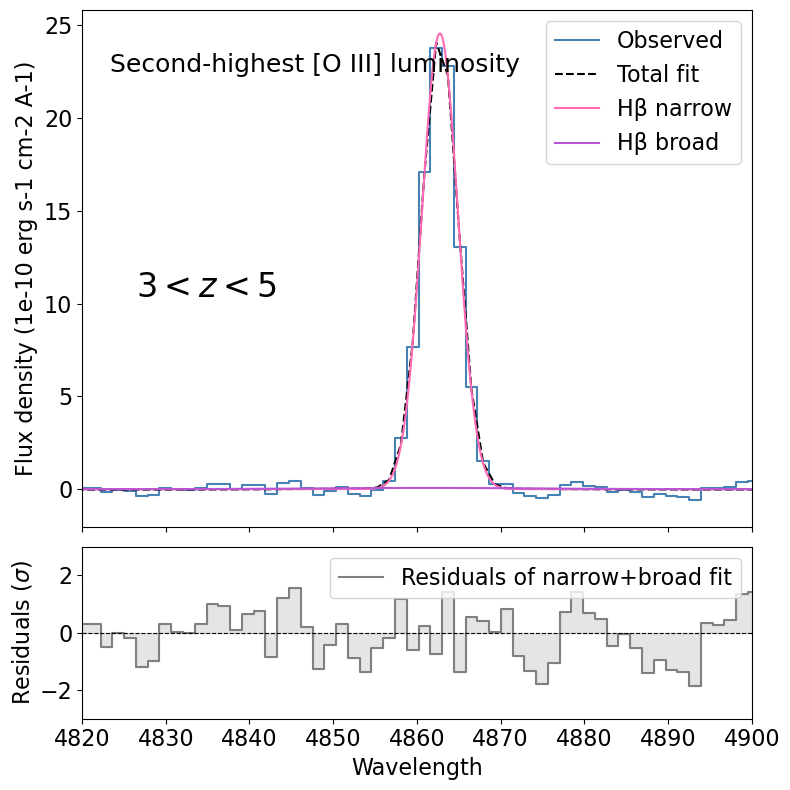}
    \end{subfigure}
    \begin{subfigure}[b]{0.33\textwidth}
        \includegraphics[width=\textwidth]{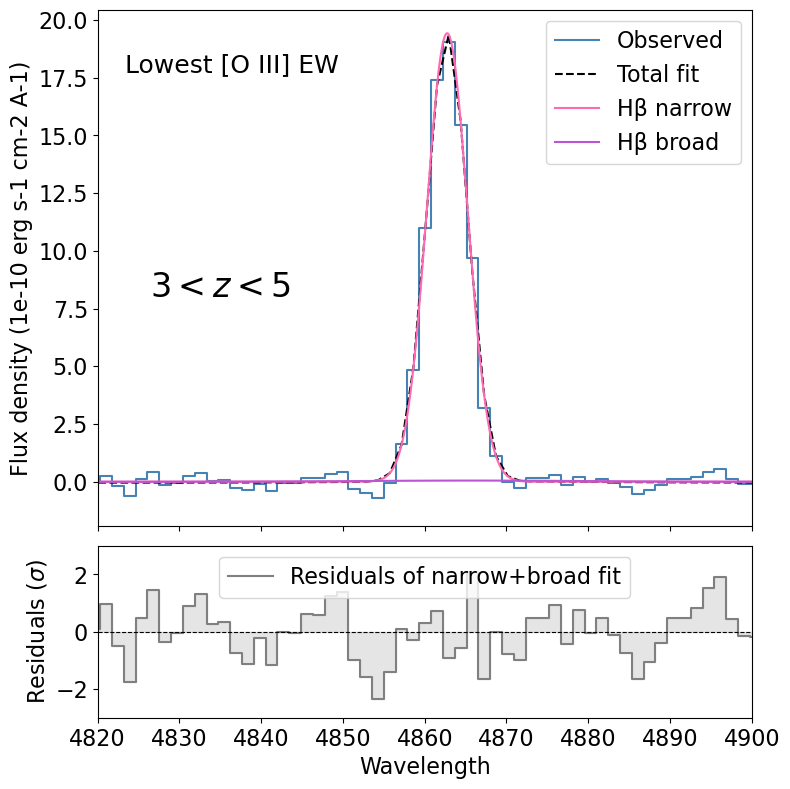}
    \end{subfigure}
    \begin{subfigure}[b]{0.33\textwidth}
        \includegraphics[width=\textwidth]{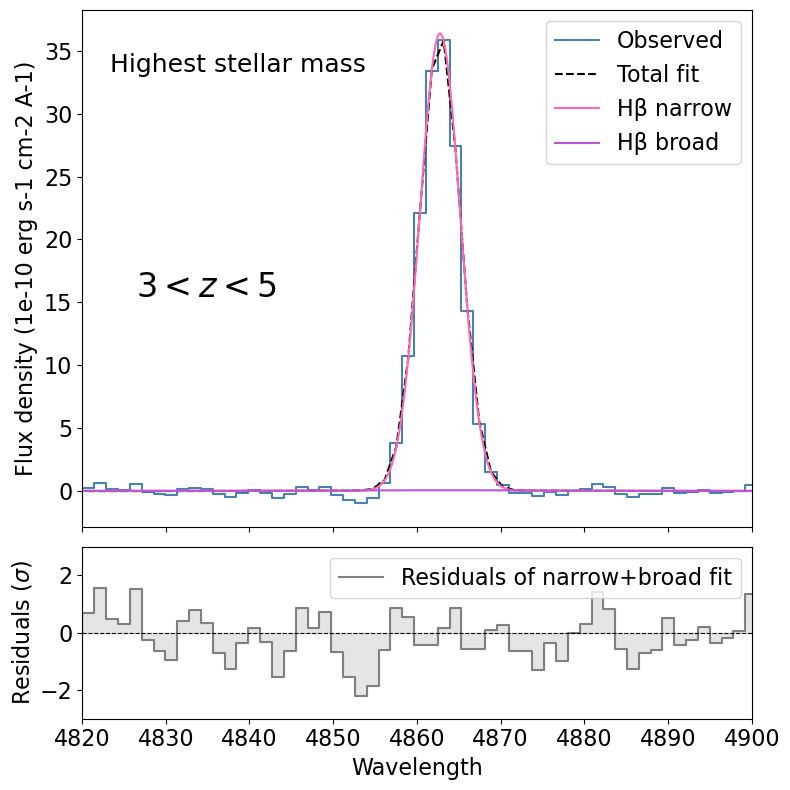}
    \end{subfigure}
    \begin{subfigure}[b]{0.33\textwidth}
        \includegraphics[width=\textwidth]{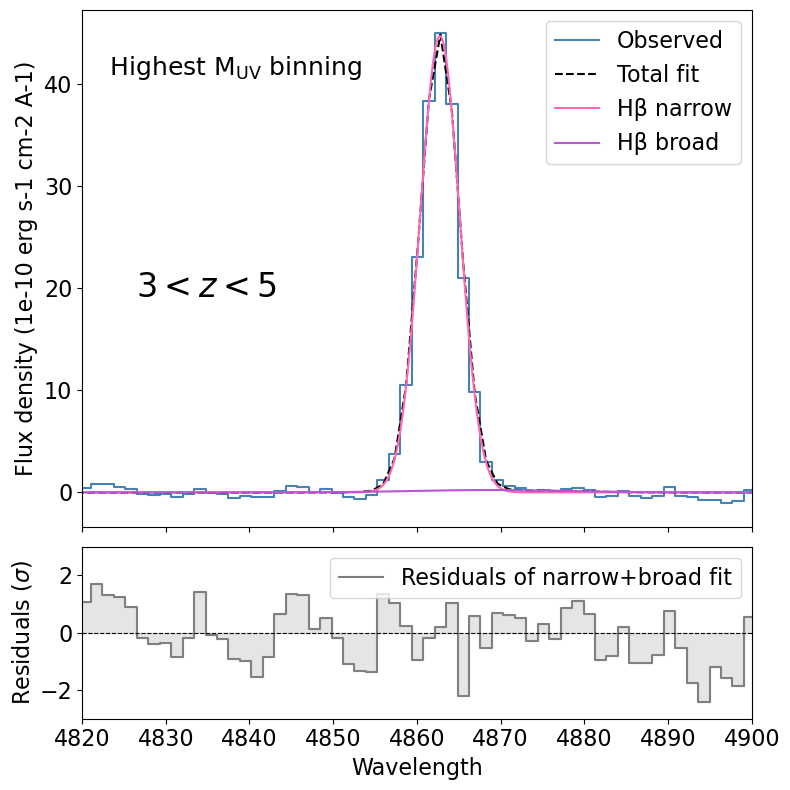}
    \end{subfigure}
    \begin{subfigure}[b]{0.33\textwidth}
        \includegraphics[width=\textwidth]{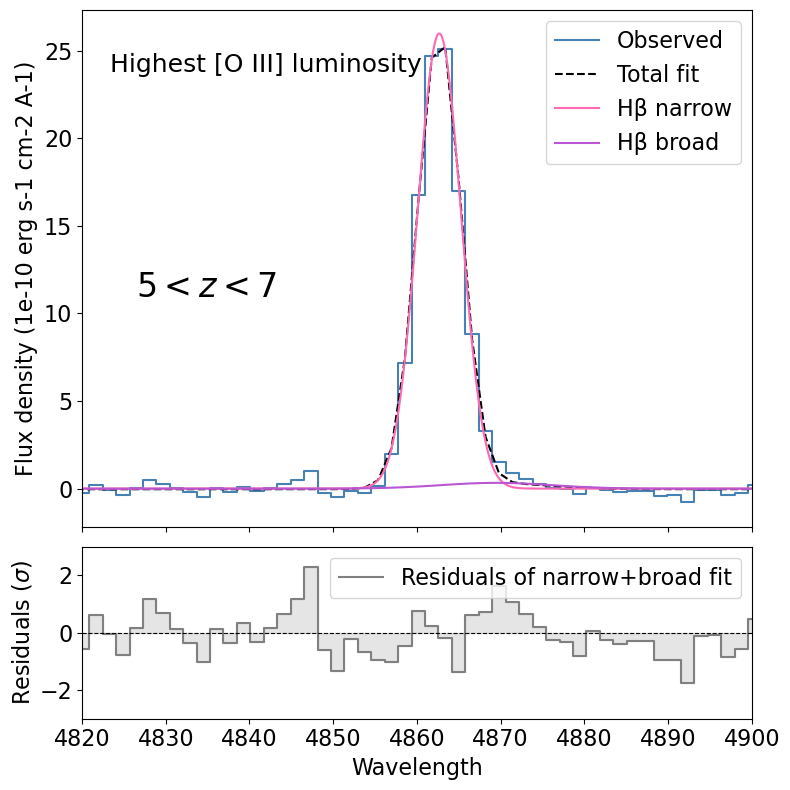}
    \end{subfigure}
    \begin{subfigure}[b]{0.33\textwidth}
        \includegraphics[width=\textwidth]{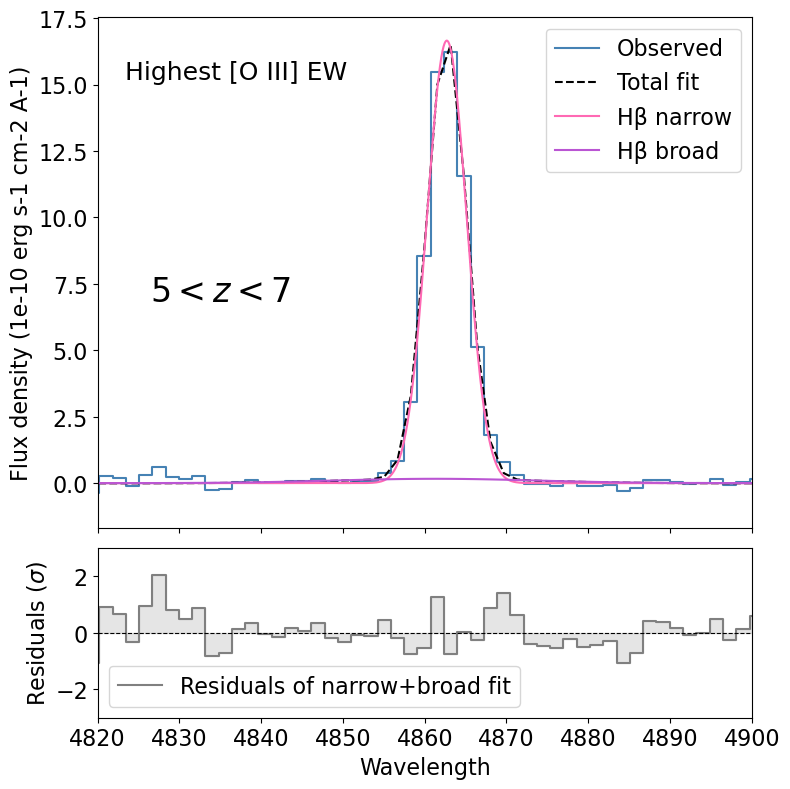}
    \end{subfigure}
    \caption{The results of fitting a broad component to H$\beta$ in the stacks where we have detected AGN activity. Left to right and top to bottom: Highest luminosity 3<z<5 ($\Delta BIC = -14.15$), second-highest luminosity 3<z<5 ($\Delta BIC = -15.26$), lowest EW 3<z<5 ($\Delta BIC = -15.28$) highest stellar mass 3<z<5 ($\Delta BIC = -15.77$), highest M$_{UV}$ 3<z<5 ($\Delta BIC = -12$), highest luminosity 5<z<7 ($\Delta BIC = -6.57$), highest EW 5<z<7 ($\Delta BIC = -12.16$). All values of $\Delta BIC$ are in favour of a narrow only model.} 
    \label{fig:no_hbeta}
\end{figure*}

We show the results of fitting a broad component to H$\beta$ in Fig.~\ref{fig:no_hbeta}. The results of $\Delta BIC = BIC_{\text{narrow only}} - BIC_{\text{narrow+broad}}$ (given in the captions) show that a model without a broad component is preferred.

\begin{figure*}
    \centering
    \begin{subfigure}[b]{0.33\textwidth}
        \includegraphics[width=\textwidth]{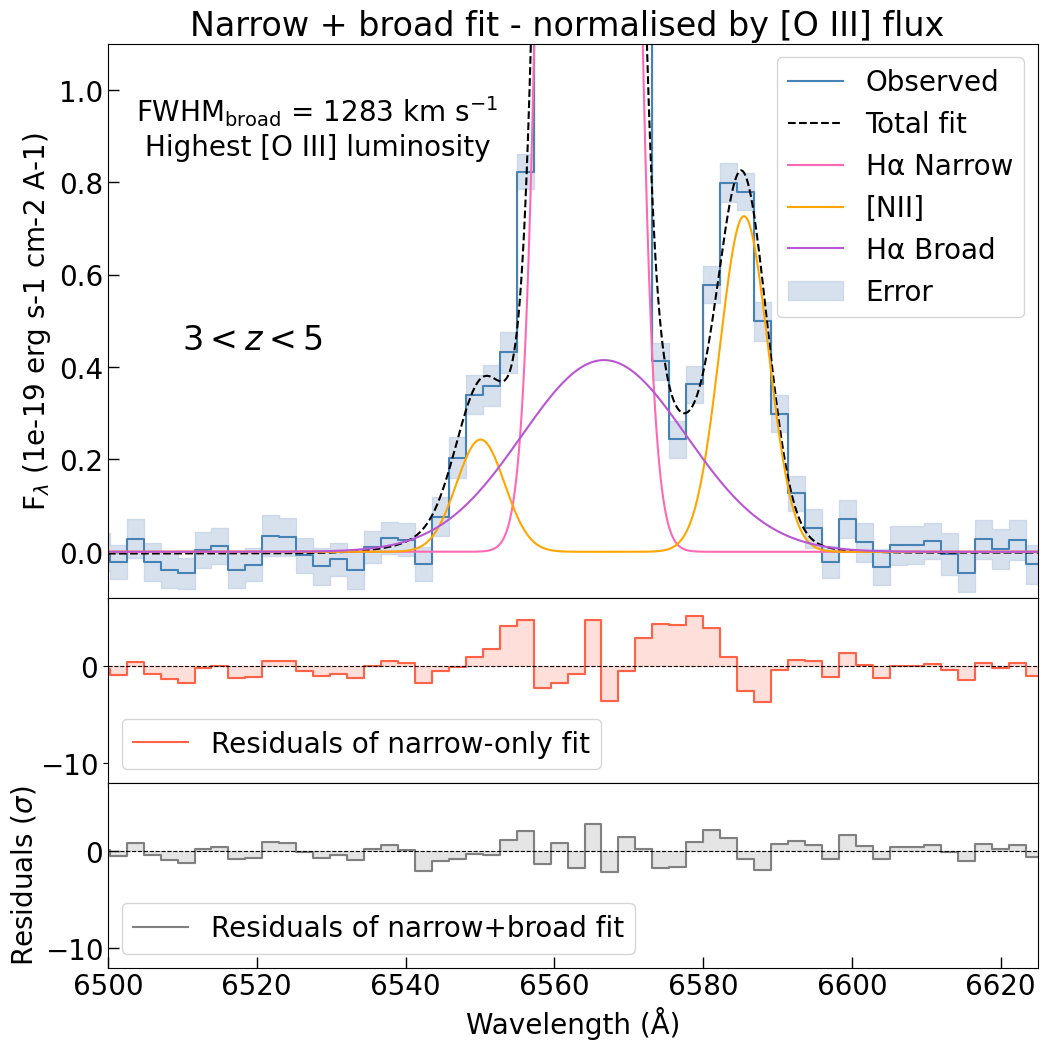}
    \end{subfigure}
    \begin{subfigure}[b]{0.33\textwidth}
        \includegraphics[width=\textwidth]{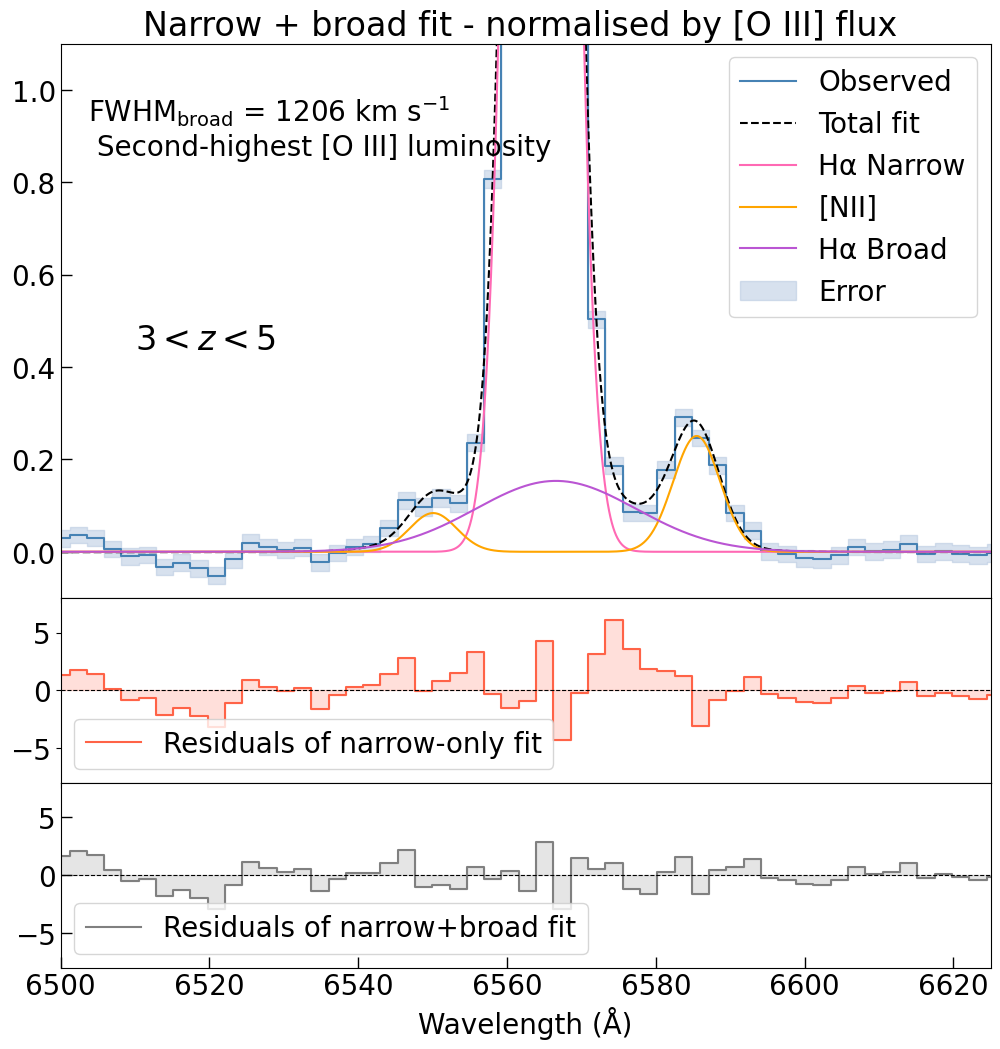}
    \end{subfigure}
    \begin{subfigure}[b]{0.33\textwidth}
        \includegraphics[width=\textwidth]{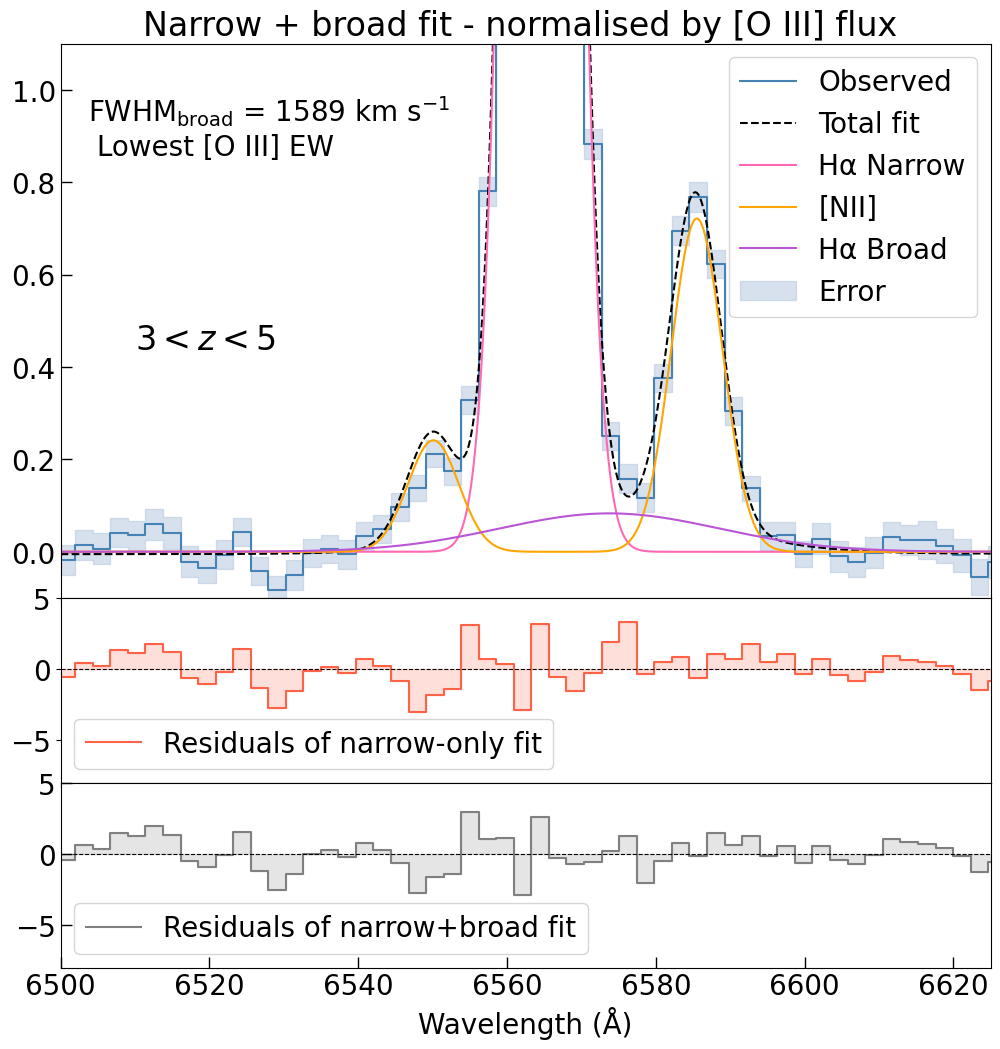}
    \end{subfigure}
    \begin{subfigure}[b]{0.33\textwidth}
        \includegraphics[width=\textwidth]{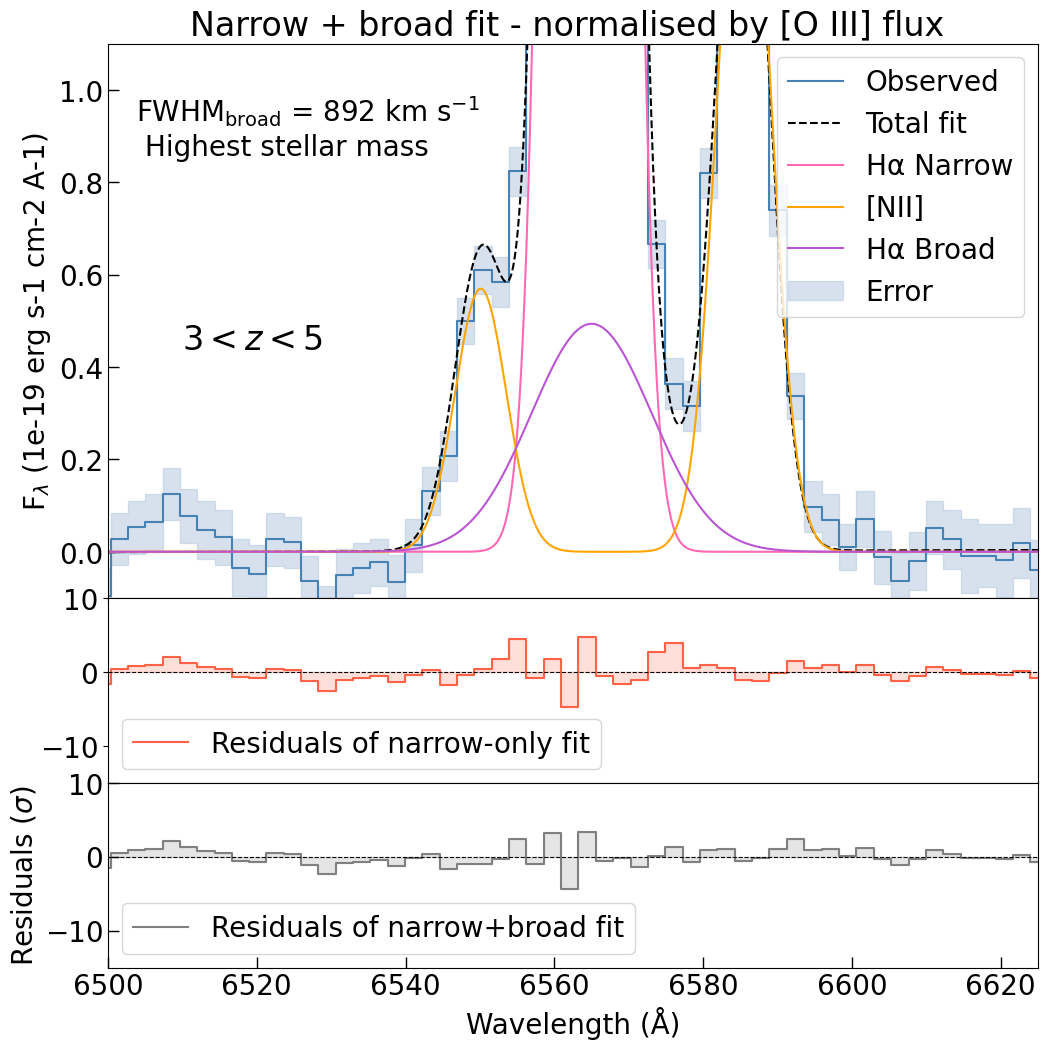}
    \end{subfigure}
    \begin{subfigure}[b]{0.32\textwidth}
        \includegraphics[width=\textwidth]{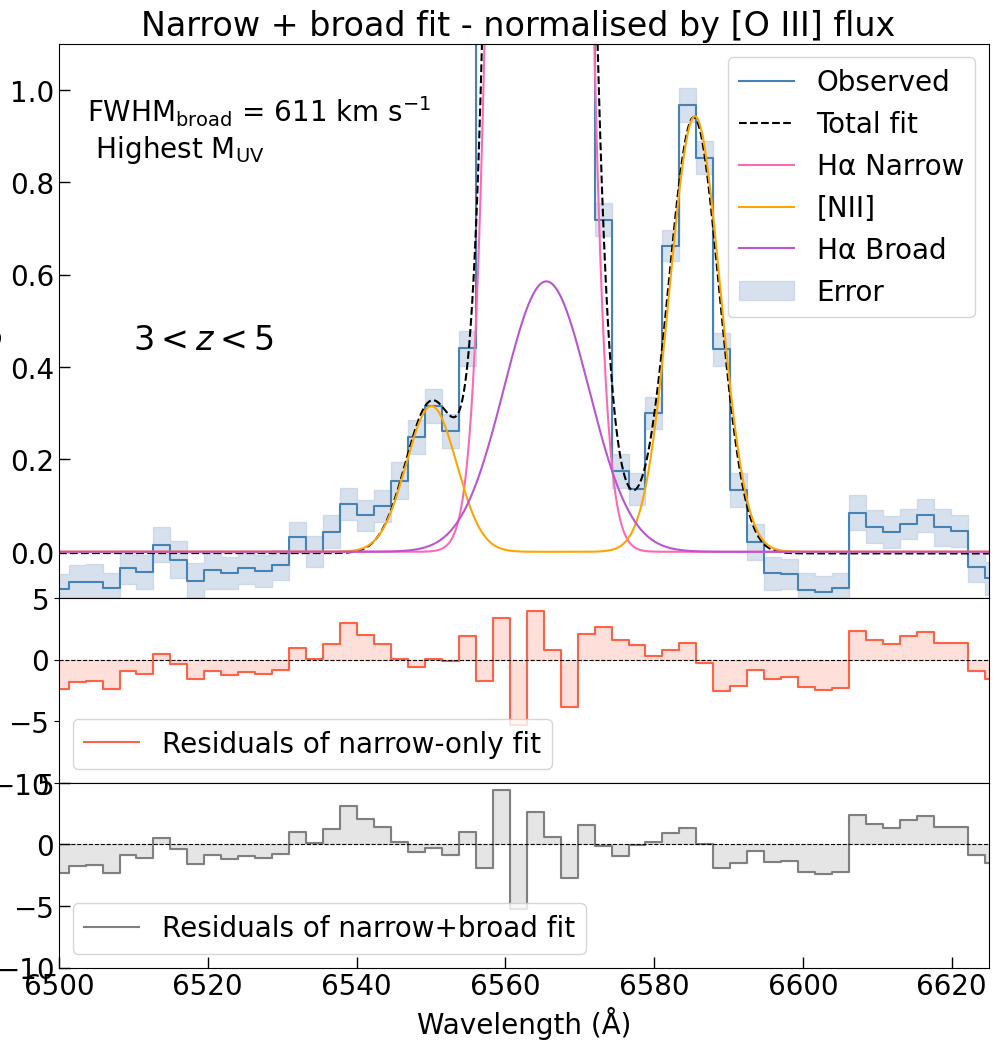}
    \end{subfigure}
        \begin{subfigure}[b]{0.33\textwidth}
        \includegraphics[width=\textwidth]{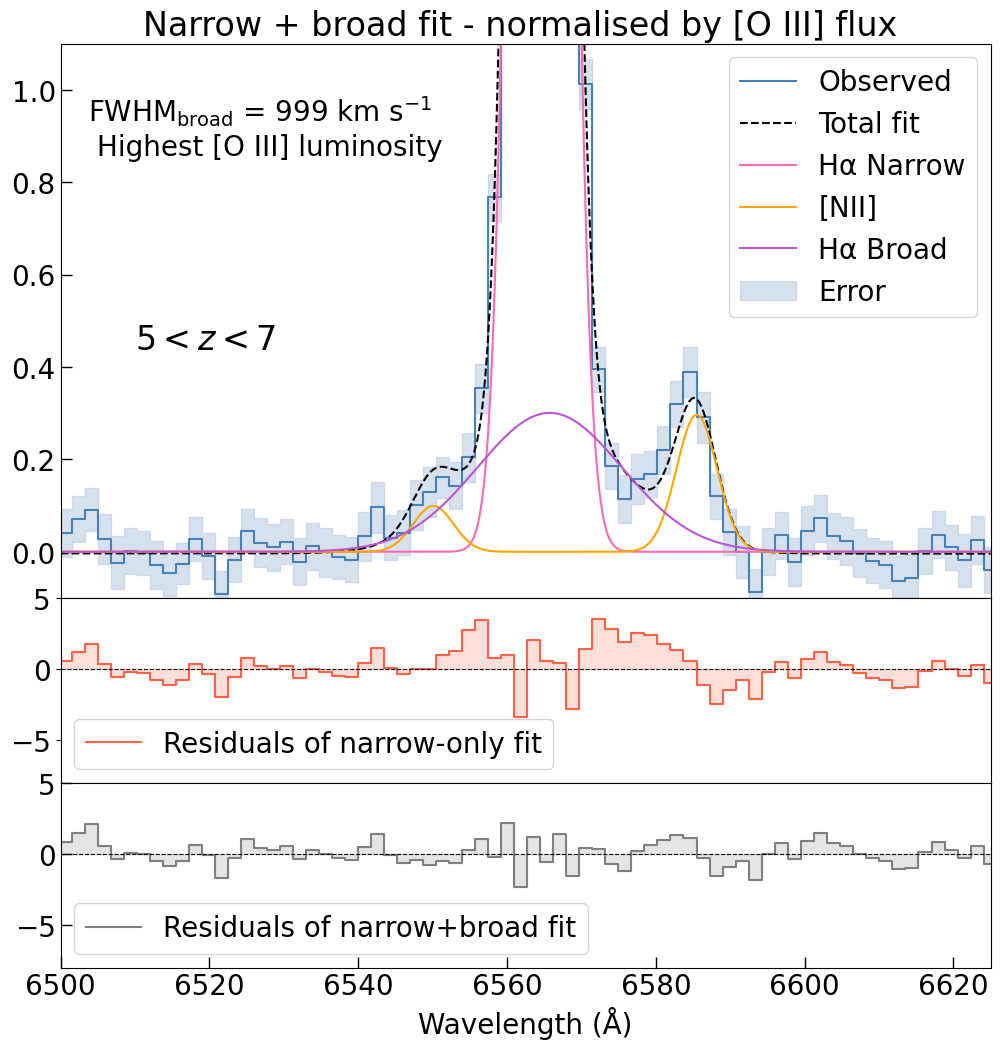}
    \end{subfigure}
    \begin{subfigure}[b]{0.33\textwidth}
        \includegraphics[width=\textwidth]{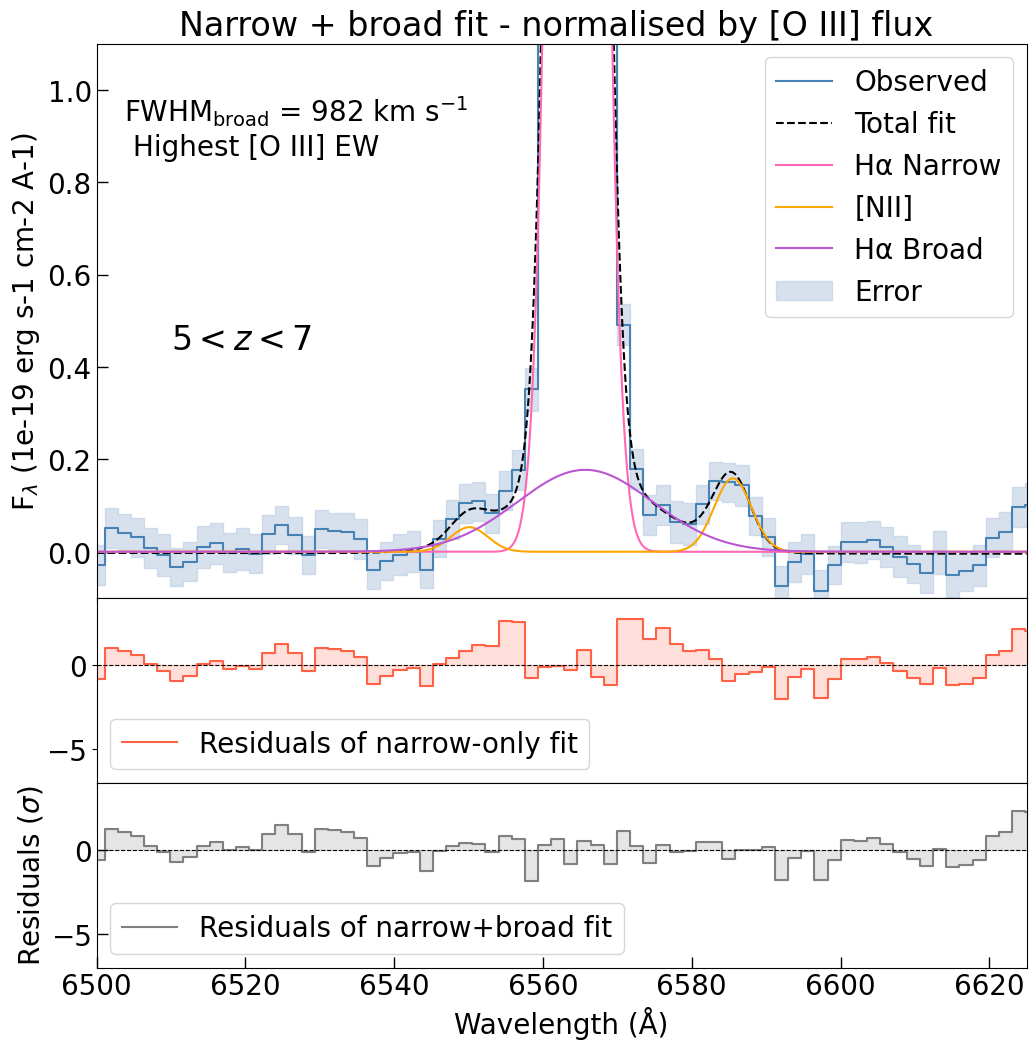}
    \end{subfigure}
    \caption{The fitted H$\alpha$ broad line of the stacks with AGN confirmed, where the stacks are normalised by \OIII flux. The broad H$\alpha$ components still appear in most of the stacks but some are slightly narrower, indicating that it is the sources with the highest \OIII flux that contributes the most to the broad components.}
    
    \label{fig:agn_norm}
\end{figure*}

We present the broad line fits to the stacks that are normalised by \OIII flux in Fig.~\ref{fig:agn_norm}. This is the same figure as Figs.~\ref{fig:highEW}, ~\ref{fig:halpha_broad1}, except the fitting has been done on the normalised stacks.
For the highest stellar mass stack, the broad H$\alpha$ emission is weakened by the normalisation, which the narrower value of FWHM$_{\text{broadH}\alpha}$ reflects. The other stacks do not exhibit significant changes to their broad H$\alpha$ profiles. This indicates that it is the sources with the highest $F_{\text{[O III]}}$ that contribute the broad component to the stack because they are weighted down from the normalised reducing the broad component in the highest stellar mass stacks, but causing no change in the other stacks with the brightest \OIII emission. This is in agreement with our results that the brightest \OIII sources are those with the broad components. 

\begin{figure*}
    \centering
    \begin{subfigure}[b]{0.33\textwidth}
        \includegraphics[width=\textwidth]{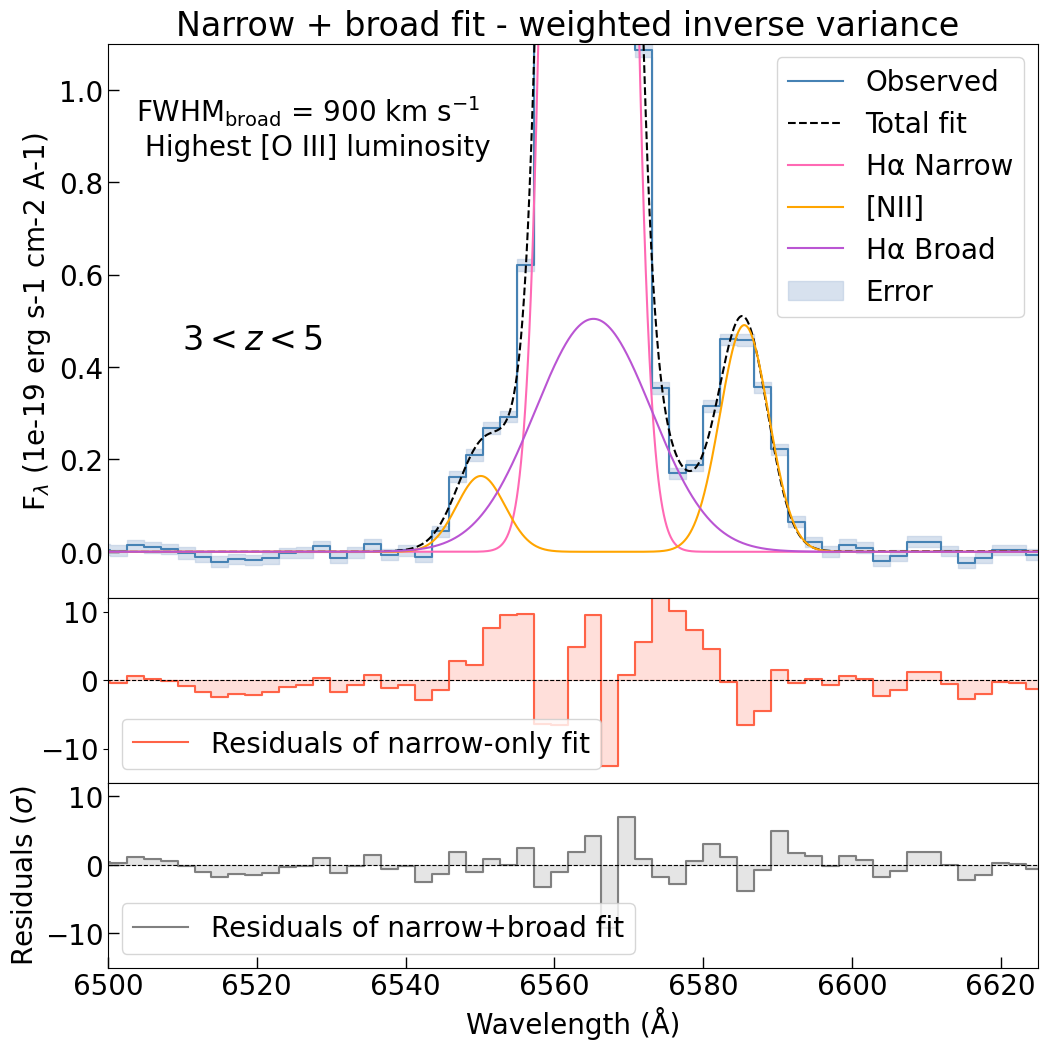}
    \end{subfigure}
    \begin{subfigure}[b]{0.33\textwidth}
        \includegraphics[width=\textwidth]{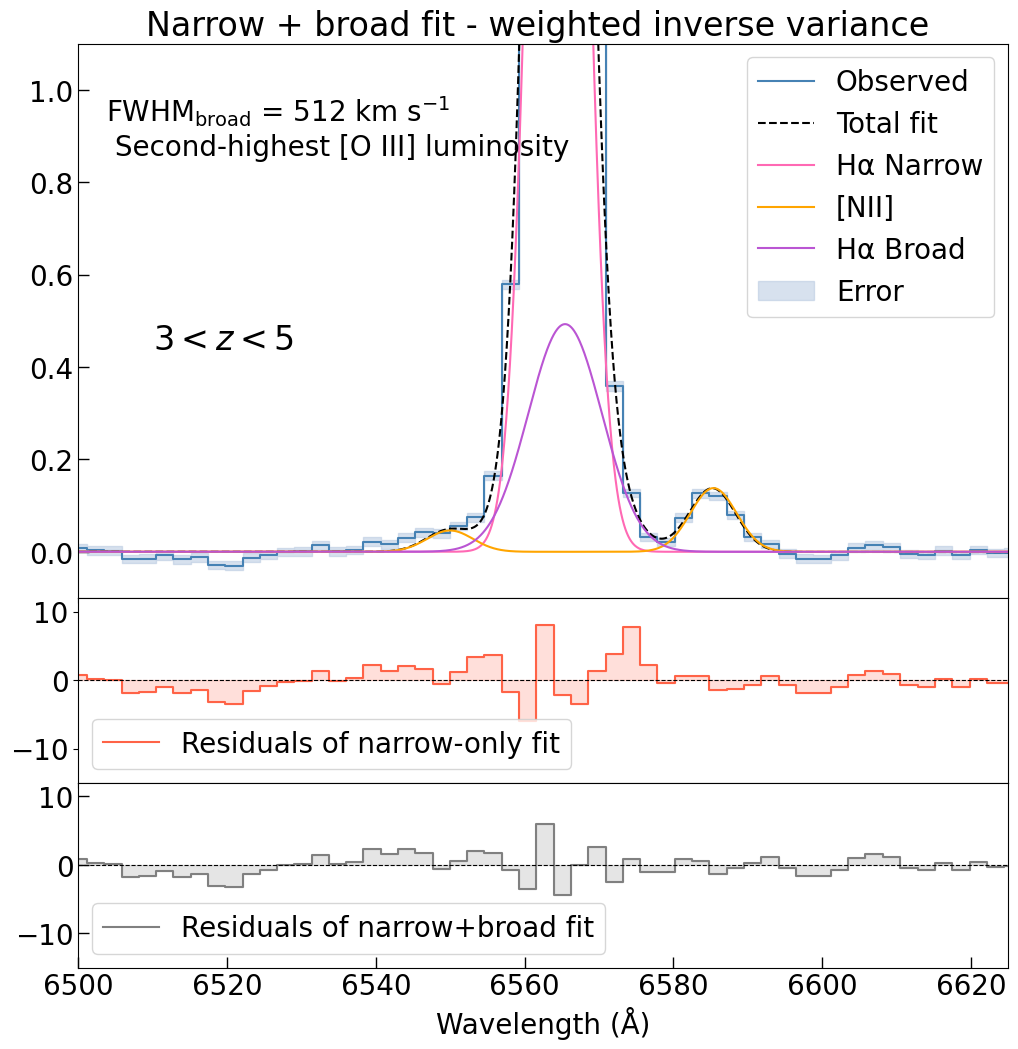}
    \end{subfigure}
    \begin{subfigure}[b]{0.33\textwidth}
        \includegraphics[width=\textwidth]{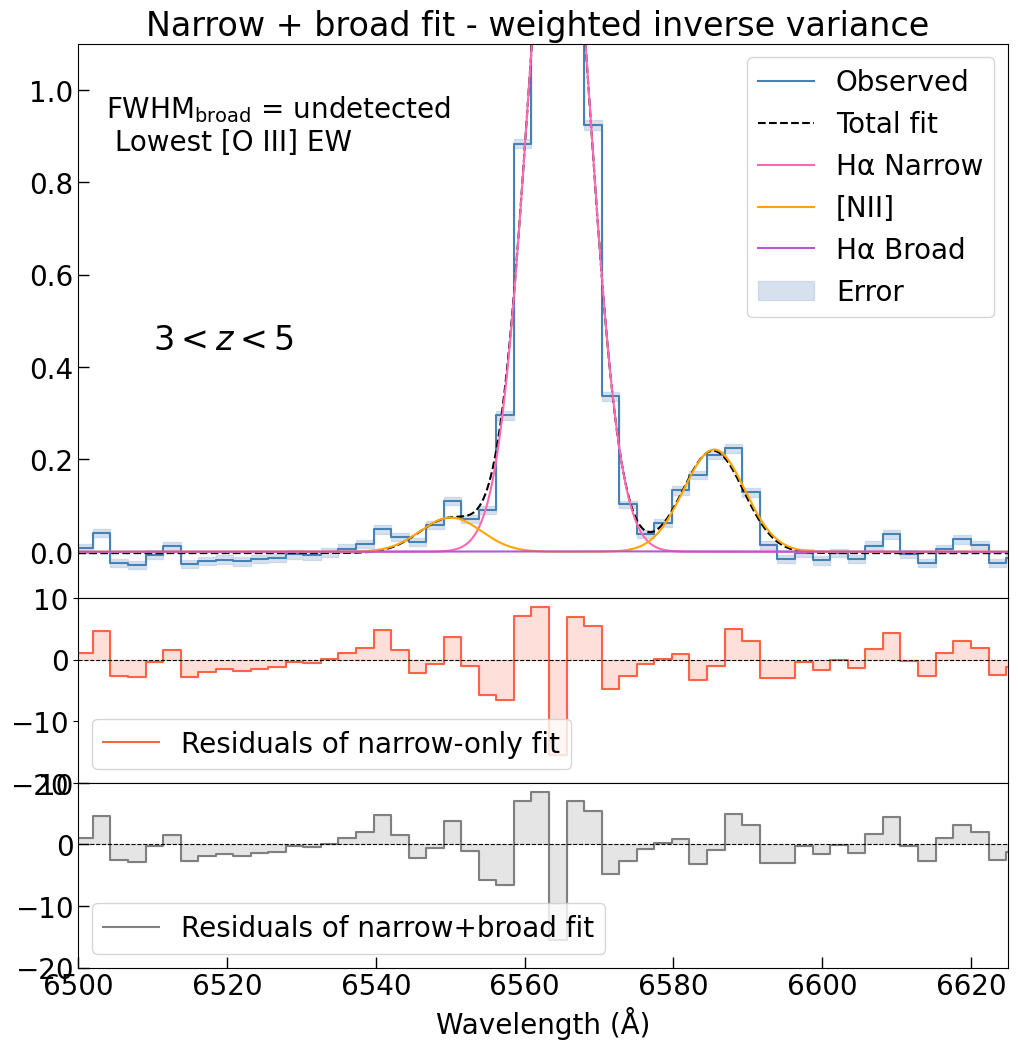}
    \end{subfigure}
    \begin{subfigure}[b]{0.33\textwidth}
        \includegraphics[width=\textwidth]{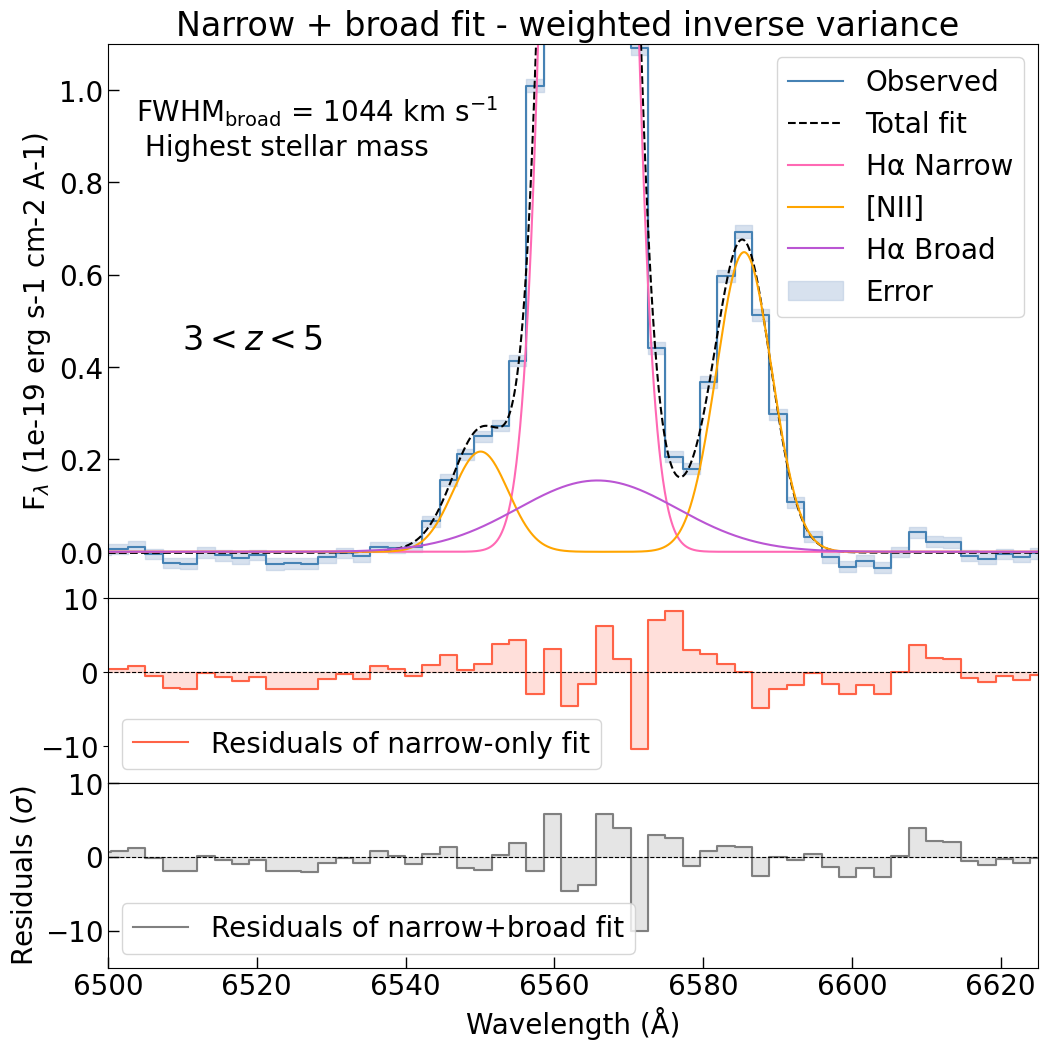}
    \end{subfigure}
    \begin{subfigure}[b]{0.32\textwidth}
        \includegraphics[width=\textwidth]{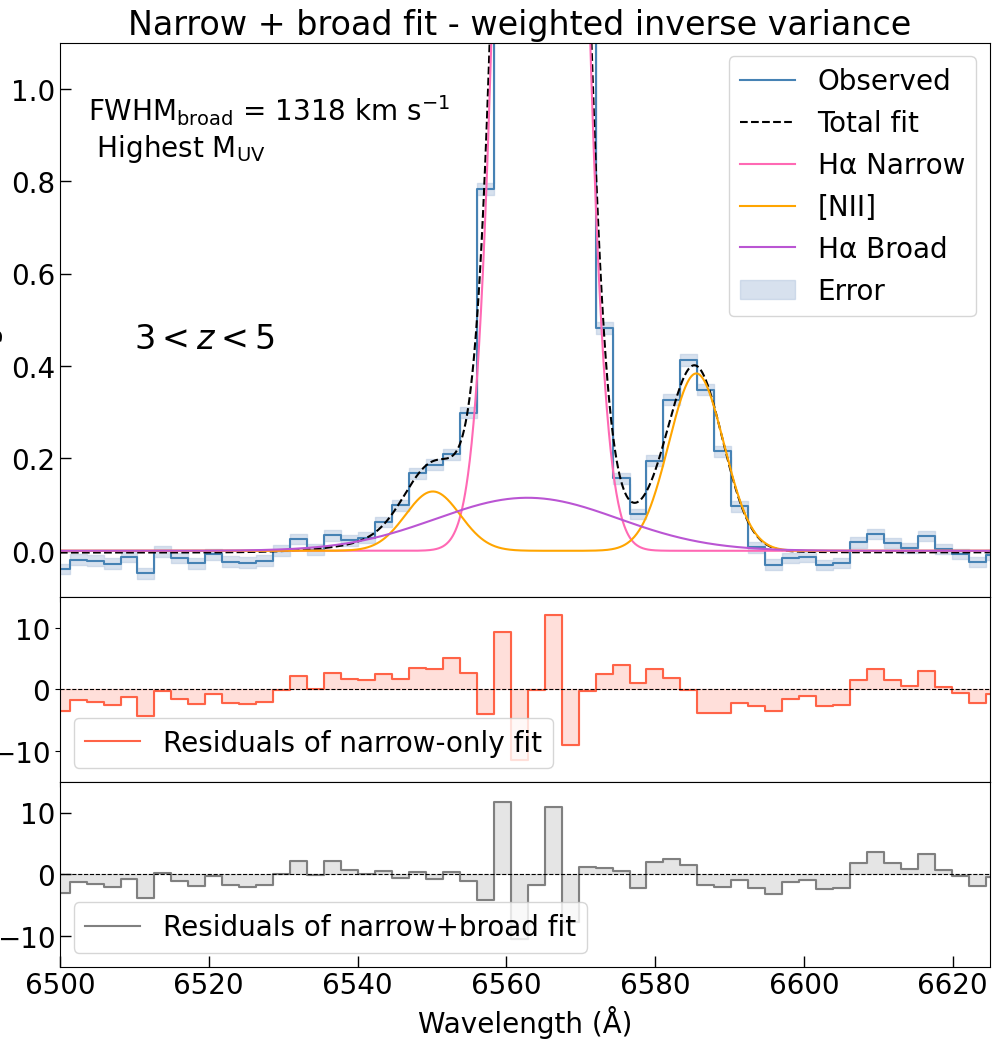}
    \end{subfigure}
    \begin{subfigure}[b]{0.33\textwidth}
        \includegraphics[width=\textwidth]{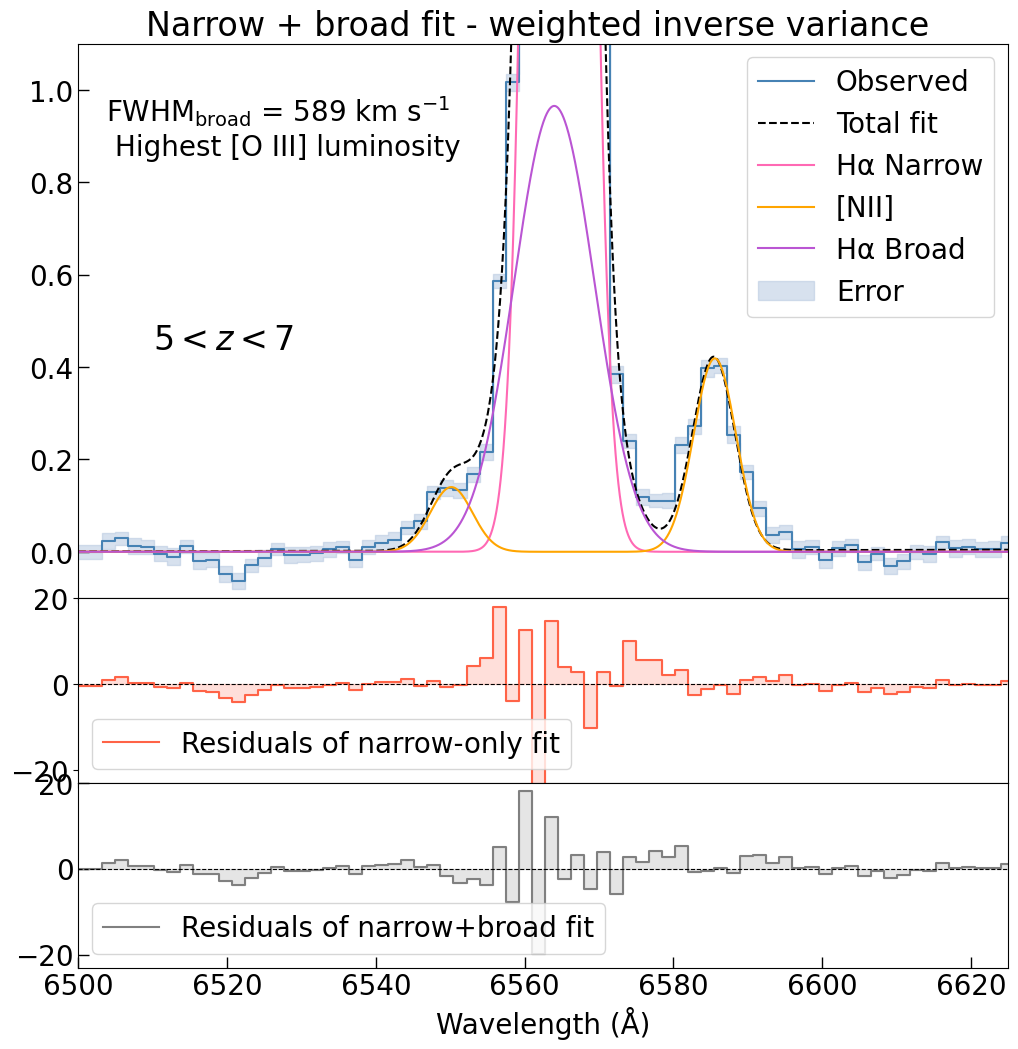}
    \end{subfigure}
    \begin{subfigure}[b]{0.33\textwidth}
        \includegraphics[width=\textwidth]{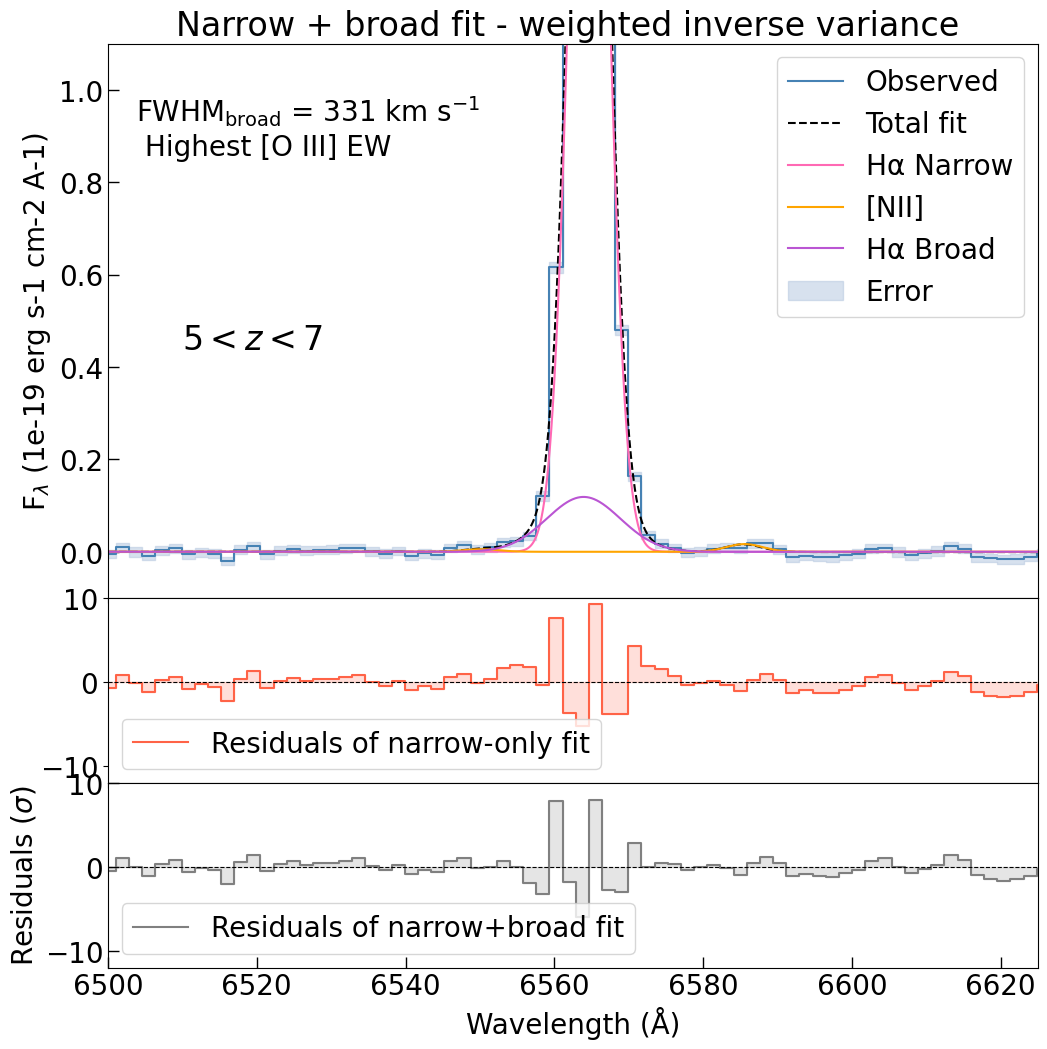}
    \end{subfigure}
    \caption{The fitted H$\alpha$ line of the stacks with AGN confirmed, where the stacks are weighted by rms$^{-2}$. Some of the broad H$\alpha$ lines are no longer detected due to increased noise in the emission line regions of the individual sources, resulting in reduced flux around the emission line regions in the stack (e.g. see Figure \ref{fig:rms_comparison}).} 
    \label{fig:agn_rms}
\end{figure*}

\begin{figure}
	\includegraphics[width=\columnwidth]{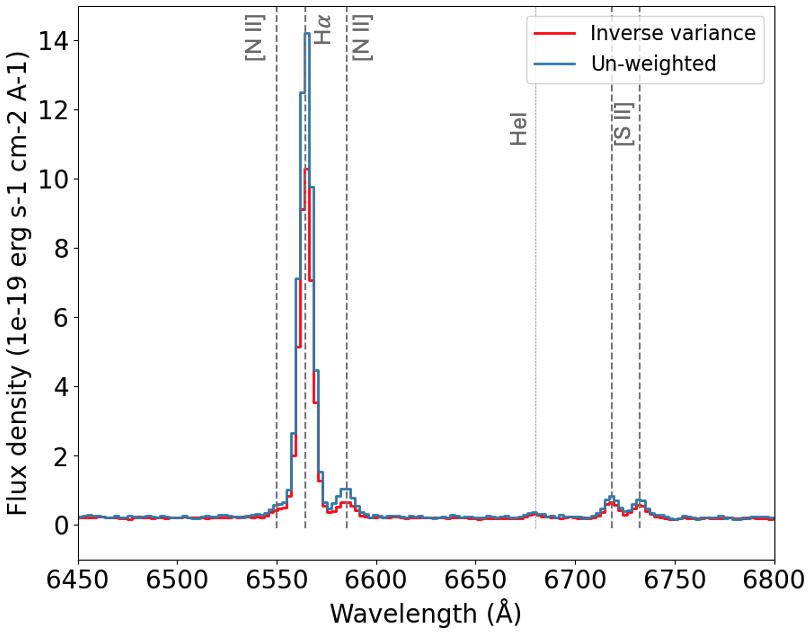}
    \caption{The highest luminosity $3<z<5$ stack weighted by inverse variance compared to the unweighted stack. The decreased flux in the emission line regions of the weighted stacks show that there is increased noise in these regions.}
    \label{fig:rms_comparison}
\end{figure}

We also present the results for the stacks weighted by rms$^{-2}$ in Figure \ref{fig:agn_rms}. These results show that the broad component is not detected as well as it is in the unweighted stacks, and some of the previously detected broad components are no longer detected, e.g. the second highest luminosity bin $3<z<5$, the lowest EW bin $3<z<5$ and the highest luminosity and EW $5<z<7$ stack. To understand why this happens, we compare a stack without any normalisation or weighting, to the same stack but weighted by the inverse variance in Figure \ref{fig:rms_comparison}. It is clear that the average flux from the stack that is unweighted and not normalised is different from the average flux from the stack that is normalised by inverse variance. This effect is most prominent around the emission line regions - the continuum values are fairly consistent. Since the noise usually increases in regions below emission lines, this could imply that the overall line fluxes in the inverse variance weighted stacks are decreased which is causing the broad component to be less detectable, rather than the broad component itself having more noise and therefore being poorly detected. Therefore, we conclude that while some of the broad line detections are not found in the inverse variance weighted stacks, this does not necessarily imply that they have too much noise to be attributed to real features of the stacks, and it is likely that they are just too faint to be detected due to the overall decrease in flux around the emission lines.

\section{Results of the stacks without broad H\texorpdfstring{$\alpha$}{alpha}}
\label{appendix:appendix_b}
Table \ref{tab:non_agn} lists the methods we used to rule out the possibility that these stacks show AGN activity. Our exclusion reasons are any of the following:

\begin{enumerate}
    \item SNR of the fitted broad H$\alpha$ component is <3,
    \item there is a broad component in \OIII with approximately the same FWHM as the broad H$\alpha$ component
    \item $\Delta BIC = BIC_{\text{narrow H}\alpha} - BIC_{\text{broad H}\alpha} <6$
\end{enumerate}

\begin{table*}
    \centering
    \caption{Results of fitting a broad component to H$\alpha$ and \OIII in the stacks that we ruled out as hosting an AGN. Column 1: Stack. Column 2: Method we used to rule out the stack. Column 3: If the method is broad H$\alpha$ counterpart in \OIII then we give $\Delta BIC_{\text{O III}} = BIC_{H\alpha \text{kinematics}} - BIC_{\text{free fit}}$ - if this is close to zero or negative it shows that the broad H$\alpha$ component also appears in \OIII, indicating outflows instead a BLR. If exclusion method is $\Delta BIC_{\text{broad H}\alpha}$ we give $\Delta BIC = BIC_{\text{narrow H}\alpha} - BIC_{\text{broad H}\alpha}$ - if this is <6 then the model does not strongly prefer the broad fit. If the method is SNR<3 we give the SNR of the fitted broad H$\alpha$ component.}
    \label{tab:non_agn}
    \resizebox{0.8\textwidth}{!}{%
    \begin{tabular}{lcccccccc}
        \hline
        Stack & Exclusion method & Results\\[0.15cm] 
        \hline
        3<z<5 second-lowest \OIII luminosity ($4.44\times 10^{41} - 8.01\times 10^{41}$ erg/s) & SNR$<3$ & 0.2$\sigma$ \\[0.15cm] 
        3<z<5 lowest \OIII luminosity ($7.92\times 10^{41} - 4.29\times 10^{41}$ erg/s) & SNR$<3$ & 1.3$\sigma$ \\[0.15cm] 
        3<z<5 highest \OIII EW (806--3327 Å) & broad H$\alpha$ counterpart in \OIII & -8\\[0.15cm] 
        3<z<5 second-highest \OIII EW (434--803 Å) & SNR$<3$ & 0.9$\sigma$ \\[0.15cm] 
        3<z<5 second lowest \OIII EW (201--430)& broad H$\alpha$ counterpart in \OIII & 0.3\\[0.15cm] 
        3<z<5 second highest stellar mass ($\log(M_{*}/M_{\odot})$=8.9--0.34) & broad H$\alpha$ counterpart in \OIII & -8\\[0.15cm] 
        3<z<5 second lowest stellar mass ($\log(M_{*}/M_{\odot})$=8.52--8.9) & SNR$<3$ & 1.2$\sigma$\\[0.15cm] 
        3<z<5 lowest stellar mass & $\Delta BIC_{\text{broad H}\alpha}$ ($\log(M_{*}/M_{\odot})$=5.97--8.51) & 4\\[0.15cm] 
        5<z<7 second highest \OIII luminosity ($1.5\times 10^{42} - 3.04\times 10^{42}$ erg/s) &  SNR$<3$ & 1.1$\sigma$ \\[0.15cm] 
        5<z<7 second lowest \OIII luminosity ($8.5\times 10^{41} - 1.5\times 10^{42}$ erg/s) & SNR$<3$ & 1.9$\sigma$ \\[0.15cm] 
        5<z<7 lowest \OIII luminosity ($7.85\times 10^{40} - 8.3\times 10^{41}$ erg/s) & SNR$<3$ & 2$\sigma$ \\[0.15cm] 
        5<z<7 second highest \OIII EW (754--1224 Å) & SNR$<3$ & 1.8$\sigma$ \\[0.15cm] 
        5<z<7 second lowest \OIII EW (499--739 Å) & SNR$<3$ & 0.3$\sigma$ \\[0.15cm] 
        5<z<7 lowest \OIII EW (67--487 Å) & broad H$\alpha$ counterpart in \OIII & -2 \\[0.15cm] 
        5<z<7 highest stellar mass ($\log(M_{*}/M_{\odot})$=9.01--10.17) & broad H$\alpha$ counterpart in \OIII & -9\\[0.15cm] 
        5<z<7 second highest stellar mass ($\log(M_{*}/M_{\odot})$=8.47--8.99) & broad H$\alpha$ counterpart in \OIII & 1.5\\[0.15cm] 
        5<z<7 second lowest stellar mass ($\log(M_{*}/M_{\odot})$=8.09--8.45) & SNR$<3$ & 2$\sigma$ \\[0.15cm] 
        5<z<7 lowest stellar mass ($\log(M_{*}/M_{\odot})$=7.01--8.08) & broad H$\alpha$ counterpart in \OIII & -55\\[0.15cm] 
        3<z<5 second-highest $M_{\rm UV}$ (-18.8-- -19.2) & SNR$<3$ & 2.1$\sigma$ \\[0.15cm] 
        3<z<5 second-lowest $M_{\rm UV}$ (-18.25-- -18.8)& SNR$<3$ & 1$\sigma$ \\[0.15cm] 
        3<z<5 lowest $M_{\rm UV}$ (-16.0-- -18.25)& SNR$<3$ & 0.1$\sigma$ \\[0.15cm]
        5<z<7 highest $M_{\rm UV}$ (-19.6-- -20.8)& SNR$<3$ & 2.2$\sigma$ \\[0.15cm]
        5<z<7 second-highest $M_{\rm UV}$ (-19.13-- -19.6)& SNR$<3$ & 1.6$\sigma$ \\[0.15cm]
        5<z<7 second-lowest $M_{\rm UV}$ (-18.57-- -19.06)& SNR$<3$ & 1.2$\sigma$ \\[0.15cm]
        5<z<7 lowest $M_{\rm UV}$ (-15.87-- -18.57)& broad H$\alpha$ counterpart in \OIII & -15\\[0.15cm] 
        
        \hline
    \end{tabular}%
    }
\end{table*}

\section{Deriving black hole masses using other stacking methods}
\label{appendix:masses}
It is important to note that normalisation and weighting methods make the recovery of the intrinsic flux of the stack more difficult and uncertain, and this information is critically needed to infer the average black hole mass and luminosity. But since it is unclear how to treat the flux in this case, we decide to compare the BH masses to see how consistent they are with our results so far. Table \ref{tab:comparing_masses} compares the black hole masses from each of the three stacking methods. The masses from the normalised stacks are all within ~0.2 dex of the masses from the unweighted and un-normalised stacks. The highest stellar mass 3<z<5 stack and the highest M$_{\rm UV}$ 3<z<5 stacks have their broad components ruled out in these stacks. These results are expected because as described in section \ref{appendix:appendix_a}, the broad components in the stacks are dominated by the sources with the brightest \OIII emission. Since most of the stacks with detected broad components are the brightest \OIII luminosity or EW, when they are normalised by the bright \OIII sources the difference is small since most of the sources in the stack have bright \OIII. The highest stellar mass and M$_{\rm UV}$ are more affected because there are likely fewer bright \OIII sources so when they are weighted down the broad component is diminished. 

The masses derived from the inverse variance weighted stacks are smaller than the those derived from the stacks that are not weighted, and as described in Section \ref{appendix:appendix_a}, some of them no longer have a detected broad H$\alpha$ component. This is also expected, because as we showed in Figure \ref{fig:rms_comparison} the overall flux is diminished around the emission line regions in the inverse variance weighted stacks, causing less flux from the broad component, thus smaller masses. Despite this, the masses are within 0.5 dex of the masses from the unweighted stacks therefore our results would not change significantly if we were to adopt the inverse variance weighted stacks for our analysis rather than the unweighted.

\begin{table*}
    \centering
    \caption{Comparing the BH masses derived from three stacking methods: unweighted and not normalised, normalised by \OIII flux, and weighted by inverse variance. Column 1: stack. Column 2: The $\log(M_{\rm BH}/M_{\odot})$ of the unweighted and unnormalised stack. Column 3: The $\log(M_{\rm BH}/M_{\odot})$ of the stack normalised by \OIII flux. Column 4: The $\log(M_{\rm BH}/M_{\odot})$ of the stack weighted by inverse variance. The lines with dashes indicate that these stacks do not have a detected broad H$\alpha$ component.}
    \label{tab:comparing_masses}
    \resizebox{1\textwidth}{!}{%
    \begin{tabular}{lccr} 
        \hline
	Stack & Unweighted, not normalised & Normalised by \OIII flux & Weighted by inverse variance\\
	\hline
	3<z<5 highest \OIII luminosity & $6.48^{+0.36}_{-0.34}$ & $6.54^{+0.36}_{-0.35}$ & $6.08^{+0.36}_{-0.34}$\\[0.15cm] 
	3<z<5 second-highest \OIII luminosity & $6.46^{+0.37}_{-0.36}$ & $6.24^{+0.39}_{-0.40}$ & --\\[0.15cm] 
	3<z<5 lowest \OIII EW & $6.48^{+0.36}_{-0.35}$ & $6.53^{+0.38}_{-0.37}$ & --\\[0.15cm] 
        3<z<5 highest stellar mass & $6.58^{+0.36}_{-0.35}$ & -- & $6.10^{+0.35}_{-0.34}$\\[0.15cm] 
        5<z<7 highest \OIII luminosity & $6.35^{+0.37}_{-0.36}$ & $6.38^{+0.36}_{-0.36}$ & --\\[0.15cm] 
        5<z<7 highest \OIII EW & $6.30^{+0.37}_{-0.36}$ & $6.24^{+0.39}_{-0.37}$ & --\\[0.15cm] 
        3<z<5 highest M$_{\rm UV}$ & $6.26^{+0.35}_{-0.35}$ & -- & $6.06^{+0.38}_{-0.38}$\\[0.15cm] 
	\hline
    \end{tabular}}
\end{table*}


\bsp	
\label{lastpage}
\end{document}